\documentclass[trackchanges,twocolumn]{aastex7}

\usepackage{longtable}
\usepackage{multirow} 
\usepackage{makecell}
\usepackage{xcolor}
\usepackage{ae,aecompl}
\usepackage{graphicx}	
\usepackage{amsmath}	
\usepackage{amssymb}	
\usepackage{color,xcolor}
\usepackage{multirow}
\usepackage{makecell}
\usepackage{caption}
\usepackage{xparse}
\usepackage{ulem}
\usepackage{placeins}  
\usepackage{subcaption}
\usepackage{booktabs}
\usepackage{orcidlink}   
\begin{document}

\title{\bf Tracking Protostellar Variability in Massive Protoclusters with ALMA: I. Insights from QUARKS and MaMMOtH}


\correspondingauthor{Yuhan Yang; Tie Liu; Sheng-Yuan Liu}
\email{yangyuhan@shao.ac.cn; liutie@shao.ac.cn; syliu@asiaa.sinica.edu.tw}

\author[orcid=0009-0000-0178-7472]{Yuhan Yang}
\affiliation{Shanghai Astronomical Observatory, Chinese Academy of Sciences, No.80 Nandan Road, Shanghai 200030, People's Republic of China}
\affiliation{School of Astronomy and Space Sciences, University of Chinese Academy of Sciences,\\
No.19A Yuquan Road, Beijing 100049, People's Republic of China}
\email{yangyuhan@shao.ac.cn}

\author[orcid=0000-0002-5286-2564]{Tie Liu} 
\affiliation{State Key Laboratory of Radio Astronomy and Technology, Shanghai Astronomical Observatory, Chinese Academy of Sciences, \\
80 Nandan Road, Shanghai 200030, People's Republic of China}
\email{liutie@shao.ac.cn}

\author[orcid=0000-0003-4603-7119]{Sheng-Yuan Liu}
\affiliation{Institute of Astronomy and Astrophysics, Academia Sinica, 11F of ASMAB, AS/NTU No.1, Sec. 4, Roosevelt Road, Taipei 10617, Taiwan}
\email{syliu@asiaa.sinica.edu.tw}

\author[0000-0002-6773-459X]{Doug Johnstone}
\affiliation{NRC Herzberg Astronomy and Astrophysics, 5071 West Saanich Rd, Victoria, BC, V9E 2E7, Canada}
\affiliation{Department of Physics and Astronomy, University of Victoria, Victoria, BC, V8P 5C2, Canada}
\email{doug.johnstone@gmail.com}

\author[0000-0002-7154-6065]{Gregory Herczeg}
\affiliation{Kavli Institute for Astronomy and Astrophysics, Peking University, Yiheyuan Lu 5, Haidian Qu, 100871 Beijing, Peoples Republic of China}
\affiliation{Department of Astronomy, Peking University, Yiheyuan 5, Haidian Qu, 100871 Beijing, China}
\email{gherczeg1@gmail.com}

\author[0000-0001-9822-7817]{Wenyu Jiao}
\affiliation{State Key Laboratory of Radio Astronomy and Technology, Shanghai Astronomical Observatory, Chinese Academy of Sciences, \\
80 Nandan Road, Shanghai 200030, People's Republic of China}
\email{astrojiao@gmail.com}

\author[]{Yu-Nung Su}
\affiliation{Institute of Astronomy and Astrophysics, Academia Sinica, 11F of ASMAB, AS/NTU No.1, Sec. 4, Roosevelt Road, Taipei 10617, Taiwan}
\email{ynsu@asiaa.sinica.edu.tw}

\author[0000-0001-7573-0145]{Xiaofeng Mai}
\affiliation{State Key Laboratory of Radio Astronomy and Technology, Shanghai Astronomical Observatory, Chinese Academy of Sciences, \\
80 Nandan Road, Shanghai 200030, People's Republic of China}
\email{maixf@shao.ac.cn}

\author[0000-0001-5950-1932]{Fengwei Xu}
\affiliation{Max Planck Institute for Astronomy, K\"onigstuhl 17, 69117 Heidelberg, Germany}
\email[]{fengwei@mpia.de, fengweilookuper@gmail.com}

\author[0000-0001-8258-9813]{D.~M.-A.~Meyer}
\affiliation{Institute of Space Sciences (ICE, CSIC), Campus UAB, Carrer de Can Magrans s/n, 08193 Barcelona, Spain}
\email{dmameyer.astro@gmail.com}

\author[0000-0002-9836-0279]{Siju Zhang}
\affiliation{Departamento de Astronom\'ia, Universidad de Chile, Casilla 36-D, Santiago, Chile}
\email{sijuzhangastro@gmail.com}

\author[0000-0002-6045-0359]{Eduard Vorobyov}
\affiliation{Institut für Astro- und Teilchenphysik, Universität Innsbruck, 
Technikerstraße 25, 6020 Innsbruck, Austria}
\affiliation{Research Institute of Physics, Southern Federal University, 
Rostov-on-Don 344090, Russia}
\email{eduard.vorobiev@univie.ac.at}

\author[0000-0002-8389-6695]{Suinan Zhang}
\affiliation{Department of Earth and Planetary Sciences, Institute of Science Tokyo, Meguro, Tokyo, 152-8551, Japan}
\affiliation{National Astronomical Observatory of Japan, 2-21-1 Osawa, Mitaka, Tokyo, 181-8588, Japan}
\email{suinan.zhang@gmail.com}

\author[0000-0003-4506-3171]{Qiuyi Luo}
\affiliation{Institute of Astronomy, Graduate School of Science, The University of Tokyo, 2-21-1 Osawa, Mitaka, Tokyo 181-0015, Japan}
\affiliation{Department of Astronomy, School of Science, The University of Tokyo, 7-3-1 Hongo, Bunkyo, Tokyo 113-0033, Japan}
\email{luoqiuyi233@gmail.com}

\author[0000-0003-1649-7958]{Guido Garay}
\affiliation{Departamento de Astronom\'ia, Universidad de Chile, Casilla 36-D, Santiago, Chile}
\affiliation{Chinese Academy of Sciences South America Center for Astronomy, National Astronomical Observatories, Chinese Academy of Sciences, Beijing, 100101, PR China}
\email{guido@das.uchile.cl}

\author[0000-0002-9836-0279]{Xi Chen}
\affiliation{Center for Astrophysics, Guangzhou University, Guangzhou 510006, People’s Republic of China}
\email{chenxi@gzhu.edu.cn}

\author[0009-0005-9867-6723]{Yunfan Jiao}
\affiliation{Shanghai Astronomical Observatory, Chinese Academy of Sciences, No.80 Nandan Road, Shanghai 200030, People's Republic of China}
\affiliation{School of Astronomy and Space Sciences, University of Chinese Academy of Sciences,\\
No.19A Yuquan Road, Beijing 100049, People's Republic of China}
\email{yunfan.astro@gmail.com}

\author[0000-0002-2826-1902]{Qilao Gu}
\affiliation{State Key Laboratory of Radio Astronomy and Technology, Shanghai Astronomical Observatory, Chinese Academy of Sciences, \\
80 Nandan Road, Shanghai 200030, People's Republic of China}
\email{qlgu@shao.ac.cn}

\author[0000-0001-7817-1975]{Yankun Zhang}
\affiliation{State Key Laboratory of Radio Astronomy and Technology, Shanghai Astronomical Observatory, Chinese Academy of Sciences, \\
80 Nandan Road, Shanghai 200030, People's Republic of China}
\email{zhangyankun@shao.ac.cn}

\author[0000-0002-8149-8546]{{Ken’ichi Tatematsu}}
\affiliation{National Astronomical Observatory of Japan, National Institutes of Natural Sciences, 2-21-1 Osawa, Mitaka, Tokyo 181-8588, Japan}
\affiliation{Astronomical Science Program, Graduate Institute for Advanced Studies, SOKENDAI, 2-21-1 Osawa, Mitaka, Tokyo 181-8588, Japan}
\email{kenichi.tatematsu@nifty.com}

\author[0000-0003-2412-7092]{{Kee-Tae Kim}}
\affiliation{Korea Astronomy and Space Science Institute (KASI), 776 Daedeokdae-ro, Yuseong-gu, Daejeon 34055, Republic of Korea}
\affiliation{University of Science and Technology, Korea (UST), 217 Gajeong-ro, Yuseong-gu, Daejeon 34113, Republic of Korea}
\email{ktkim@kasi.re.kr}

\author[0000-0001-7575-5254]{Andrey M. Sobolev}
\affiliation{State Key Laboratory of Radio Astronomy and Technology, Xinjiang Astronomical Observatory, CAS,
150 Science 1-Street, Urumqi, Xinjiang, 830011, P. R. China}
\email{andrej.sobolev@gmail.com}

\author[]{Sergey Parfenov}
\affiliation{Ural Federal University, 19 Mira Street, 620002 Ekaterinburg, Russia}
\email{sergey.parfenov@urfu.ru}

\author[0000-0002-3773-7116]{Dmitry A. Ladeyschikov}
\affiliation{Ural Federal University, 19 Mira Street, 620002 Ekaterinburg, Russia}
\email{dmitry.ladeyschikov@urfu.ru}

\author[0000-0002-9574-8454]{Leonardo Bronfman}
\affiliation{Departamento de Astronom\'ia, Universidad de Chile, Casilla 36-D, Santiago, Chile}
\email{leo@das.uchile.cl}

\begin{abstract}
Millimeter/Submillimeter variability is often attributed to dynamical disk-mediated accretion yet detection is limited to low-mass protostars in nearby clouds. Recent observations have also revealed significant (sub)millimeter variability in high-mass protostars, but the confirmed cases are scarce and lack systematic monitoring. In this work, we analyzed multi-epoch Atacama Large Millimeter/submillimeter Array (ALMA) Band 6 (1.3\,mm) continuum observations of 22 massive protoclusters, with epoch separations ranging from a few hours to more than two years, while achieving a consistent angular resolution of $\sim\rm 0.3$ \arcsec. These data allow us to track variability of protostars across a broader mass range and in an environment markedly different from nearby clouds. Using a custom processing pipeline of data reduction, image alignment, and relative flux calibration, we achieve high-precision flux measurements and, for the first time, investigate millimeter variability in massive protoclusters using interferometric data from the statistical point of view. Applying the {\it astrodendro} algorithm, we identified 383 condensations and tracked their variations in their peak intensities. Standard deviation analysis and difference maps reveal five variable sources, corresponding to a lower limit of 1.3\% on the variable fraction. Among these, I13111--6228 stands out as it hosts a hypercompact H{\sc ii} region that an increase of approximately 68\% in continuum peak intensity over one year, with an uncertainty of 2\%. 


\end{abstract}

\keywords{stars: formation --- stars: protostars --- (sub)millimeter: stars}


\section{Introduction} 
\label{sec:intro}

Understanding how protostars gain the bulk of their mass during the main accretion phase while they are still deeply embedded in their natal envelopes remains a fundamental challenge in star formation (e.g. \citealt{2009Evans,2018Kristensen}). Early estimates of protostellar accretion rates, inferred from luminosity under the assumption of steady accretion, were insufficient to explain the observed stellar masses within the expected formation timescales (approximately 0.1 Myr; \citealt{1990Kenyon,2010Dunham}). Although extended formation timescales and improved sensitivity have partially alleviated this ``luminosity problem'' (e.g., \citealt{2011Offner,2023Fischer}), the observed protostellar luminosities still vary by 3--4 orders of magnitude. This wide range has motivated a growing interest in time-variable accretion as a key mechanism for mass assembly. 

For low-mass young stellar objects, temporal photometric variability has been observed not only in the optical and infrared toward FU Orionis and EX Lupi-type objects in the optically revealed stage \citep{2014Audard,2025OYCAT,2025Contreras}, but also in the (sub)millimeter wavelengths at much earlier, embedded (Class 0, Class I) protostellar phases \citep{2015Safron,2024Mairs,2025Sheehan,2025Laznevoi}. These outburst phenomena can be attributed to various gravitational instabilities in circumstellar disks, and, in particular, to gravitational instability in which the disk fragments into dense clumps that subsequently migrate inward and trigger episodic accretion onto the central protostar. Numerical simulations suggest such variable accretion events occurs not only in low-mass star formation (e.g., \citealt{2005Vorobyov,2011Machida,2012Nayakshin,2015Vorobyov,2016Hosokawa}) but also in the high-mass star regime (e.g., \citealt{2017Meyer,2018Meyer}). Further studies suggest that massive protostars also undergo episodic accretion events that produce substantial luminosity bursts, and may contribute up to $\sim60\%$ of their zero-age main-sequence mass, with burst durations ranging from several years to nearly a century \citep[e.g.,][]{2019Meyer,2021Elbakyan}. 

Massive protoclusters, which are forming compact clusters of protostars including high-mass ones with characteristic sizes of $0.1$--$1$\,pc \citep[e.g.,][]{2007Cyganowski,2013Palau,2014Monge,2024Xu}, are promising targets to study such variability in environments markedly different from nearby clouds. Their members typically span a wide range of evolutionary stages, from starless cores through hot molecular cores to ultracompact H{\sc ii} regions \citep[e.g.,][]{2017ApJ...849...25L,2021MNRAS.505.2801L,2025ApJS..280...33Y}, providing an ideal laboratory for investigating accretion-driven variability across different stages of massive star formation.

Recent observations have directly confirmed powerful luminosity outbursts in massive star-forming regions, widely interpreted as episodes of enhanced accretion.
The earliest confirmed examples are S255IR-NIRS3 
\citep{2017Caratti, 2018Cesaroni, 2018Szymczak, 2019Liu, 2020Liu, 2020Uchiyama} 
and NGC~6334I-MM1 
\citep{2017Hunter,2018Hunter,2021Hunter,2018Brogan,2018MacLeod,2021Chibueze}. Multi-wavelength observations showed that the continuum emission from millimeter to infrared wavelengths increased by a factor of 5.5 and 16.3. These events are interpreted as episodic accretion bursts of the internal massive young stellar objects (MYSOs), which are likely triggered by disk fragmentation, followed by inward migration of the clumps and their tidal destruction. \citep[e.g.,][]{2017Meyer,2021Elbakyan}. Such behavior is consistent with the theoretical burst mode of accretion proposed for massive star formation \citep{2021Meyer,2022Meyer}. 

Notably, both the S255IR-NIRS3 and NGC 6334I-MM1 outbursts were accompanied by flaring of the 6.7 GHz methanol (CH$_3$OH) maser \citep{2015Fujisawa,2018Szymczak,2018MacLeod,2019Brogan}. Since Class II CH$_3$OH masers, including the 6.7~GHz transition, are radiatively pumped by far-infrared photons \citep{2003Minier,2008Xu}, the observed maser flaring indicates an increase in the far-infrared thermal radiation field within the envelope surrounding the MYSO, implying enhanced heating of the circumstellar material. Adding to this evidence are confirmed cases like G358.93-0.03~MM1 \citep{2021Stecklum,2020Burns,2023Burns}, G24.33+0.14 \citep{2022Hirota}, G323.46-0.08 \citep{2024Wolf}, and M17~MIR \citep{2021Chen,2024Zhou,2025Chen}. 

These findings further support the idea that episodic accretion may play a crucial role in the formation of massive stars. By assembling a substantial fraction of stellar mass in short-lived accretion bursts, this process modulates radiative feedback and the thermal and ionization structure of the circumstellar environment, producing the large luminosity spread and characteristic multi-wavelength variability observed in massive protostars. Despite their increasing number, accretion bursts in massive protostars remain rare and largely serendipitous, hindering efforts to statistically constrain the nature of episodic accretion in the high-mass regime. Therefore, systematic, long-term monitoring is essential to understand their occurrence rate, properties, and driving mechanisms.

The JCMT Transient Survey \citep{2017Herczeg, 2018Johnstone,2024Mairs} observed six fields targeting at intermediate- to high-mass star-forming regions. However, the relatively large beam size imposes limitations on detailed variability monitoring in such distant regions \citep{2025Chen,2024Park,Wang_prep,Zhang_prep}. Previous systematic interferometric studies have investigated (sub)millimeter variability in nearby low-mass star-forming regions using facilities such as the Submillimeter Array (SMA), the Combined Array for Research in Millimeter-wave Astronomy (CARMA), the NOrthern Extended Millimeter Array (NOEMA) and ALMA \citep[e.g.,][]{2018Liu,2019Francis,2022Francis,2020Wendeborn,2023Vargas}, demonstrating the feasibility of interferometric monitoring of variability. Yet such efforts remain confined to nearby molecular clouds, and a systematic interferometric search for variability across multiple massive protoclusters has not yet been undertaken.

In this paper, we present the first systematic investigation of continuum variability in massive protoclusters at 1.3 mm wavelength, based on high-resolution interferometric data obtained from two recent ALMA projects: the Querying Underlying mechanisms of massive star formation with ALMA-Resolved gas Kinematics and Structures (QUARKS) survey \citep{Liu2024QUARKS-I} and the Massive Star-Forming Regions with Variable Methanol Masers: Observations at High Angular Resolution (MaMMOtH) survey \citep{Liu_prep}. Our analysis includes multi-epoch observations of 22 distant massive protoclusters. Each source has observations from at least two epochs separated by more than one year, while across the full dataset the epoch separations range from only a few hours to more than two years. Specifically, this work expands upon previous studies of (sub)millimeter continuum variability by searching for variable sources across a wider range of masses within more massive, active, and complex star-forming environments. 

\begin{figure*}[bht!]
    \centering
    \includegraphics[angle=0, width=0.85\textwidth]{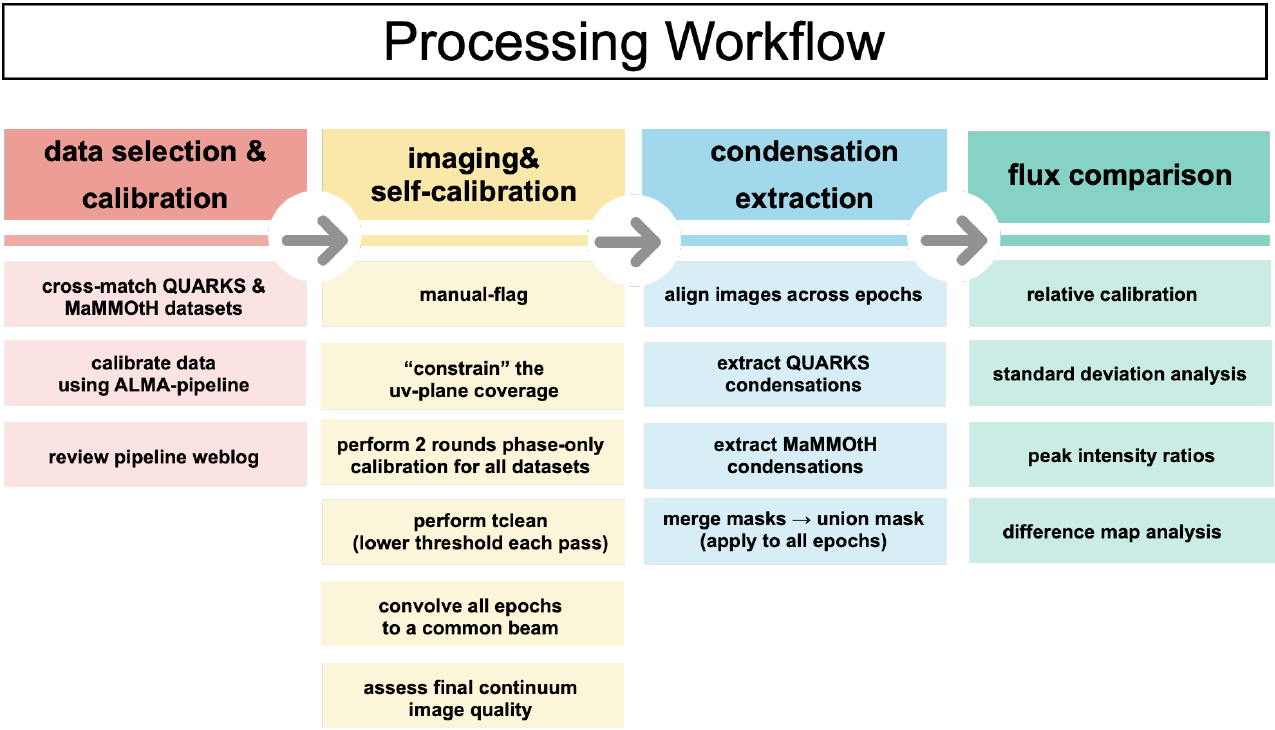}  
    \caption{
    Processing workflow. Schematic overview of the four main processing steps adopted in this study. The process begins with data selection and calibration, including dataset cross-matching and ALMA pipeline calibration to prepare the data for further steps (Sect.~\ref{sec:obs}). In the second step, imaging and self-calibration, we perform spectral Line flagging and self-calibration (Sect.~\ref{sub:data_flag_selfcalibration}), followed by smoothing all epochs of each source to a common beam to produce consistent continuum images across epochs (Sect.~\ref{sub:continuum_imaging}). In the third step, condensation extraction, involves aligning images across epochs, identifying compact condensations in both datasets, and applying a unified mask for consistent source extraction (Sect.~\ref{sub:core_extraction}). Finally, in flux comparison, we perform relative calibration (Sect.~\ref{sub:relative_calibration}), track the peak intensity evolution of each condensation across epochs (Sect.~\ref{sub:variables_analysis}), and evaluate variability using both the standard deviation method (Sect.~\ref{sub:SD_analysis}) and the difference map analysis (Sect.~\ref{sub:difference_map}).
    }
    \label{fig:workflow}    
\end{figure*}


The paper is structured as follows. In Sect.~\ref{sec:obs}, we describe source selection and present the sample. We also provide details of the corresponding ALMA observations. In Sect.~\ref{sec:data_reduction_methods}, we outline each step of the data reduction process for the multi-epoch continuum maps. This includes calibration, identification of line-free channels for continuum determination, self-calibration, and smoothing all maps to a common synthesized beam to ensure consistent angular resolution across epochs. In Sect.~\ref{sec:results_and_analysis}, we present the procedures for source identification and flux extraction across multiple epochs. We also describe the relative flux calibration methods applied to ensure consistency, followed by an analysis of the resulting millimeter variability. In Sect.~\ref{sec:discussion}, we discuss the results, outline the current limitations of this study, and present our plans for future investigations. Finally, in Sect.~\ref{sec:Conclusion}, we summarize our main findings and conclusions. 

\section{Observations and Sample Selections} 
\label{sec:obs}
A schematic overview of the full processing workflow is provided in Figure~\ref{fig:workflow}. This study uses data from two independent ALMA projects: QUARKS and MaMMOtH. The QUARKS Survey (Project IDs: 2019.1.00685.S and 2021.1.00095.S; PIs: Lei Zhu, Guido Garay, and Tie Liu) investigates 139 massive star-forming clumps that have IRAS colors similar to those of ultracompact H\,\textsc{ii} regions, using ALMA Band~6 observations ($\lambda \approx 1.3$\,mm) across 156 pointings. The QUARKS observations began in late October 2021 using both the ALMA 12-m array configurations (C-2 and C-5) and the Atacama Compact Array (ACA) 7-m antennas. The ACA observations were completed in late May 2022 \citep{2024Xu}, followed by C-2 and C-5 executions that continued through June 2024. The temporal coverage and per-epoch sensitivity of the QUARKS observations are sufficient to enable internal variability comparisons within the survey. As a follow-up to the ALMA Three-millimeter Observations of Massive Star-forming regions survey \citep[ATOMS;][]{Liu2020ATOMS-I}, QUARKS aims to statistically characterize key processes of star formation (e.g., fragmentation, outflows, disks) within an unbiased sample of protoclusters. Detailed descriptions of survey design, target selection, and observation strategy are provided in \citet{Liu2024QUARKS-I}.

The MaMMOtH Survey (Project IDs: 2021.1.00311.S and 2022.1.00974.S; PI: Sheng-Yuan Liu) targets 169 massive star-forming regions associated with Class~II CH$_3$OH masers. The sample is primarily drawn from two well-established catalogs of monitoring observations of 6.7~GHz methanol masers \citep{2004Goedhart, 2018Szymczak}, and is supplemented by two additional well-known maser sources, G352.630-1.067 \citep{2019Chen} and G353.273$+$0.641 \citep{2013Motogi}, based on previous detections of Class~II CH$_3$OH maser activity. Observations were conducted using ALMA Band~6 with the 12-m array configurations (C-4 and C-5), complemented by Band~7 observations with the ACA. The ACA observations began in October 2021 and were completed in mid-May 2023, while the C-4 executions were carried out from April 2022 to April 2024, followed by the C-5 observations that continued from May 2022 through June 2024. The primary goal of the MaMMOtH Survey is to establish a statistical baseline for the study of millimeter continuum variability in massive star-forming regions associated with 6.7 GHz methanol masers, and to examine its connection to the physical and chemical conditions of the star-forming environment. The survey description, including full observational details, will be introduced in Liu et al. (in prep.).

The ALMA Band~6 receivers in dual-polarization mode were used to conduct both the QUARKS and MaMMOtH surveys. For the QUARKS survey, four spectral windows (SPWs) were configured, each with a bandwidth of 1.875\,GHz and a velocity resolution of $\sim1.3\,\mathrm{km\,s^{-1}}$. The central frequencies of the SPWs were set at approximately 217.92, 220.32, 231.37, and 233.52\,GHz. Detailed information on the spectral setup and targeted molecular lines can be found in Table~2 of \citet{Liu2024QUARKS-I}. For the MaMMOtH survey, four spectral windows were similarly configured, each with a bandwidth of 1.875\,GHz and a velocity resolution of $\sim1.3\,\mathrm{km\,s^{-1}}$. The central frequencies of the SPWs were set at approximately 217.63, 220.00, 231.05, and 232.87\,GHz.

Benefiting from the comparable frequency coverage and matched observational setups (ALMA C-5 and $\sim 0.3\arcsec$ resolution) of the two projects, we constructed a multi-epoch sample of 22 massive protoclusters from the QUARKS and MaMMOtH surveys. We cross-matched the MaMMOtH and QUARKS source lists by requiring that their pointing centers differ by less than 60\arcsec, and this yielded 22 massive protoclusters with overlapping coverage. Among these cross-matched sources, eight massive protoclusters have at least two epochs from the two surveys that are separated by more than one year. In addition, the QUARKS survey alone provides 14 sources with two epochs separated by more than one year. Taken together, these 22 massive protoclusters constitute the largest sample to date of sources with observational epochs separated by more than one year across the two projects. Additional shorter-interval epochs are included when the observational setups were consistent. The molecular gas reservoirs of these protoclusters range from 68 to 7585 $M_{\odot}$, and they are located at distances of 1.4 to 11.6 kpc \citep{Liu2024QUARKS-I,2024Xu}. A summary of the observational parameters, including observing dates, calibrators, and baseline ranges, is provided in Appendix~\ref{appendix:observing_parameters}.

\section{Data Reduction Methods}
\label{sec:data_reduction_methods}

\subsection{Spectral Line Flagging and Self-calibration}
\label{sub:data_flag_selfcalibration}

The data were initially calibrated using the ALMA pipeline within the Common Astronomy Software Applications package (CASA, version~6.5.4.9; \citealt{2022PASP..134k4501C}). This process included bandpass, gain, and flux calibration, employing the calibrators listed in Table~\ref{appendix:observing_parameters}. The output of this procedure consists of calibrated measurement sets for each individual execution blocks. Each calibrated measurement set was subsequently split into individual files corresponding to the science targets, using the four scientific SPWs. For each source, the four SPWs were combined to improve the continuum sensitivity.

Line emission frequency ranges were flagged to clearly separate continuum emission from spectral line features. To ensure consistency across multiple epochs and surveys, a uniform flagging strategy was adopted for the spectral line frequency ranges. Specifically, prominent spectral line features within the four SPWs were first identified in a reference epoch based on visual inspection of the visibility data. The flagged frequency ranges were then manually adjusted to include additional line emission present in other epochs but absent in the reference epoch. This procedure was repeated iteratively until all visible spectral line features were consistently flagged across all epochs.

\begin{figure*}[ht!]
    \centering
    \includegraphics[angle=0, width=0.75\textwidth]{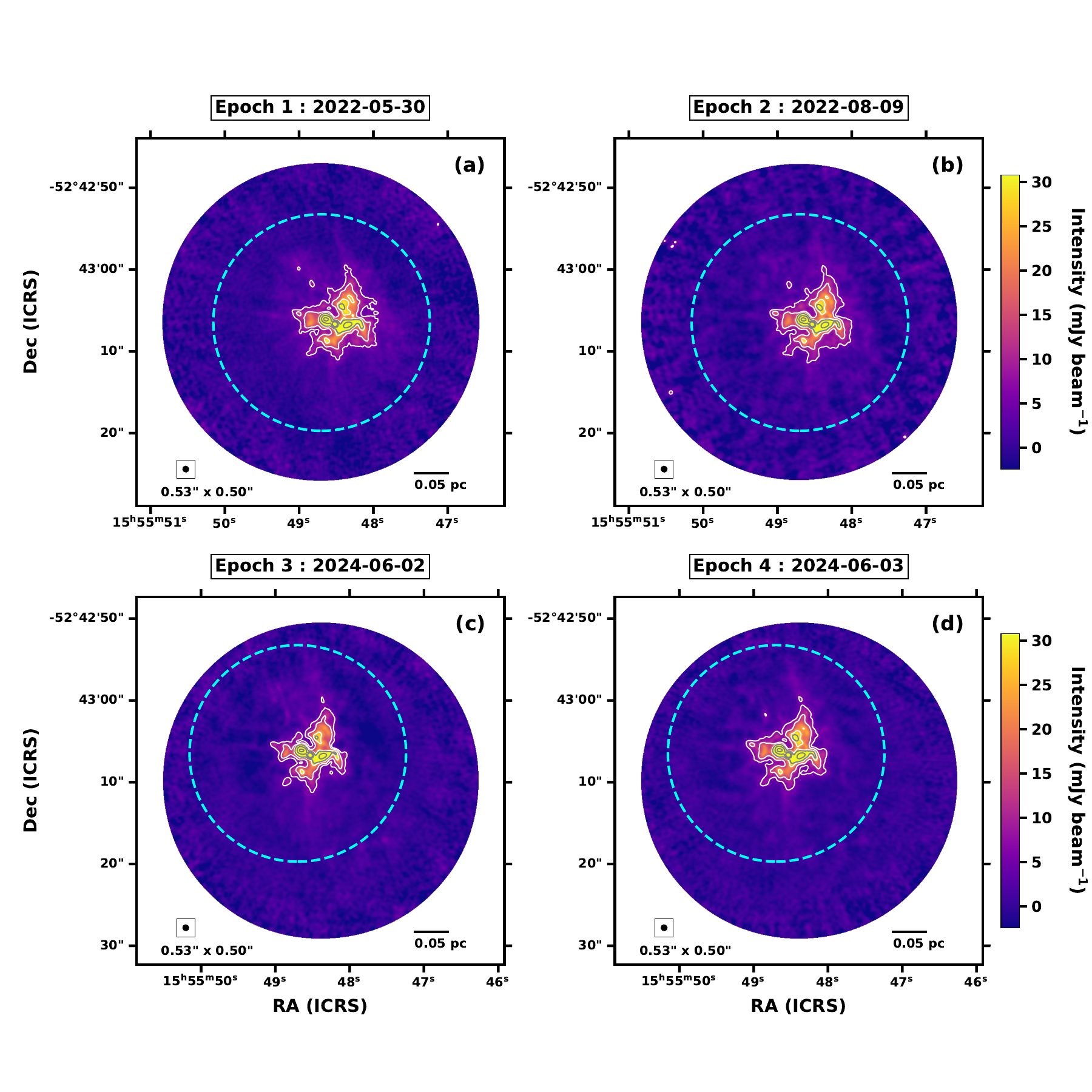}  
    \caption{
    ALMA Band 6 continuum images of I15520--5234 obtained from four epochs spanning over two years (with the observing dates indicated in the figure). The cyan dashed circle in each panel marks the 0.5 primary beam FWHM region of the MaMMOtH survey ($\sim$13.24$\arcsec$), which is also used as the reference area for image alignment. Upper panels (a, b): MaMMOtH 1.3\,mm continuum emission. Lower panels (c, d): QUARKS 1.3\,mm continuum emission. For consistency, all contour levels in the four panels are defined using the rms noise level measured from the first-epoch image (2022~May~30), where $\sigma_{\rm rms} = 1.23~\mathrm{mJy~beam^{-1}}$. White contours show [5, 10, 20]$\times\sigma_{\rm rms}$, while gray contours show [25, 50, 80]$\times\sigma_{\rm rms}$. The synthesized beam of each epoch is shown in the lower-left corner of each panel, and a 0.05\,pc scale bar is indicated in the lower-right corner.
    }

    \label{fig:continuum}    
\end{figure*}

Self-calibration is a technique used to correct visibility phases and/or amplitudes by comparing the observed visibilities with a model of the source itself (e.g., \citealt{selfcal}). Following standard interferometric guidelines (e.g., \citealt{1999ASPC..180.....T}), phase-only self-calibration is applicable when the target is detected with a signal-to-noise ratio (S/N) > 3 within a solution interval shorter than the timescale of significant phase variation across all baselines to a single antenna. Given the brightness of our targets, self-calibration was applied to the full dataset to enhance image quality and dynamic range. Two rounds of phase-only self-calibration were performed: the first with \textit{solint = ``inf''} and the second with \textit{solint = ``int''}. After each iteration, deeper cleaning was performed to refine the source model.

\subsection{Continuum Imaging}
\label{sub:continuum_imaging}

To minimize discrepancies in \textit{uv}-plane sampling arising from differences in ALMA configurations and the Earth’s rotation among epochs, and to ensure reliable flux comparisons, we constrained the \texttt{uvrange} parameter during imaging within the \textit{tclean} task. Based on the \texttt{Amplitude v.s. UVWave} distributions of each epoch, we determined the common overlapping \textit{uv}-range for each source by selecting the maximum shared \textit{uv} coverage across all available epochs. The same \textit{uv}-range was used across epochs for a given source, but it could differ between sources. This procedure matches the baseline coverage in the \textit{uv}-plane, thereby improving the cross-epoch consistency of the images. Our method provides a simple solution for the current data. Previous studies \citep[e.g.,][]{2019Francis,2020Francis,2022Francis} suggest that a more careful consideration of the \textit{uv}-plane would be necessary in certain contexts, such as when the differences in array configuration between epochs are significant.

\begin{figure*}[!ht]
    \centering
    \includegraphics[angle=0, width=0.8\textwidth]{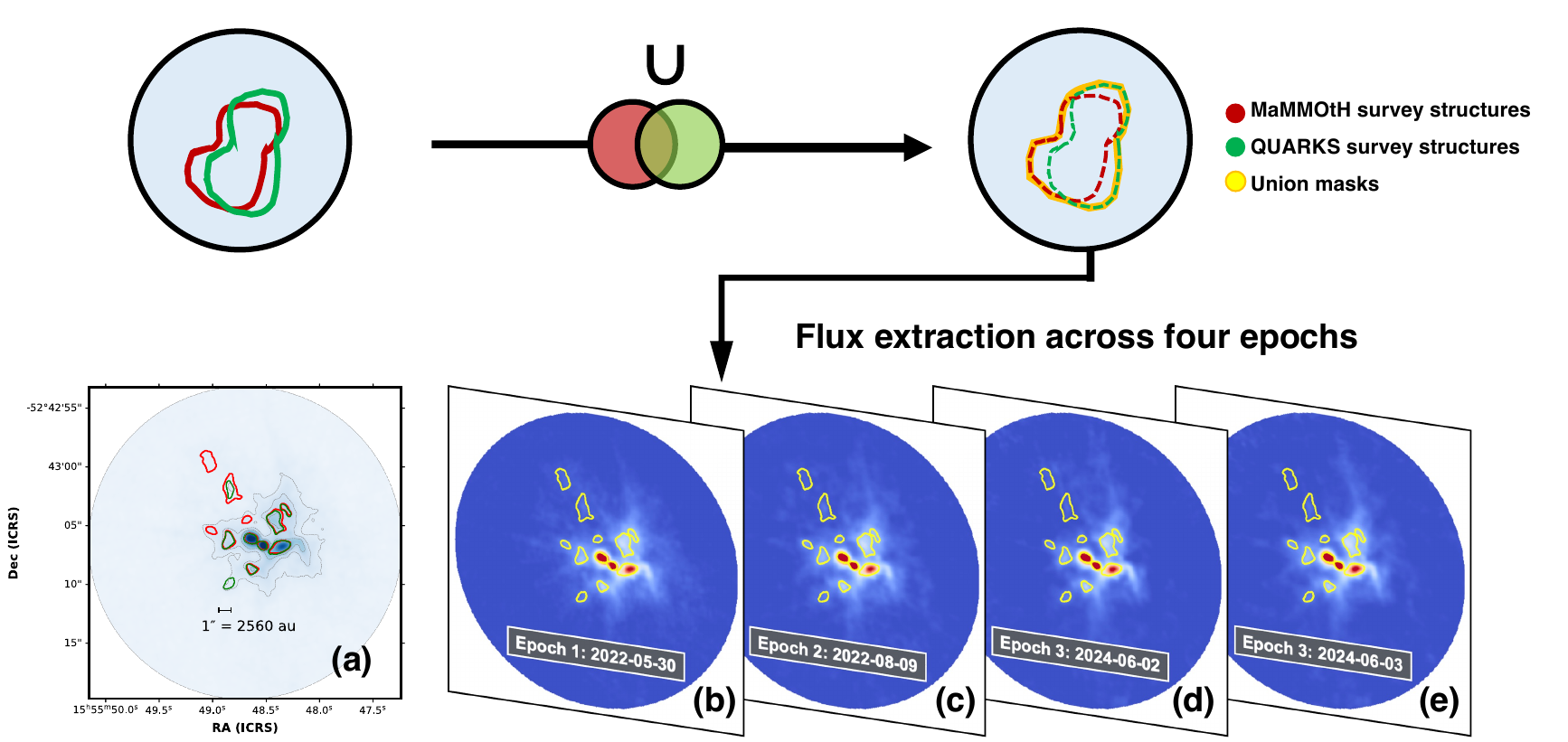}  
    \caption{
    \textit{Top:} Cartoon illustrating the extraction of structures and their fluxes. 
    Red and green contours represent condensations independently identified in the MaMMOtH and QUARKS surveys using the \texttt{astrodendro} algorithm. 
    The union of these two sets defines a common mask, shown in yellow, which is consistently applied across all observing epochs. \textit{Bottom:} Application to I15520--5234.  (a) 1.3\,mm continuum image before primary beam correction. The dashed black contours indicate levels of [10, 24, 48, 96]$\times\sigma_{\rm rms}$, where $\sigma_{\rm rms}=0.58~\mathrm{mJy~beam^{-1}}$. Red and green contours trace condensations identified independently in the MaMMOtH and QUARKS datasets. (b–e) Continuum images after primary beam correction from 2022~May~30, 2022~August~09, 2024~June~02, and 2024~June~03. The yellow contour traces the union mask used to extract peak intensities for the flux comparison analysis in Sect.~\ref{sub:SD_analysis}. 
    }
    \label{fig:core}    
\end{figure*}

Initial continuum images for each epoch were generated using the \textit{tclean} task in CASA 6.6.1.17 with \texttt{briggs} robust weighting of 0.5. The Multi-scale Multi-Frequency Synthesis (\textit{mtmfs}) deconvolution algorithm was adopted with \textit{nterm} of 2. This algorithm improves wideband imaging by simultaneously modeling both spectral and spatial structures, restoring extended emission, and enhancing overall image fidelity. \setcounter{footnote}{0} During the cleaning process, masks were automatically generated using the \texttt{auto-multithresh} algorithm, with input parameters recommended by the guides\footnote{\url{https://casaguides.nrao.edu/index.php/Automasking_Guide_CASA_6.6.1}} for the 12-m array (C-5). The image size was set to [900, 900] pixels with a pixel scale of 0.05\arcsec, and the \textit{pblimit} parameter was set to 0.2.

An iterative cleaning strategy was adopted, involving three successive runs of \textit{tclean} with progressively lower threshold values. The initial \texttt{threshold} was set to approximately five times the rms of the dirty image. After the first round of cleaning and phase-only self-calibration, the \texttt{threshold} was reduced to about three times the rms. Following the second round of cleaning and the second phase-only self-calibration, the final \texttt{threshold} was set to 2–3 times the rms.

A standardized beam-matching and image-smoothing procedure was applied to ensure uniform angular resolution for sources observed across multiple epochs and surveys. After imaging, a common restoring beam was computed using the \texttt{radio\_beam}\footnote{\url{https://radio-beam.readthedocs.io/en/latest/}} package. Each image was subsequently convolved to the common beam using the \texttt{imsmooth} task. The target beam was matched to the largest synthesized beam among the continuum images. The resulting values of $\theta_{\mathrm{conv}}$ range from $\sim 0.27^{\prime\prime}$ to $0.68^{\prime\prime}$. This procedure ensures uniform angular resolution across epochs, while the use of a slightly larger common beam reduces noise in individual maps and facilitates consistent measurements of flux variability. In total, 56 continuum images were generated for 22 massive protoclusters, which were subsequently used for the epoch-by-epoch flux analysis. 

\section{Results and Analysis}
\label{sec:results_and_analysis} 
\subsection{Source Extraction Strategy}
\label{sub:core_extraction}

To ensure consistent and reliable comparisons of continuum emission across epochs, we developed a unified source extraction strategy that includes image alignment, structure identification, and flux measurement. Figure~\ref{fig:continuum} presents the after primary beam correction continuum images of the source I15520--5234 at four epochs. This target, observed over a span of more than two years, is among the most comprehensively sampled sources in our study and serves as a representative example of the source extraction.

As a first step, we visually inspected the continuum images using the CARTA image viewer\footnote{CARTA: Cube Analysis and Rendering Tool for Astronomy, \url{https://cartavis.org/}}. This visual inspection was used only to account for differences in the phase center settings between the QUARKS and MaMMOtH surveys and to define a common cutout region that encompasses all major source structures across epochs, rather than for the actual image alignment. Taking the first epoch as a reference, we identified the morphological center of the source and defined a circular region centered at this position with an initial radius equal to 0.5 times the primary beam FWHM ($\sim$13.24$\arcsec$). We verified that this region fully enclosed the source emission in all epochs; if not, the radius was incrementally increased until full coverage was achieved. In this case, the initial radius was sufficient, as indicated by the cyan circle in Figure~\ref{fig:continuum}. All images were then cropped to this common region using the \texttt{Cutout2D}\footnote{\url{https://docs.astropy.org/en/stable/api/astropy.nddata.Cutout2D.html}} package, ensuring a consistent field of view across epochs.

During the construction of difference maps (Sect.~\ref{sub:difference_map}), systematic relative offsets between epochs were identified, which manifested as positive–negative residual patterns (see Figure~\ref{fig:alignment}). For observations within the same survey, the offsets are more plausibly associated with variations in phase calibration quality between epochs. This motivated the use of a quantitative image alignment procedure. Image alignment was performed using a sub-pixel phase cross-correlation method. For each source, the first epoch was selected as the reference frame, and relative positional shifts between the reference image and subsequent epochs were measured using a phase cross-correlation algorithm implemented in the \texttt{scikit-image}\footnote{\url{https://scikit-image.org/docs/0.25.x/api/skimage.registration.html}} package. An upsampling factor of 100 was adopted, corresponding to a positional accuracy of $\sim$0.01 pixel. The measured shifts were then applied via spline interpolation to align all images to the common reference frame. The results are summarized in Table~\ref{tab:offsets}, which lists the offsets $\Delta x$ and $\Delta y$ (in pixel units) relative to the reference epoch. Difference maps were subsequently inspected to confirm that source structures were well matched and that no systematic offsets remained.

Figure\,\ref{fig:core} illustrates our source extraction procedure, taking I15520--5234 as an example. To avoid elevated edge noise introduced by primary beam correction, structure identification was performed on the continuum images uncorrected for the primary beam. Structures were extracted independently for the MaMMOtH and QUARKS datasets. For each survey, multiple observations of the same source were combined to improve sensitivity. We used the {\it astrodendro} \footnote{\url{http://www.dendrograms.org/}} algorithm to decompose the emission hierarchically. The highest hierarchical level is a ``leaf" (i.e., a structure with no substructure), corresponding to what we define as a condensation. The three key  parameters are \textit{min\_value} (the minimum pixel intensity to be considered), \textit{min\_delta} (the minimum height for any local maximum to be defined as an independent entity), and \textit{min\_npix} (the minimum number of pixels for a leaf to be defined as an independent entity). To ensure consistency across the two surveys, we adopted a uniform set of parameters: \textit{min\_value} = $5\,\sigma_{\mathrm{rms}}$, \textit{min\_delta} = $2$–$3\,\sigma_{\mathrm{rms}}$, and \textit{min\_npix} equals the number of pixels corresponding to the beam area. Here, $\sigma_{\mathrm{rms}}$ denotes the rms noise level of the continuum image. For multi-epoch sources, we averaged the images prior to structure identification and determined $\sigma_{\mathrm{rms}}$ from the averaged image before primary-beam correction. In the example shown in Figure\,\ref{fig:core}, we used \textit{min\_delta} = $3\,\sigma_{\mathrm{rms}}$, with $\sigma_{\mathrm{rms}}= 1.02\,\mathrm{mJy~beam^{-1}}$ for MaMMOtH and $= 1.07\,\mathrm{mJy~beam^{-1}}$ for QUARKS, and \textit{min\_npix} = 119.

In Figure~\ref{fig:core}, structures extracted from the MaMMOtH and QUARKS datasets are shown with red and green contours. Although the condensations detected in the two surveys do not always match, all structures from both datasets are retained in the final union mask, shown in yellow contour. We constructed this final mask by taking the union of the individual masks of identified regions in different epochs. This union mask was then applied to each epoch for flux extraction. To ensure that the measured fluxes represent the true emission distribution, this step was performed after primary beam correction. We identified a total of 383 condensations within the 22 protoclusters. The peak intensities and coordinates of all condensations are summarized in Appendix~\ref{appendix:peakintensity_parameters}, where the source names indicate the corresponding region (i.e., the massive protocluster) in which each condensation is located. The peak intensity is the peak value within the dendrogram contour, which is a good approximation of the total integrated intensity for point sources. The complete table is available online. For each protocluster, condensation IDs are assigned starting from 1 and are ordered according to their spatial distribution in the map images, following a right-to-left and bottom-to-top sequence.

\subsection{Relative Calibration}
\label{sub:relative_calibration}
After aligning the observations and extracting the sources, we derived and applied a relative flux calibration factor to each dataset in order to accurately track the peak intensity variations of a given object across all epochs. This approach allows for more robust measurements of intrinsic variability within each field. Our procedure generally follows the method outlined by \citet{2017Mairs}. The relative flux calibration procedure consists of the following four steps:

\begin{figure}[ht!]
    \centering
    \begin{subfigure}{\linewidth}
        \centering
        \includegraphics[width=0.9\linewidth]{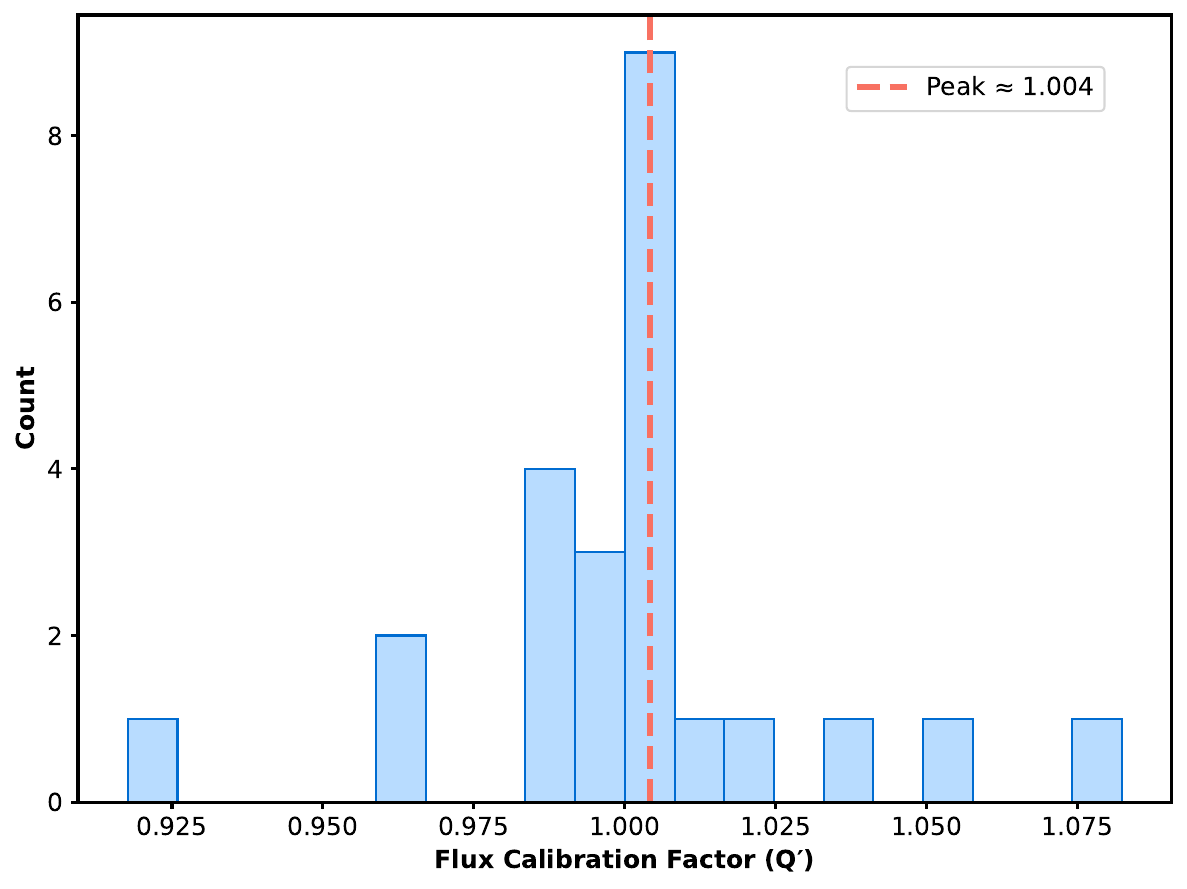}
    \end{subfigure}
    \vskip\baselineskip
    \begin{subfigure}{\linewidth}
        \centering
        \includegraphics[width=0.9\linewidth]{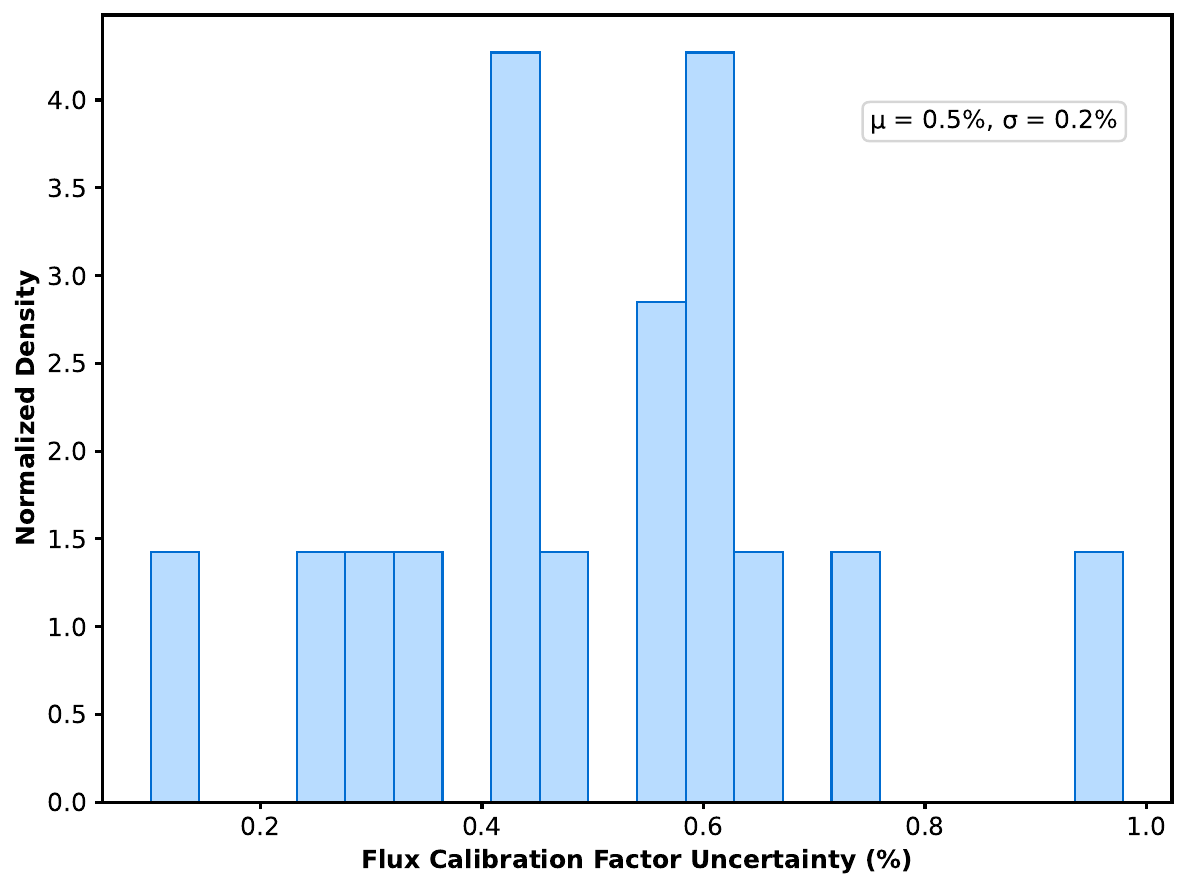}
    \end{subfigure}
    \caption{Relative flux calibration overview. Top: distribution of flux calibration factors. Bottom: corresponding uncertainties.}
    \label{fig:Qfig}
\end{figure}

\begin{enumerate}
    \item 
    We first divided the 22 protoclusters into six groups based on their observation dates. Protoclusters were assigned to the same group only if they were observed on exactly the same set of dates, and each group shared the same set of ALMA calibrator sources. As a result, a group may contain multiple protoclusters, while each protocluster belongs to only one group. This grouping ensures consistent temporal sampling and enables consistent relative flux calibration across all epochs within each group.
    
    \item
    At different observing epochs, even for the same field or source, the measured continuum flux may vary due to non-astrophysical effects such as changes in instrumental response, atmospheric conditions, and calibration uncertainties. To mitigate these systematic effects, we identify a subset of sources that can serve as relative flux calibrators. These sources are assumed to have intrinsically constant (or nearly constant) emission over the timescales probed by our observations. Specifically, within each group, we selected candidate calibrator sources with $\mathrm{S/N} > 30$ in all available epochs, ensuring that their measured peak intensities are robust against noise fluctuations. For each candidate, we then calculated the mean peak intensity over the valid epochs and normalized the peak intensity of each epoch by this mean. We adopted a threshold of 0.1 for this normalized standard deviation as a compromise between retaining a sufficient number of calibrator candidates and ensuring calibration reliability. Sources with normalized standard deviations below this threshold were classified as stable and used as relative flux calibrators.

    \item 
    We then construct a normalized peak intensity ratio defined as:
    \begin{equation}
    r_i^{(s)} = \frac{I_{\max,i}^{(s)}}{\langle I_{\max}^{(s)} \rangle_t},
    \end{equation}
    where $I_{\max,i}^{(s)}$ is the peak intensity of source $s$ measured within the dendrogram contour at epoch $i$, and $\langle I_{\max}^{(s)} \rangle_t$ denotes the mean over time of the peak intensity of that source, averaged over all valid epochs in the group. The flux calibration factor for each epoch $i$ was then obtained by averaging the mean flux normalized ratios across all stable calibrators in the group:
    \begin{equation}
    Q_i' = \left\langle r_i^{(s)} \right\rangle.
    \end{equation}
    This procedure assumes that, although individual calibrator sources may exhibit small residual fluctuations, their ensemble-averaged flux remains constant over time, allowing robust estimation of systematic calibration offsets. We note, however, with a limited number of calibrator sources in each group, the uncertainty on the ensemble mean may still represent a significant source of error.

    \item 
    The derived flux calibration factors \( Q_i' \) spanned a range from 0.918 to 1.082. These factors were applied to all sources observed in the corresponding epoch \( i \) within the same group, and are shown in the top panel of Figure~\ref{fig:Qfig}. Specifically, for each epoch $i$, we divided both the peak intensity and the rms values of all sources by $Q_i'$. This operation harmonized the flux scale across epochs while preserving the relative S/N. The bottom panel of Figure~\ref{fig:Qfig} presents the distribution of the corresponding flux calibration uncertainties. The mean calibration uncertainty was $\mu = 0.5\%$ with a standard deviation of $\sigma = 0.2\%$. These results indicate that the relative flux calibration achieved better than 1\% accuracy and remained stable across all epochs.
\end{enumerate}

\subsection{Searching for (Sub)Millimeter Variables}
\label{sub:variables_analysis}
Investigating flux variability in our survey is confronted by several key challenges. Each ALMA target was observed at only a small number of epochs (2–4 per protocluster), and the background rms varies across different protoclusters. To address these limitations, we employed two complementary methods to cross-validate variability among our samples (383 condensations). First, we adopted the methodology outlined by \citet{2018Johnstone}, originally developed for the JCMT Transient Survey. Although that survey benefits from more extensive temporal coverage and highly uniform observing conditions, which result in greater statistical robustness, our use of consistent data reduction and relative flux calibration still allows us to adopt the same framework as a practical indicator for identifying candidate variable sources. Second, to further increase the reliability of our results, we applied a difference map analysis to independently verify variability in the identified candidates.

\subsubsection{Standard Deviation Analysis}
\label{sub:SD_analysis}
For each detected condensation, we extract the peak intensity measured in each epoch and compute two key statistical quantities: the mean of the peak intensity over time, $\langle I_{\max} \rangle_t$, and the corresponding standard deviation ($\mathrm{SD}$). The uncertainty in measuring the peak intensity of these sources is primarily influenced by two factors: the background rms for faint condensations, relative calibration accuracy for bright condensations. We defined the fiducial standard deviation for each source, \( \mathrm{SD}_{\mathrm{fid}}(i) \), which represents the expected measurement uncertainty in peak intensity, as:

\begin{equation}
\mathrm{SD}_{\mathrm{fid}}(i) =
\sqrt{
\sigma_{\mathrm{rms}, i}^2 +
\left( \varepsilon_{\mathrm{cal}} \cdot \langle I_{\max} \rangle_t \right)^2
},
\label{eq:SDfid}
\end{equation}

where \( \sigma_{\mathrm{rms}, i} \) denotes the background rms associated with the source \( i \). The term \( \varepsilon_{\mathrm{cal}} \) represents the expected relative flux calibration uncertainty, which we adopt to be 0.5\% (see Sect.~\ref{sub:relative_calibration} for details), while \( \langle I_{\max} \rangle_t \) denotes the time-averaged peak intensity of the source.

\begin{figure}[tb]
    \centering
    \includegraphics[angle=0, width=0.45\textwidth]{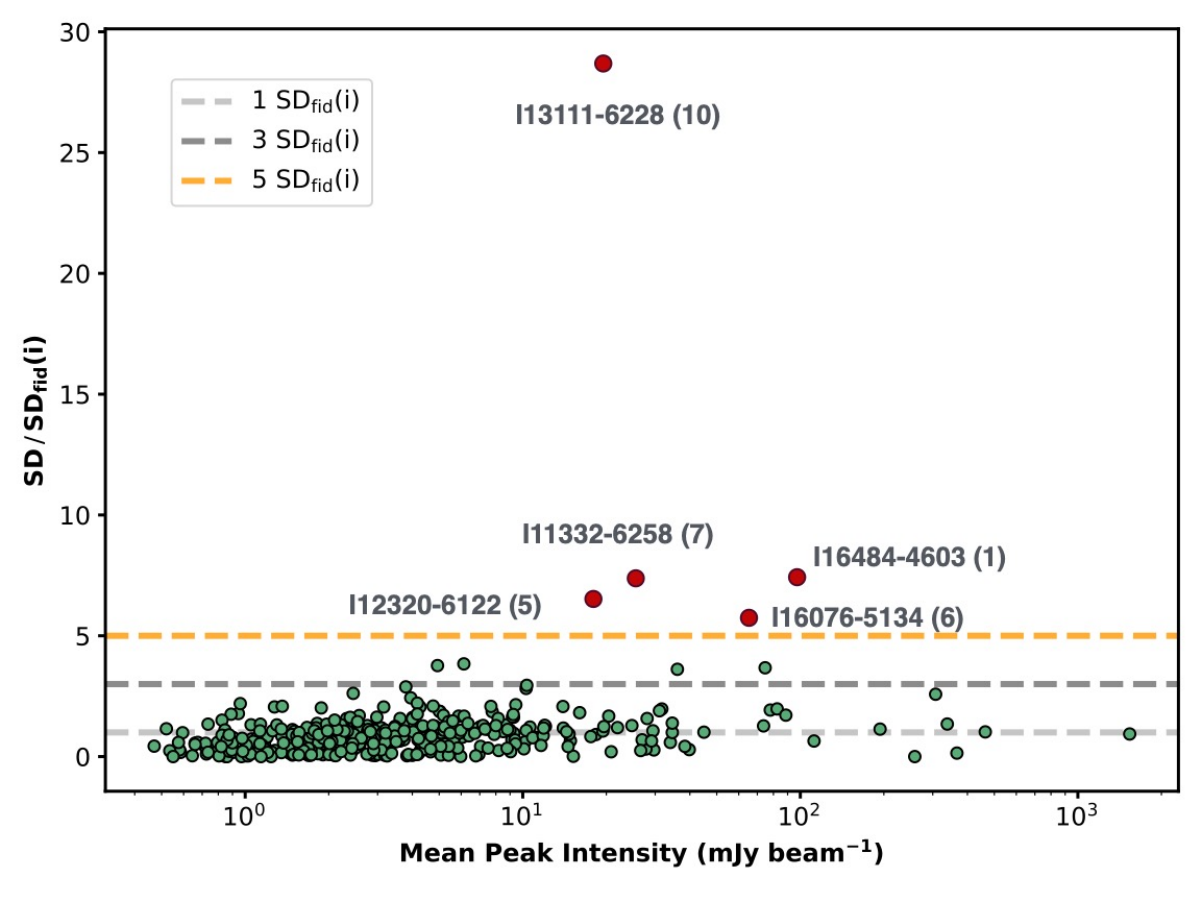} 
    \caption
    {Normalized standard deviation of peak intensity ($\mathrm{SD}/\mathrm{SD}_{\mathrm{fid}}$) as a function of the time-averaged mean peak intensity for condensations across 22 protoclusters. Horizontal dashed lines indicate levels at 1, 3, and 5 times the fiducial expectation ($\mathrm{SD}/\mathrm{SD}_{\mathrm{fid}} = 1, 3, 5$). Sources exceeding the 5 times fiducial level are highlighted in red and flagged as candidate variables. The numbers in parentheses next to the source names indicate the condensation IDs, consistent with those listed in Table~\ref{table:variable} and Table~\ref{table:epoch_data}.}
    \label{fig:SDfig}    
\end{figure}

Figure~\ref{fig:SDfig} shows the normalized standard deviation ($\mathrm{SD}/\mathrm{SD}_{\mathrm{fid}}$) as a function of the mean peak intensity for each condensation. In this work, we adopt a conservative variability threshold of $\mathrm{SD}/\mathrm{SD}_{\mathrm{fid}} > 5$, beyond which deviations are unlikely to result from calibration uncertainties or stochastic noise. The JCMT Transient Survey employed a lower threshold of 2 \citep{2018Johnstone, 2025Chen}; however, given our smaller number of epochs, we adopted a more stringent criterion to minimize false positives and ensure that only the most statistically significant outliers are flagged as candidate variables.

\setlength{\tabcolsep}{1.5pt}
\begin{deluxetable*}{cccccccccc}
\tabletypesize{\scriptsize}
\tablewidth{0pt}
\tablecaption{Candidate Variables and Their Statistical Variability \label{table:variable}}
\tablehead{
\colhead{Source Name} &
\colhead{Condensation ID$^a$} &
\colhead{SD/SD$_{\mathrm{fid}}$} &
\multicolumn{3}{c}{Fractional Difference (\%)$^b$} &
\colhead{$\langle I_{\max} \rangle_t$} &
\colhead{Fractional Amplitude (\%)} &
\colhead{Type$^c$} &
\colhead{$D$ $^d$}
\\
\cline{4-6}
\colhead{} &
\colhead{} &
\colhead{} &
\colhead{Epoch 2-1} &
\colhead{Epoch 3-1} &
\colhead{Epoch 4-1} &
\colhead{(mJy\,beam$^{-1}$)} &
\colhead{} &
\colhead{} &
\colhead{(kpc)}
}
\colnumbers
\startdata
I11332-6258 & 7  & 7.4  & 11.7 & 14.4 & \nodata & 25.55 & 13.3 & hot core        & 1.40 \\
I12320-6122 & 5  & 6.5  & 16.2 & \nodata & \nodata & 17.97 & 15.0 & young protostar       & 4.17 \\
I13111-6228 & 10 & 28.7 & 68.3 & \nodata & \nodata & 19.51 & 51.1 & HC H{\sc ii} region & 2.97 \\
I16076-5134 & 6  & 5.7  & -0.8 & 7.8   & 5.3     & 65.33 & 8.3  & hot core        & 5.31 \\
I16484-4603 & 1  & 7.4  & -7.3 & -9.3  & -10.4   & 97.37 & 11.1 & hot core        & 2.17 \\
\enddata
\tablecomments{Peak intensities are used after applying the relative calibration.}
\tablenotetext{a}{Corresponds to the ``Condensation ID'' entries listed in Table~\ref{table:epoch_data}.}
\tablenotetext{b}{Corresponds to the ``Obs.~Date'' entries listed in Table~\ref{tab:almaobs}.}
\tablenotetext{c}{See Sect.~\ref{sub:variables_properties} for a discussion of the source classifications.}
\tablenotetext{d}{Distances are adopted from \citet{Liu2024QUARKS-I}.}
\end{deluxetable*}

Applying this criterion, we identify five candidate variables among the 383 detections with multi-epoch measurements. Their quantitative variability metrics are summarized in Table\,\ref{table:variable}. 

The fractional difference is calculated as $(I_{\max,i} - I_{\max,j}) / I_{\max,j}$, where $I_{\max,i}$ and $I_{\max,j}$ are the peak intensities at two different epochs. Multiple pairs of epochs (e.g., 2--1, 3--1, and 4--1) are used to evaluate the variability across the monitoring period. The observed fractional amplitude is defined as $(\max_t(I_{\max}) - \min_t(I_{\max})) / \langle I_{\max} \rangle_t$, where $\max_t(I_{\max})$ and $\min_t(I_{\max})$ denote the maximum and minimum peak intensities measured across all observing epochs, and $\langle I_{\max} \rangle_t$ is the mean peak intensity over time.

Among them, condensation 10 in I13111--6228 is particularly notable, exhibiting a normalized standard deviation of $\mathrm{SD}/\mathrm{SD}_{\mathrm{fid}} \approx 28.7$, far above the adopted threshold. This is the first detection of millimeter variability toward this protocluster. The other four candidates show more moderate excess deviations, with $\mathrm{SD}/\mathrm{SD}_{\mathrm{fid}}$ ranging from $\sim$5.7 to 7.4.

\begin{figure}[tb]
    \centering
    \includegraphics[angle=0, width=0.45\textwidth]{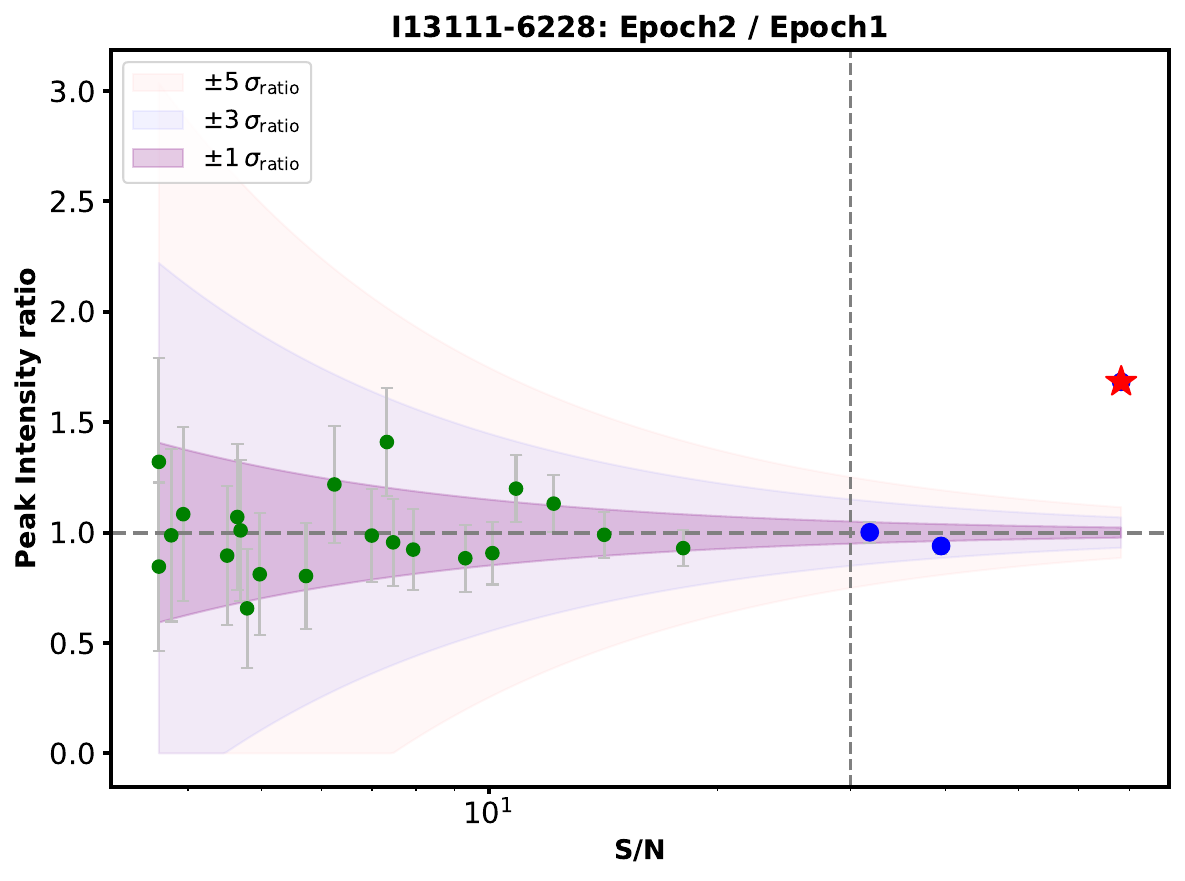} 
    \caption{Peak intensity ratio versus S/N at the reference epoch for I13111--6228. Each point represents an individual condensation extracted from the I13111--6228. The shaded regions denote $\pm1$, $\pm3$, and $\pm5\sigma_{\mathrm{ratio}}$. Points with S/N greater than 30 are shown in blue, while those with S/N less than 30 are shown in green. The red star marks a deviation beyond $5\sigma_{\mathrm{ratio}}$.}
    \label{fig:ratiofig}    
\end{figure}

Although the fiducial standard deviation analysis provides a global identification of candidate variables, it does not explicitly capture the epoch-to-epoch flux behavior of individual condensations. To complement this approach, we extend the same fiducial standard deviation model by propagating uncertainties and computing the peak intensity ratio between epochs, referenced to epoch~1 for consistency (see Appendix~\ref{appendix:ratio_method}). We find that the variable candidates identified with this ratio-based method are fully consistent with those selected from the fiducial standard deviation analysis.

As an illustration, Figure~\ref{fig:ratiofig} shows the source I13111--6228, in which condensation~10 exhibits the most significant intensity variation among all candidates. This condensation is clearly offset from the fiducial model, well exceeding the $5\sigma_{\mathrm{ratio}}$ threshold, and is marked with a red star. For reference, the points with $\mathrm{S/N}<30$ are colored green, reflecting their larger noise-dominated uncertainties, while those with higher S/N are shown in blue and provide more reliable measurements. Additional ratio plots for representative candidates are shown in Appendix~\ref{appendix:ratio_method}, and the complete set of figures is available online.

\begin{figure*}[tb]
    \centering
    \includegraphics[angle=0, width=1\textwidth]{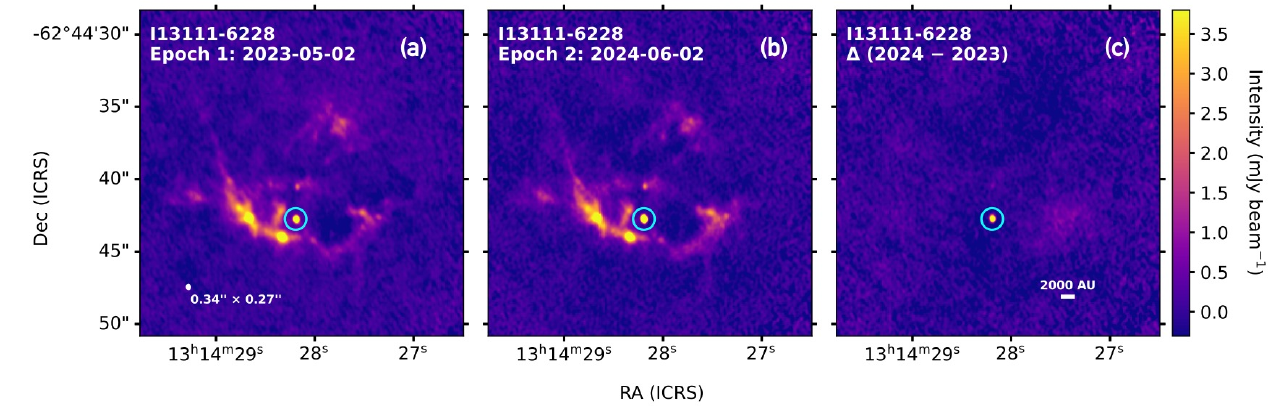} 
    \caption{
    1.3\,mm continuum images of I13111--6228 observed with ALMA in the QUARKS survey at two epochs, along with their difference map. (a) Image from Epoch~1 observed on 2023~May~02. (b) Image from Epoch~2 observed on 2024~June~02. (c) Difference map between panels~(a) and~(b), produced after aligning the two images for visual comparison. The cyan circle marks a radius of 0.75\arcsec\ centered on the residual peak in the difference map. All three panels share the same color scale and the same synthesized beam of 0.34\arcsec\ $\times$ 0.27\arcsec, which is shown as the white filled ellipse in the lower left corner. The rms noise level in the difference map is $\sigma_{\rm rms} =0.22~\mathrm{mJy~beam^{-1}}$, and the residual peak intensity reaches $10.73~\mathrm{mJy~beam^{-1}}$, yielding S/N $\approx 48$.
    }
    \label{fig:image_I13111_6228}    
\end{figure*}

\subsubsection{Difference Maps Analysis}
\label{sub:difference_map}
To further distinguish genuine variability among the five candidate variables identified by statistical analysis, we performed a difference map analysis. For each candidate variables, we measured the peak residual intensity at the location of the condensation in the difference map. A candidate is considered a robust variable if the residual peak exceeds a significance threshold of five times the rms noise level and the residual emission is spatially compact and coincident with the source position. For the five candidates in this study, the S/N of the residual peaks ranges from $\sim$ 6 to 48, indicating that all of them are genuine variable sources.

Here we present the results for I13111--6228, the most significant case in our sample. The difference maps for the remaining four variable sources are shown in Appendix~\ref{appendix:difference_map}. Panels (a) and (b) of Figure~\ref{fig:image_I13111_6228} show the 1.3\,mm continuum emission toward I13111--6228 observed with ALMA in May 2023 and June 2024. The two datasets were processed with identical calibration and imaging procedures to ensure a fair comparison. Panel (c) presents the difference map formed by subtracting panel (a) from panel (b). A compact residual feature is detected at the position of condensation~10, while no significant residuals appear toward the other condensations. The peak intensity of condensation~10 increases from $14.54\,\rm mJy\,beam^{-1}$ in 2023 to $ 24.47\,\rm mJy\,beam^{-1}$ in 2024 after relative calibration, corresponding to a $\sim$ 68\% rise with an uncertainty of 2\%. The rms noise level of the difference map is $0.22\,\rm mJy\,beam^{-1}$, and the residual peak reaches $10.73\,\rm mJy\,beam^{-1}$ (S/N~$\approx$~48). The large, isolated residual confirms that condensation~10 in I13111--6228 is a genuinely variable source.

\section{Discussion} 
\label{sec:discussion} 
As detailed in Sect.~\ref{sub:variables_analysis}, we identified five variable sources from the total sample of 383 condensations by applying a conservative variability metric to their peak intensities. Among these, I13111--6228 stands out, exhibiting a 68\% increase in continuum peak intensity over one year, with an uncertainty of 2\%. The identification is based on a standard deviation analysis and is further confirmed through continuum image difference maps, from which the variability amplitudes are quantified. In the following, we briefly discuss the implications of these findings and outline directions for future work.

\subsection{Detection Fraction and Sampling Biases}
\label{sub:detection_rate}
In this work, we derive a millimeter continuum variability detection fraction of $\sim$1.3\% within our ALMA sample. This value should be interpreted as a lower limit on the intrinsic incidence of continuum variability, as it is strongly dependent on observational factors such as temporal sampling, sensitivity, angular resolution, and sample selection.

The sample analyzed in this work is drawn from two surveys with distinct target selection strategies. The QUARKS survey was designed to investigate star formation in a broadly unbiased sample of infrared-bright massive protoclusters that mostly host UC H{\sc ii} region candidates, whereas the MaMMOtH survey specifically targets regions associated with 6.7\,GHz methanol masers to study millimeter continuum variability. Consequently, the sample considered in this work is biased toward massive star-forming systems that are more likely to host actively accreting protostars or UC H{\sc ii} regions.

The detection fraction is measured from 383 compact condensations identified across 22 massive protoclusters observed with ALMA Band~6 (1.3\,mm) at an angular resolution of $\sim0.3\arcsec$. Fourteen protoclusters were observed by both QUARKS and MaMMOtH, with eight additional regions observed exclusively by QUARKS. The protoclusters span molecular gas masses of $\sim$68--7585$M_{\odot}$ and distances from 1.4 to 11.6 kpc \citep{Liu2024QUARKS-I, 2024Xu}; see Table~\ref{tab:almaobs}.

\begin{figure}[!tb]
    \centering
    \includegraphics[angle=0, width=0.45\textwidth]{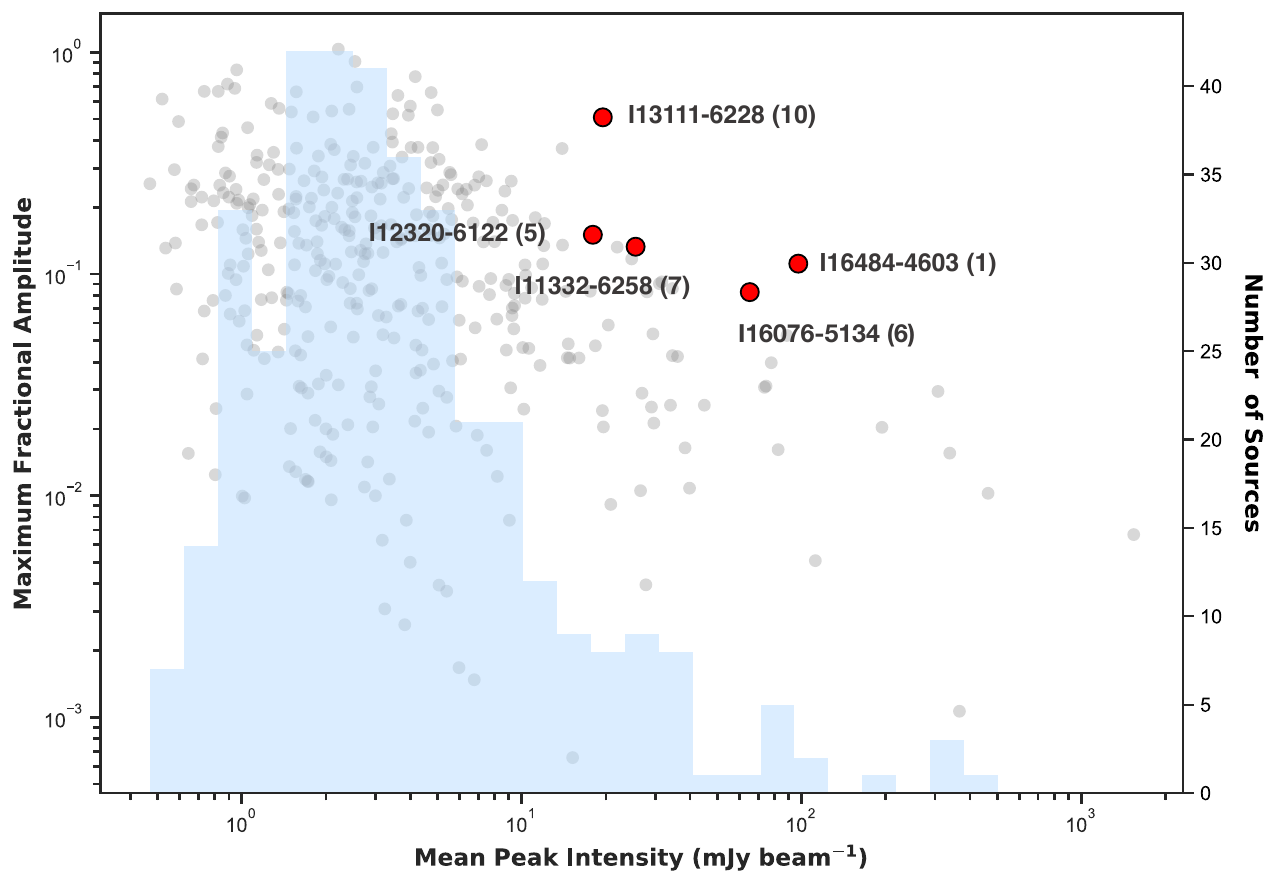} 
    \caption{Scatter plot of the fractional amplitude as a function of the mean peak intensity for all condensations with valid multi-epoch measurements. Each point represents a single condensation, where the mean peak intensity is calculated by averaging the peak intensities over all available observing epochs for that source. The background histogram shows the distribution of the source-averaged peak intensity, with one entry per source. Gray points denote individual condensations, while the red circles highlight the detected variable sources. The numbers in parentheses next to the source names indicate the condensation IDs, consistent with those listed in Table~\ref{table:variable} and Table~\ref{table:epoch_data}.}
    \label{fig:fractional_amp}    
\end{figure}

All sources have at least two observing epochs separated by more than one year. Temporal sampling represents a further limitation: most condensations are observed in only two epochs, and at most four, which is insufficient to trace long-term variability. As a result, the intrinsic timescales of continuum variability cannot be robustly constrained with the current data.

Figure~\ref{fig:fractional_amp} presents the fractional amplitude as a function of the mean peak intensity for all condensations with valid multi-epoch millimeter measurements. Each point represents a single condensation, where the mean peak intensity is calculated by averaging the peak intensities over all available observing epochs. The background histogram shows the distribution of this source-averaged mean peak intensity across the full sample, providing context for the underlying brightness distribution. All five detected variable sources have source-averaged mean peak intensities exceeding 15~mJy~beam$^{-1}$ and are located among the brightest condensations in our sample.

The JCMT Transient Survey provides a useful point of reference for assessing the impact of observational strategy on variability detection. It has monitored submillimeter continuum variability since 2015 with a typical monthly cadence. Focusing primarily on nearby ($\lesssim$500~pc) Gould Belt clouds dominated by low-mass star formation, the survey reports that more than 30\% of bright sub-millimeter continuum sources (peak brightness $\gtrsim$0.5~Jy~beam$^{-1}$ with a beam size of 14.6\arcsec) exhibit significant secular variability at 850~$\mu$m \citep{2021ApJ...920..119L}.

Overall, the observed 1.3\% detection fraction primarily reflects observational limitations and sample selection effects. It should therefore be regarded as an observationally defined lower limit on the incidence of millimeter continuum variability under the conditions of this study. Therefore, long-term monitoring observations as the JCMT Transient Survey did are needed to more quantitatively evaluate the variability of protostars within massive proto-clusters.

\subsection{Classification of Variable Sources}
\label{sub:variables_properties}

\begin{figure}[!tb]
    \centering
    \includegraphics[angle=0, width=0.45\textwidth]{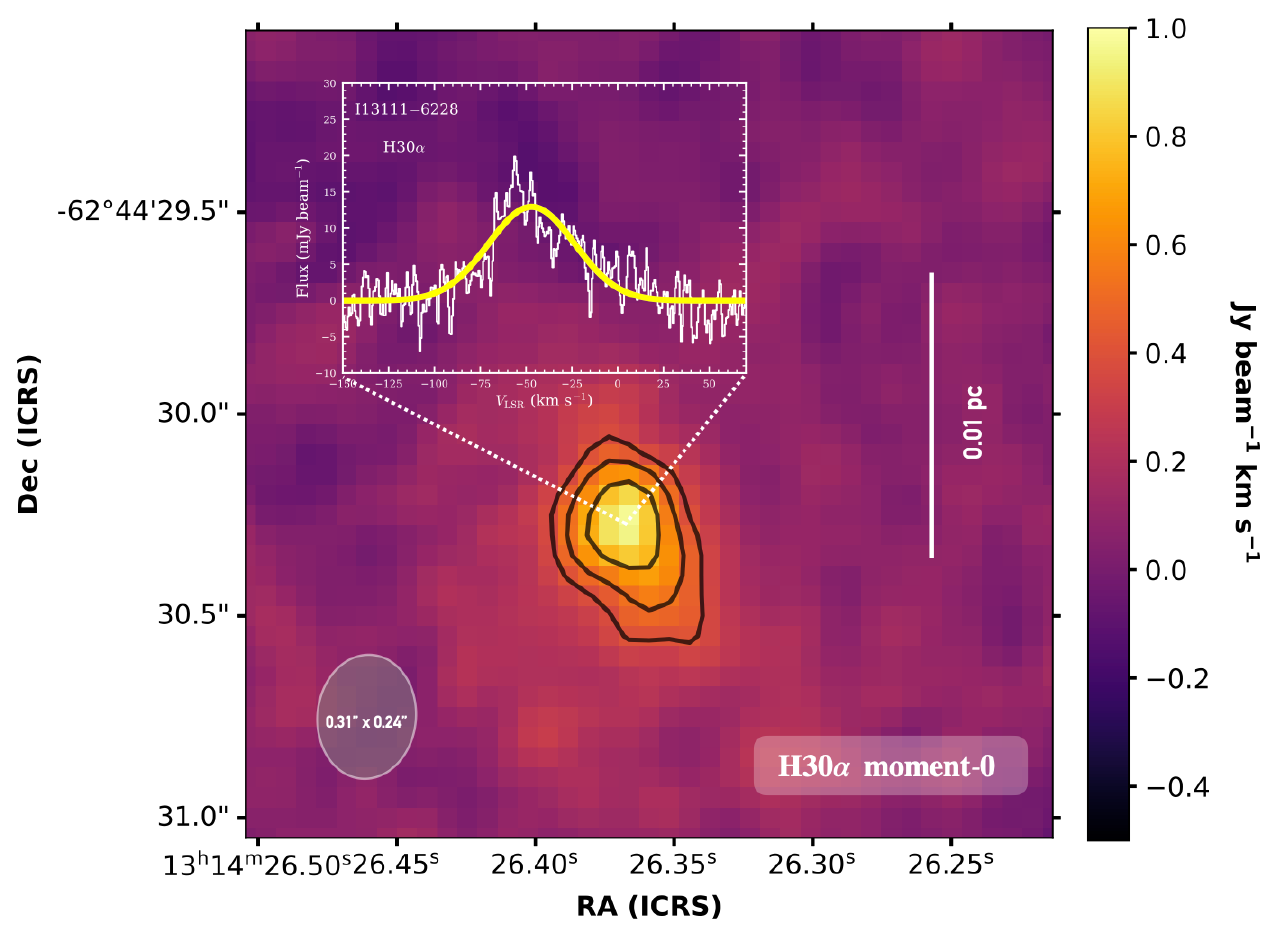} 
    \caption{
    Integrated intensity map of H30$\alpha$ emission toward condensation 10 in I13111--6228, with the corresponding spectrum shown in the inset. The black contours are at levels of [3, 4, 5]$\times\sigma_{\rm rms}$, where $\sigma_{\rm rms} = 0.13~\mathrm{Jy~beam^{-1}~km~s^{-1}}$. 
    The synthesized beam is shown in the lower-left corner, and a 0.01\,pc scale bar is indicated toward the right side of the panel.
    The inset shows the H30$\alpha$ spectrum extracted at the peak position, with a Gaussian fit overlaid in yellow.
    }
    \label{fig:H30alpha}    
\end{figure}

To examine the physical nature of the detected variable sources, we used molecular line emission from four spectral windows in the cleanest dataset from the QUARKS survey, combining continuum data from C-5 (TM1, $\sim0.3$\,\arcsec), C-2 (TM2, $\sim1$\,\arcsec), and ACA 7-m array ($\sim5$\,\arcsec). Condensations exhibiting H30$\alpha$ emission are classified as H\textsc{ii} regions, while those without detectable H30$\alpha$ but showing strong CH$_3$CN emission are classified as hot cores. Hypercompact (HC) H{\sc ii} regions are characterized by extremely compact sizes ($d \sim 0.03$\,pc) and high electron densities ($n_{\mathrm{e}} \sim 10^6$\,cm$^{-3}$; \citealt{2005Kurtz}). They also typically exhibit broader radio recombination line (RRL) FWHMs (40--100~km~s$^{-1}$) than ultracompact (UC) H{\sc ii} regions, which usually show linewidths of 25--30~km~s$^{-1}$ \citep{2019Yang}.

\begin{figure*}[tb]    
\centering
\includegraphics[angle=0, width=1.0\textwidth]{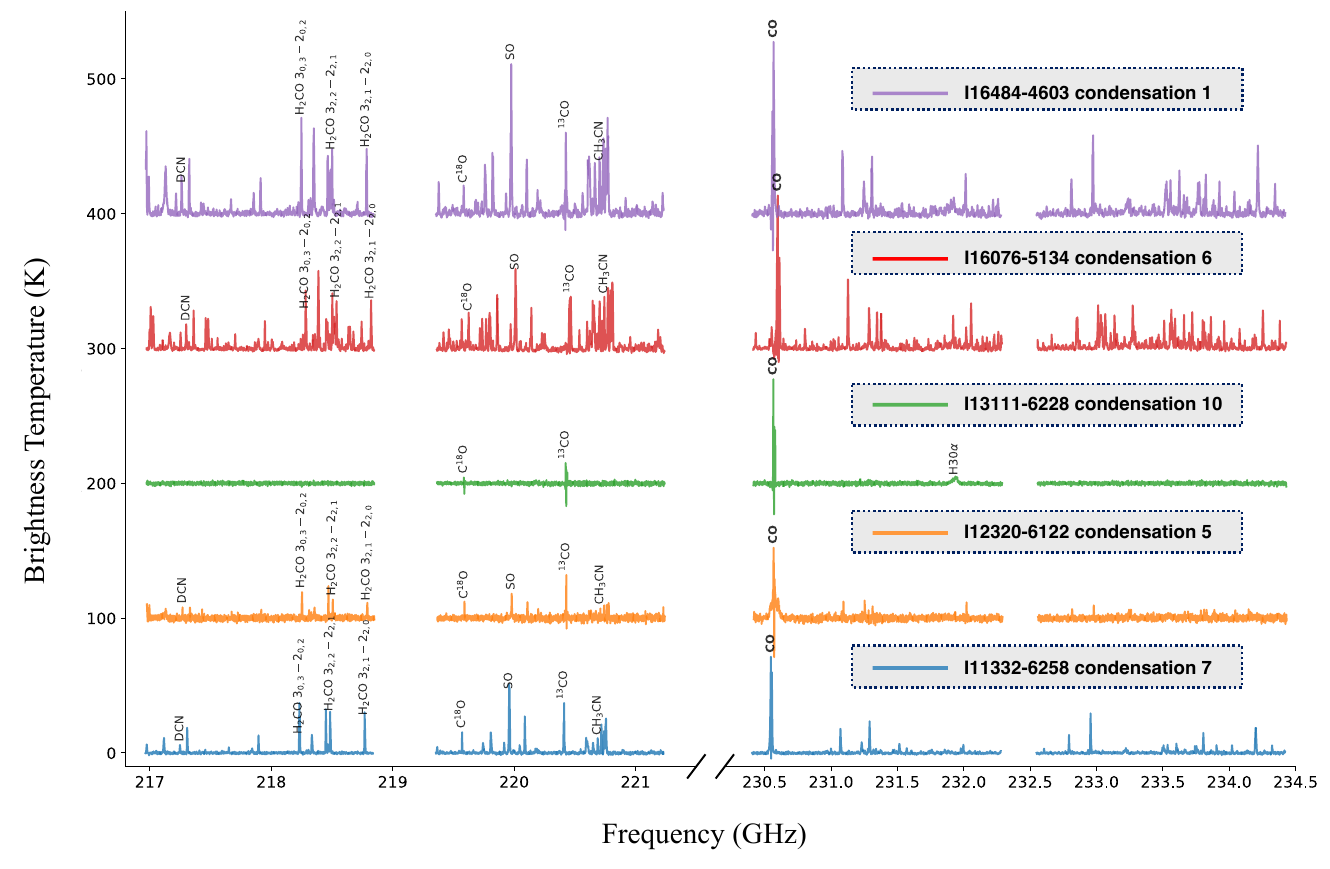} 
\caption{
    Demonstration of the spectra from the four spectral windows of the QUARKS dataset for five variable sources: I16484–4603 condensation~1 (purple), I16076–5134 condensation~6 (red), I13111–6228 condensation~10 (green), I12320–6122 condensation~5 (orange), and I11332–6258 condensation~7 (blue). Several strong molecular transitions are labeled in the spectra.
    }
    \label{fig:5variables}    
\end{figure*}

\begin{figure*}[!tb]
    \centering
    \includegraphics[angle=0, width=1\textwidth]{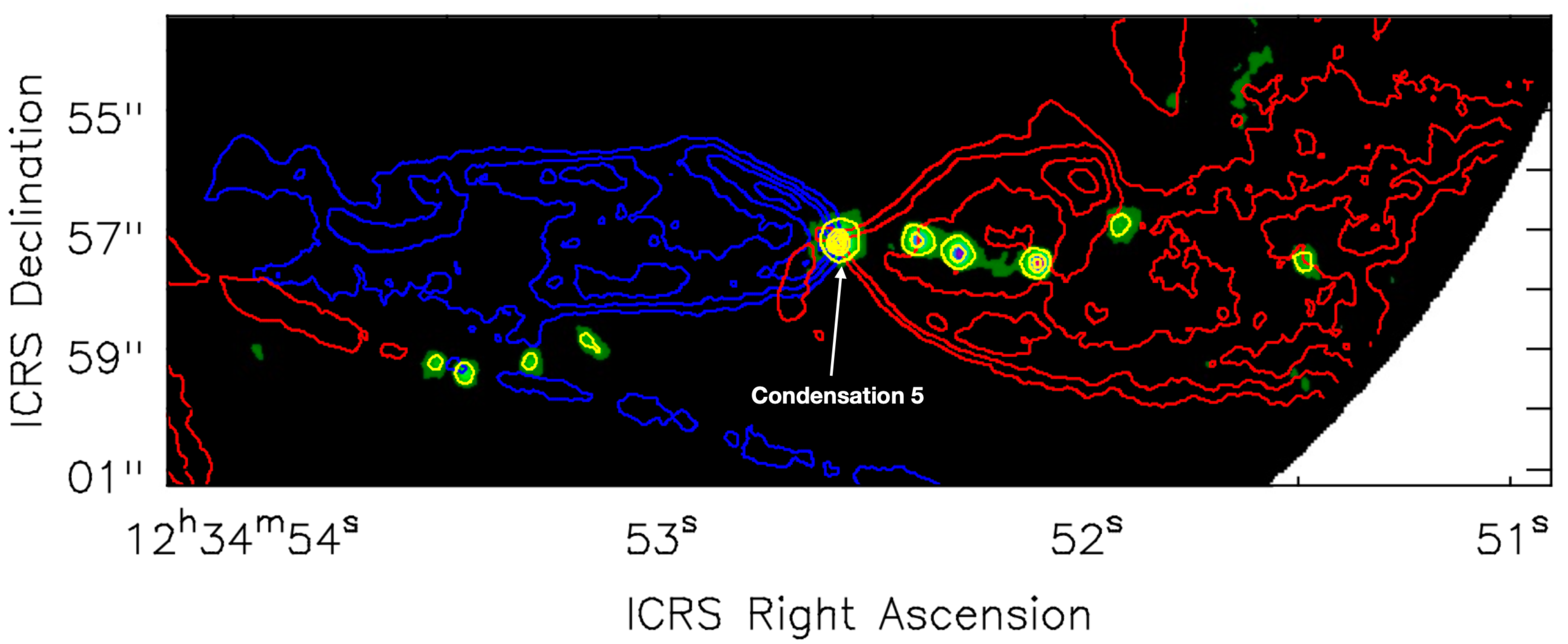} 
    \caption{CO outflows in condensation 5 of I12320–6122: 1.3 mm continuum (color image) with yellow contours at [0.1, 0.3, 0.5, 0.7, 0.9]$\times I_{\rm peak}$, where peak intensity $I_{\rm peak} = 0.016~\mathrm{Jy~beam^{-1}}$. Blue contours: high-velocity blueshifted outflow  (-108 to -58 $\mathrm{km~s^{-1}}$), with contour levels [0.3, 0.5, 0.7, 0.9]$\times S_{\rm peak}$ (peak integrated intensity $S_{\rm peak} = 2.68~\mathrm{Jy~beam^{-1}~km~s^{-1}}$). Red contours: high-velocity redshifted outflow (-38 to 12 $\mathrm{km~s^{-1}}$), with contour levels [0.3, 0.5, 0.7, 0.9]$\times S_{\rm peak}$ ($S_{\rm peak} = 4.23~\mathrm{Jy~beam^{-1}~km~s^{-1}}$).
    }
    \label{fig:outflow}    
\end{figure*}

Among the five variable sources, the most significant variability is observed in condensation~10 within I13111--6228, which is located in the G305 massive star-forming complex at a distance of $\sim2.97$\,kpc \citep{Liu2024QUARKS-I}. It has a molecular gas reservoir of $\sim760\,M_\odot$ on clump scale (radius $\sim0.75$\,pc; \citealt{Liu2024QUARKS-I, 2024Xu}). Figure~\ref{fig:H30alpha} presents the integrated intensity (moment~0) map of the H30$\alpha$ emission, with the corresponding spectrum shown in the inset. Given its compact physical size ($< 0.01$~pc) and broad recombination line ($> 40$~km~s$^{-1}$), we classify this source as a (HC) H{\sc ii} region.  
We note that radio stars (e.g., MWC~349) can also exhibit broad recombination lines and significant variability \citep{2001Gordon}, our target could also be one of its kind. Although the exact nature of this source remains intriguing and requires further observations for confirmation, we tentatively classify it as a (HC) H{\sc ii} region in this study.

Figure~\ref{fig:5variables} shows the spectra for the five variable sources extracted from the four spectral windows of the QUARKS dataset. CH$_3$CN $K=4$ emission is detected toward the other four variable sources. Among them, condensation~7 in I11332--6258, condensation~6 in I16076--5134, and condensation~1 in I16484--4603 exhibit rich complex molecular line emission and are thus classified as hot molecular cores.

Although CH$_3$CN line emission is also detected toward condensation~5 in I12320--6122, its spectrum is significantly less line-rich compared to the other three hot molecular cores. As shown in Figure~\ref{fig:outflow}, this source, however, drives a highly energetic and collimated bipolar outflow. Based on this evidence, we classify it as a high-mass protostellar object (HMPO), which represents an earlier evolutionary phase than the hot core stage. In addition to I12320--6122, both I11332--6258 condensation~7 and I16484--4603 condensation~1 are also associated with bipolar molecular outflows \citep{Jiao_prep}. While condensation~6 in I16076--5134 is associated with a candidate explosive outflow \citep{2022ApJ...937...51G}. The physical properties and variability mechanisms of these sources will be examined in further detail in our forthcoming studies.

\subsection{Crowding Effect and the Need for Interferometric Monitoring}
\label{sub:confusion_effect}

\begin{figure*}[tb]    
\centering
\includegraphics[angle=0, width=0.8\textwidth]{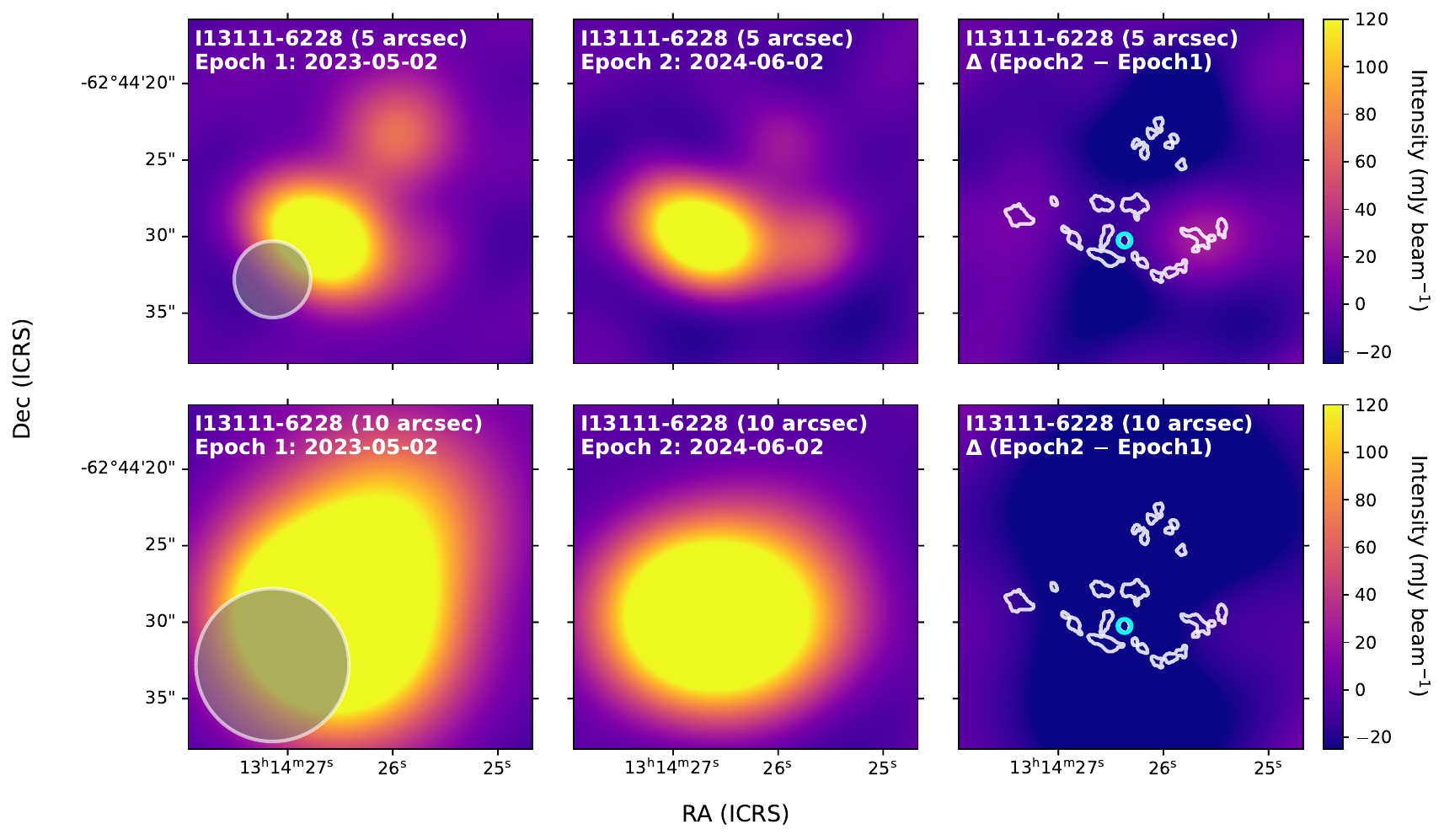} 
\caption{
    ``Crowding Effect'' test for I13111--6228. Top panels: ALMA 1.3\,mm images from 2023 May and 2024 June convolved to 5.0\arcsec, and their difference map (right). Bottom panels: same data convolved to 10.0\arcsec. The beam size is shown in the lower-left panel of the first column of each row. White contours in the third column indicate the union mask used for flux extraction, and the cyan circle marks the variable source condensation~10.
    }
    \label{fig:beam_dilution_test}    
\end{figure*}

The ability to detect (sub)millimeter continuum variability in massive protoclusters is also fundamentally limited by sensitivity and angular resolution. Single-dish telescopes (e.g., JCMT, IRAM 30-m) have beam sizes of several to tens of arcseconds, encompassing multiple protostellar cores and diffuse emission within a single field of view. Consequently, intrinsic variations from individual sources are spatially averaged, leading to severe beam dilution. Previous JCMT studies have emphasized this limitation, showing that high-resolution and high-sensitivity observations can detect flux variability and identify the embedded young stellar objects responsible for such events more effectively \citep{2019Park}. 

Here, we demonstrate how this ``crowding effect'' impacts the detectability of continuum variability. We convolved the ALMA data of 13111–6228 to lower resolutions of 5.0\arcsec\ and 10.0\arcsec\ (see Figure~\ref{fig:beam_dilution_test}), roughly corresponding to the beams of a big 30-50-m class single-dish telescope at 1.3\,mm wavelengths. The beam sizes are indicated in the lower-left corner of the first column. The third-column maps display the union mask (white contours) derived from the high-resolution data, with the cyan contour marking the variable condensation~10 identified at 0.3\arcsec. At 5\arcsec\ resolution, the compact cores blend into two broad emission peaks, fully erasing all substructure. Condensation 10 becomes indistinguishable within the merged emission, and no flux variation can be recognized. When the image is further smoothed to 10.0\arcsec, the emission becomes completely unresolved and no significant variation is detectable. This demonstrates that continuum variability would be strongly confused within a single-dish beam, and the existing single-dish telescopes (e.g., JCMT, IRAM 30-m) are not suitable for monitoring (sub)millimeter variability of protostars in distant massive protoclusters. 

In contrast, high-resolution interferometric observations are able to resolve individual protostars and measure their flux variations in protoclusters at large distances. Therefore, building on our methodology and findings, the next step is to use the growing ALMA multi-epoch archive for a larger variability search.


\section{Conclusion} 
\label{sec:Conclusion}
We conducted a systematic multi-epoch ALMA Band 6 ($\lambda \approx 1.3\,\mathrm{mm}$) continuum study of 22 massive protoclusters, covering timescales from hours to two years, to investigate millimeter variability. Using a dedicated processing pipeline that incorporates data reduction, image alignment, and relative flux calibration, we achieved high-precision flux measurements for 383 compact condensations.

Standard deviation analysis and difference maps identified five variable sources, corresponding to a lower limit on the variable fraction of 1.3\%. Among them, condensation 10 in I13111--6228 shows the most significant variation, with its peak intensity increasing by 68\% over one year, with an uncertainty of 2\%, well above the statistical thresholds. The five detected variable sources all exhibit variability on timescales longer than one year, with no significant variations observed on shorter timescales ($\sim$ hours). The physical properties of these variable sources will be analyzed in detail in upcoming studies.

Our study provides new insights into millimeter continuum variability by establishing a statistical framework for detecting variability in massive protoclusters and extending the scope of such studies toward more massive, active, and complex star-forming environments. Future surveys with larger samples, more extensive epoch coverage, higher resolution, and longer baselines will be crucial for constraining the occurrence rate, amplitude, and physical drivers of millimeter variability, resolving small-scale structure within condensations, and clarifying the role of episodic accretion in massive star formation.

\section*{Acknowledgements}

T.L. acknowledges the supports by the National Key R\&D Program of China no. 2022YFA1603100, National Science and Technology Major Project 2024ZD1100601,  National Natural Science Foundation of
China (NSFC) through grants no. 12073061 and no. 12122307, the Tianchi Talent Program of Xinjiang Uygur Autonomous Region and the Tianshan Talent Training Program 2024TSYCTD0013. 

QY-L acknowledges the support by JSPS KAKENHI Grant Number JP23K20035.

G.G. acknowledges support from the ANID Basal project FB210003.

S.Z. acknowledges support from the NAOJ ALMA Scientific Research Grant Code 2025-29B.

D.J.\ is supported by NRC Canada and by an NSERC Discovery Grant.

W.J. acknowledges support from the Shanghai Post-doctoral Excellence Program. 

L.B.acknowledges support from the ANID Basal project FB210003.

DAL and SYP were supported by the Ministry of Science and Higher Education of the Russian Federation (theme No.~FEUZ-2026-0012).

This work has been supported by the grant PID2024-155316NB-I00 funded by MICIU /AEI /10.13039/501100011033 / FEDER, UE and CSIC PIE 202350E189. This work was also supported by the Spanish program Unidad de Excelencia María de Maeztu financed by MCIN/AEI/10.13039/501100011033, and by the MaX-CSIC Excellence Award MaX4-SOMMA-ICE.

This paper makes use of the following ALMA data: ADS/JAO.ALMA\#2019.1.00685.S, 2021.1.00095.S, 021.1.00311.S and 2022.1.00974. ALMA is a partnership of ESO (representing its member states), NSF (USA) and NINS (Japan), together with NRC (Canada), NSTC and ASIAA (Taiwan), and KASI (Republic of Korea), in cooperation with the Republic of Chile. The Joint ALMA Observatory is operated by ESO, AUI/NRAO and NAOJ.

We are grateful to the anonymous referee for valuable comments that significantly improved the quality of this paper.

\facilities{ALMA}

\software{astropy \citep{2013A&A...558A..33A,2018AstropyCollaboration,2022AstropyCollaboration},  
    CASA \citep{2022PASP..134k4501C}, 
    CARTA \citep{2021ascl.soft03031C},
    Astrodendro \citep{2008ApJ...679.1338R}
          }


\bibliography{ref}

@ARTICLE{2021ApJ...920..119L,
       author = {{Lee}, Yong-Hee and {Johnstone}, Doug and {Lee}, Jeong-Eun and {Herczeg}, Gregory and {Mairs}, Steve and {Contreras-Pe{\~n}a}, Carlos and {Hatchell}, Jennifer and {Naylor}, Tim and {Bell}, Graham S. and {Bourke}, Tyler L. and {Broughton}, Colton and {Francis}, Logan and {Gupta}, Aashish and {Harsono}, Daniel and {Liu}, Sheng-Yuan and {Park}, Geumsook and {Plovie}, Spencer and {Moriarty-Schieven}, Gerald H. and {Scholz}, Aleks and {Sharma}, Tanvi and {Teixeira}, Paula Stella and {Wang}, Yao-Te and {Aikawa}, Yuri and {Bower}, Geoffrey C. and {Vivien Chen}, Huei-Ru and {Bae}, Jaehan and {Baek}, Giseon and {Chapman}, Scott and {Ping Chen}, Wen and {Du}, Fujun and {Dutta}, Somnath and {Forbrich}, Jan and {Guo}, Zhen and {Inutsuka}, Shu-ichiro and {Kang}, Miju and {Kirk}, Helen and {Kuan}, Yi-Jehng and {Kwon}, Woojin and {Lai}, Shih-Ping and {Lalchand}, Bhavana and {Lane}, James M.~M. and {Lee}, Chin-Fei and {Liu}, Tie and {Morata}, Oscar and {Pearson}, Samuel and {Pon}, Andy and {Sahu}, Dipen and {Shang}, Hsien and {Stamatellos}, Dimitris and {Tang}, Shih-Yun and {Xu}, Ziyan and {Yoo}, Hyunju and {Rawlings}, Jonathan M.~C.},
        title = "{The JCMT Transient Survey: Four-year Summary of Monitoring the Submillimeter Variability of Protostars}",
      journal = {\apj},
         year = 2021,
        month = oct,
       volume = {920},
       number = {2},
          eid = {119},
        pages = {119},
          doi = {10.3847/1538-4357/ac1679},
archivePrefix = {arXiv},
       eprint = {2107.10750},
 primaryClass = {astro-ph.SR},
       adsurl = {https://ui.adsabs.harvard.edu/abs/2021ApJ...920..119L}
}

@ARTICLE{2022ApJ...937...51G,
       author = {{Guzm{\'a}n Ccolque}, Estrella and {Fern{\'a}ndez-L{\'o}pez}, Manuel and {Zapata}, Luis A. and {Baug}, Tapas},
        title = "{Possible Explosive Dispersal Outflow in IRAS 16076-5134 Revealed with ALMA}",
      journal = {\apj},
         year = 2022,
        month = oct,
       volume = {937},
       number = {2},
          eid = {51},
        pages = {51},
          doi = {10.3847/1538-4357/ac8c35},
archivePrefix = {arXiv},
       eprint = {2208.12317},
 primaryClass = {astro-ph.GA},
       adsurl = {https://ui.adsabs.harvard.edu/abs/2022ApJ...937...51G}
}

@ARTICLE{2025ApJS..280...33Y,
       author = {{Yang}, Dongting and {Liu}, Hong-Li and {Liu}, Tie and {Liu}, Xunchuan and {Xu}, Fengwei and {Qin}, Sheng-Li and {Tej}, Anandmayee and {Garay}, Guido and {Zhu}, Lei and {Mai}, Xiaofeng and {Jiao}, Wenyu and {Zhang}, Siju and {Dib}, Sami and {Stutz}, Amelia M. and {Palau}, Aina and {Sanhueza}, Patricio and {Zavagno}, Annie and {Yang}, A.~Y. and {Tang}, Xindi and {Tang}, Mengyao and {Zhang}, Yichen and {Garc{\'\i}a}, Pablo and {Zhang}, Tianwei and {Saha}, Anindya and {Li}, Shanghuo and {Goldsmith}, Paul F. and {Bronfman}, Leonardo and {Lee}, Chang Won and {Taniguchi}, Kotomi and {Das}, Swagat Ranjan and {Gorai}, Prasanta and {Hoque}, Ariful and {Chen}, Li and {Kou}, Zhiping and {Zhou}, Jianjun and {Zhang}, Yankun and {T{\'o}th}, L. Viktor and {Baug}, Tapas and {Shen}, Xianjin and {Li}, Chuanshou and {Zou}, Jiahang and {Das}, Ankan and {Nazeer}, Hafiz and {Dewangan}, L.~K. and {Hwang}, Jihye and {Chibueze}, James O.},
        title = "{The ALMA-QUARKS Survey. III. Clump-to-core Fragmentation and Searches for High-mass Starless Cores}",
      journal = {\apjs},
         year = 2025,
        month = sep,
       volume = {280},
       number = {1},
          eid = {33},
        pages = {33},
          doi = {10.3847/1538-4365/adf847},
archivePrefix = {arXiv},
       eprint = {2508.03229},
 primaryClass = {astro-ph.GA},
       adsurl = {https://ui.adsabs.harvard.edu/abs/2025ApJS..280...33Y}
}

@ARTICLE{2021MNRAS.505.2801L,
       author = {{Liu}, Hong-Li and {Liu}, Tie and {Evans}, II, Neal J. and {Wang}, Ke and {Garay}, Guido and {Qin}, Sheng-Li and {Li}, Shanghuo and {Stutz}, Amelia and {Goldsmith}, Paul F. and {Liu}, Sheng-Yuan and {Tej}, Anandmayee and {Zhang}, Qizhou and {Juvela}, Mika and {Li}, Di and {Wang}, Jun-Zhi and {Bronfman}, Leonardo and {Ren}, Zhiyuan and {Wu}, Yue-Fang and {Kim}, Kee-Tae and {Lee}, Chang Won and {Tatematsu}, Ken'ichi and {Cunningham}, Maria R. and {Liu}, Xun-Chuan and {Wu}, Jing-Wen and {Hirota}, Tomoya and {Lee}, Jeong-Eun and {Li}, Pak-Shing and {Kang}, Sung-Ju and {Mardones}, Diego and {Ristorcelli}, Isabelle and {Zhang}, Yong and {Luo}, Qiu-Yi and {Toth}, L. Viktor and {Yi}, Hee-weon and {Yun}, Hyeong-Sik and {Peng}, Ya-Ping and {Li}, Juan and {Zhu}, Feng-Yao and {Shen}, Zhi-Qiang and {Baug}, Tapas and {Dewangan}, L.~K. and {Chakali}, Eswaraiah and {Liu}, Rong and {Xu}, Feng-Wei and {Wang}, Yu and {Zhang}, Chao and {Li}, Jinzeng and {Zhang}, Chao and {Zhou}, Jianwen and {Tang}, Mengyao and {Xue}, Qiaowei and {Issac}, Namitha and {Soam}, Archana and {{\'A}lvarez-Guti{\'e}rrez}, Rodrigo H.},
        title = "{ATOMS: ALMA three-millimeter observations of massive star-forming regions - III. Catalogues of candidate hot molecular cores and hyper/ultra compact H II regions}",
      journal = {\mnras},
         year = 2021,
        month = aug,
       volume = {505},
       number = {2},
        pages = {2801-2818},
          doi = {10.1093/mnras/stab1352},
archivePrefix = {arXiv},
       eprint = {2105.03554},
 primaryClass = {astro-ph.GA},
       adsurl = {https://ui.adsabs.harvard.edu/abs/2021MNRAS.505.2801L}
}

@ARTICLE{2017ApJ...849...25L,
       author = {{Liu}, Tie and {Lacy}, John and {Li}, Pak Shing and {Wang}, Ke and {Qin}, Sheng-Li and {Zhang}, Qizhou and {Kim}, Kee-Tae and {Garay}, Guido and {Wu}, Yuefang and {Mardones}, Diego and {Zhu}, Qingfeng and {Tatematsu}, Ken'ichi and {Hirota}, Tomoya and {Ren}, Zhiyuan and {Liu}, Sheng-Yuan and {Chen}, Huei-Ru and {Su}, Yu-Nung and {Li}, Di},
        title = "{ALMA Reveals Sequential High-mass Star Formation in the G9.62+0.19 Complex}",
      journal = {\apj},
         year = 2017,
        month = nov,
       volume = {849},
       number = {1},
          eid = {25},
        pages = {25},
          doi = {10.3847/1538-4357/aa8d73},
archivePrefix = {arXiv},
       eprint = {1705.04907},
 primaryClass = {astro-ph.GA},
       adsurl = {https://ui.adsabs.harvard.edu/abs/2017ApJ...849...25L}
}

@ARTICLE{Liu2020ATOMS-I,
       author = {{Liu}, Tie and {Evans}, Neal J. and {Kim}, Kee-Tae and {Goldsmith}, Paul F. and {Liu}, Sheng-Yuan and {Zhang}, Qizhou and {Tatematsu}, Ken'ichi and {Wang}, Ke and {Juvela}, Mika and {Bronfman}, Leonardo and {Cunningham}, Maria R. and {Garay}, Guido and {Hirota}, Tomoya and {Lee}, Jeong-Eun and {Kang}, Sung-Ju and {Li}, Di and {Li}, Pak-Shing and {Mardones}, Diego and {Qin}, Sheng-Li and {Ristorcelli}, Isabelle and {Tej}, Anandmayee and {Toth}, L. Viktor and {Wu}, Jing-Wen and {Wu}, Yue-Fang and {Yi}, Hee-weon and {Yun}, Hyeong-Sik and {Liu}, Hong-Li and {Peng}, Ya-Ping and {Li}, Juan and {Li}, Shang-Huo and {Lee}, Chang Won and {Shen}, Zhi-Qiang and {Baug}, Tapas and {Wang}, Jun-Zhi and {Zhang}, Yong and {Issac}, Namitha and {Zhu}, Feng-Yao and {Luo}, Qiu-Yi and {Soam}, Archana and {Liu}, Xun-Chuan and {Xu}, Feng-Wei and {Wang}, Yu and {Zhang}, Chao and {Ren}, Zhiyuan and {Zhang}, Chao},
        title = "{ATOMS: ALMA Three-millimeter Observations of Massive Star-forming regions - I. Survey description and a first look at G9.62+0.19}",
      journal = {\mnras},
         year = 2020,
        month = aug,
       volume = {496},
       number = {3},
        pages = {2790-2820},
          doi = {10.1093/mnras/staa1577},
archivePrefix = {arXiv},
       eprint = {2006.01549},
 primaryClass = {astro-ph.GA},
       adsurl = {https://ui.adsabs.harvard.edu/abs/2020MNRAS.496.2790L}
}

@ARTICLE{Liu2024QUARKS-I,
       author = {{Liu}, Xunchuan and {Liu}, Tie and {Zhu}, Lei and {Garay}, Guido and {Liu}, Hong-Li and {Goldsmith}, Paul and {Evans}, Neal and {Kim}, Kee-Tae and {Liu}, Sheng-Yuan and {Xu}, Fengwei and {Lu}, Xing and {Tej}, Anandmayee and {Mai}, Xiaofeng and {Bronfman}, Leonardo and {Li}, Shanghuo and {Mardones}, Diego and {Stutz}, Amelia and {Tatematsu}, Ken'ichi and {Wang}, Ke and {Zhang}, Qizhou and {Qin}, Sheng-Li and {Zhou}, Jianwen and {Luo}, Qiuyi and {Zhang}, Siju and {Cheng}, Yu and {He}, Jinhua and {Gu}, Qilao and {Li}, Ziyang and {Zhang}, Zhenying and {Zhang}, Suinan and {Saha}, Anindya and {Dewangan}, Lokesh and {Sanhueza}, Patricio and {Shen}, Zhiqiang},
        title = "{The ALMA-QUARKS Survey. I. Survey Description and Data Reduction}",
      journal = {Research in Astronomy and Astrophysics},
         year = 2024,
        month = feb,
       volume = {24},
       number = {2},
          eid = {025009},
        pages = {025009},
          doi = {10.1088/1674-4527/ad0d5c},
archivePrefix = {arXiv},
       eprint = {2311.08651},
 primaryClass = {astro-ph.GA},
       adsurl = {https://ui.adsabs.harvard.edu/abs/2024RAA....24b5009L}
}

@ARTICLE{2013A&A...558A..33A,
       author = {{Astropy Collaboration} and {Robitaille}, Thomas P. and {Tollerud}, Erik J. and {Greenfield}, Perry and {Droettboom}, Michael and {Bray}, Erik and {Aldcroft}, Tom and {Davis}, Matt and {Ginsburg}, Adam and {Price-Whelan}, Adrian M. and {Kerzendorf}, Wolfgang E. and {Conley}, Alexander and {Crighton}, Neil and {Barbary}, Kyle and {Muna}, Demitri and {Ferguson}, Henry and {Grollier}, Fr{\'e}d{\'e}ric and {Parikh}, Madhura M. and {Nair}, Prasanth H. and {Unther}, Hans M. and {Deil}, Christoph and {Woillez}, Julien and {Conseil}, Simon and {Kramer}, Roban and {Turner}, James E.~H. and {Singer}, Leo and {Fox}, Ryan and {Weaver}, Benjamin A. and {Zabalza}, Victor and {Edwards}, Zachary I. and {Azalee Bostroem}, K. and {Burke}, D.~J. and {Casey}, Andrew R. and {Crawford}, Steven M. and {Dencheva}, Nadia and {Ely}, Justin and {Jenness}, Tim and {Labrie}, Kathleen and {Lim}, Pey Lian and {Pierfederici}, Francesco and {Pontzen}, Andrew and {Ptak}, Andy and {Refsdal}, Brian and {Servillat}, Mathieu and {Streicher}, Ole},
        title = "{Astropy: A community Python package for astronomy}",
      journal = {\aap},
         year = 2013,
        month = oct,
       volume = {558},
          eid = {A33},
        pages = {A33},
          doi = {10.1051/0004-6361/201322068},
archivePrefix = {arXiv},
       eprint = {1307.6212},
 primaryClass = {astro-ph.IM},
       adsurl = {https://ui.adsabs.harvard.edu/abs/2013A&A...558A..33A}
}

@ARTICLE{2022PASP..134k4501C,
       author = {{CASA Team} and {Bean}, Ben and {Bhatnagar}, Sanjay and {Castro}, Sandra and {Donovan Meyer}, Jennifer and {Emonts}, Bjorn and {Garcia}, Enrique and {Garwood}, Robert and {Golap}, Kumar and {Gonzalez Villalba}, Justo and {Harris}, Pamela and {Hayashi}, Yohei and {Hoskins}, Josh and {Hsieh}, Mingyu and {Jagannathan}, Preshanth and {Kawasaki}, Wataru and {Keimpema}, Aard and {Kettenis}, Mark and {Lopez}, Jorge and {Marvil}, Joshua and {Masters}, Joseph and {McNichols}, Andrew and {Mehringer}, David and {Miel}, Renaud and {Moellenbrock}, George and {Montesino}, Federico and {Nakazato}, Takeshi and {Ott}, Juergen and {Petry}, Dirk and {Pokorny}, Martin and {Raba}, Ryan and {Rau}, Urvashi and {Schiebel}, Darrell and {Schweighart}, Neal and {Sekhar}, Srikrishna and {Shimada}, Kazuhiko and {Small}, Des and {Steeb}, Jan-Willem and {Sugimoto}, Kanako and {Suoranta}, Ville and {Tsutsumi}, Takahiro and {van Bemmel}, Ilse M. and {Verkouter}, Marjolein and {Wells}, Akeem and {Xiong}, Wei and {Szomoru}, Arpad and {Griffith}, Morgan and {Glendenning}, Brian and {Kern}, Jeff},
        title = "{CASA, the Common Astronomy Software Applications for Radio Astronomy}",
      journal = {\pasp},
         year = 2022,
        month = nov,
       volume = {134},
       number = {1041},
          eid = {114501},
        pages = {114501},
          doi = {10.1088/1538-3873/ac9642},
archivePrefix = {arXiv},
       eprint = {2210.02276},
 primaryClass = {astro-ph.IM},
       adsurl = {https://ui.adsabs.harvard.edu/abs/2022PASP..134k4501C}
}

@software{2021ascl.soft03031C,
       author = {{Comrie}, Angus and {Wang}, Kuo-Song and {Hsu}, Shou-Chieh and {Moraghan}, Anthony and {Harris}, Pamela and {Pang}, Qi and {Pi{\'n}ska}, Adrianna and {Chiang}, Cheng-Chin and {Simmonds}, Rob and {Chang}, Tien-Hao and {Jan}, Hengtai and {Lin}, Ming-Yi},
        title = "{CARTA: Cube Analysis and Rendering Tool for Astronomy}",
 howpublished = {Astrophysics Source Code Library, record ascl:2103.031},
         year = 2021,
        month = mar,
          eid = {ascl:2103.031},
       adsurl = {https://ui.adsabs.harvard.edu/abs/2021ascl.soft03031C}
}

@ARTICLE{2008ApJ...679.1338R,
       author = {{Rosolowsky}, E.~W. and {Pineda}, J.~E. and {Kauffmann}, J. and {Goodman}, A.~A.},
        title = "{Structural Analysis of Molecular Clouds: Dendrograms}",
      journal = {\apj},
         year = 2008,
        month = jun,
       volume = {679},
       number = {2},
        pages = {1338-1351},
          doi = {10.1086/587685},
archivePrefix = {arXiv},
       eprint = {0802.2944},
 primaryClass = {astro-ph},
       adsurl = {https://ui.adsabs.harvard.edu/abs/2008ApJ...679.1338R}
}

@ARTICLE{selfcal,
       author = {{Richards}, A.~M.~S. and {Moravec}, E. and {Etoka}, S. and {Fomalont}, E.~B. and {P{\'e}rez-S{\'a}nchez}, A.~F. and {Toribio}, M.~C. and {Laing}, R.~A.},
        title = "{Self-calibration and improving image fidelity for ALMA and other radio interferometers}",
      journal = {arXiv e-prints},
         year = 2022,
        month = jul,
          eid = {arXiv:2207.05591},
        pages = {arXiv:2207.05591},
          doi = {10.48550/arXiv.2207.05591},
archivePrefix = {arXiv},
       eprint = {2207.05591},
 primaryClass = {astro-ph.IM},
       adsurl = {https://ui.adsabs.harvard.edu/abs/2022arXiv220705591R}
}

@PROCEEDINGS{1999ASPC..180.....T,
        title = "{Synthesis Imaging in Radio Astronomy II}",
    booktitle = {Synthesis Imaging in Radio Astronomy II},
         year = 1999,
       editor = {{Taylor}, G.~B. and {Carilli}, C.~L. and {Perley}, R.~A.},
       series = {Astronomical Society of the Pacific Conference Series},
       volume = {180},
        month = jan,
       adsurl = {https://ui.adsabs.harvard.edu/abs/1999ASPC..180.....T}
}

@ARTICLE{astrodendro,
       author = {{Rosolowsky}, E.~W. and {Pineda}, J.~E. and {Kauffmann}, J. and {Goodman}, A.~A.},
        title = "{Structural Analysis of Molecular Clouds: Dendrograms}",
      journal = {\apj},
         year = 2008,
        month = jun,
       volume = {679},
       number = {2},
        pages = {1338-1351},
          doi = {10.1086/587685},
archivePrefix = {arXiv},
       eprint = {0802.2944},
 primaryClass = {astro-ph},
       adsurl = {https://ui.adsabs.harvard.edu/abs/2008ApJ...679.1338R}
}

@ARTICLE{2017Caratti,
       author = {{Caratti o Garatti}, A. and {Stecklum}, B. and {Garcia Lopez}, R. and {Eisl{\"o}ffel}, J. and {Ray}, T.~P. and {Sanna}, A. and {Cesaroni}, R. and {Walmsley}, C.~M. and {Oudmaijer}, R.~D. and {de Wit}, W.~J. and {Moscadelli}, L. and {Greiner}, J. and {Krabbe}, A. and {Fischer}, C. and {Klein}, R. and {Iba{\~n}ez}, J.~M.},
        title = "{Disk-mediated accretion burst in a high-mass young stellar object}",
      journal = {Nature Physics},
         year = 2017,
        month = mar,
       volume = {13},
       number = {3},
        pages = {276-279},
          doi = {10.1038/nphys3942},
archivePrefix = {arXiv},
       eprint = {1704.02628},
 primaryClass = {astro-ph.SR},
       adsurl = {https://ui.adsabs.harvard.edu/abs/2017NatPh..13..276C}
}

@ARTICLE{2019Liu,
       author = {{Liu}, Sheng-Yuan and {Su}, Yu-Nung and {Zinchenko}, Igor and {Wang}, Kuo-Song and {Wang}, Yuan},
        title = "{A Submillimeter Burst of S255IR SMA1: The Rise and Fall of its Luminosity}",
      journal = {Submillimeter Array Newsletter},
         year = 2019,
        month = jan,
       volume = {27},
        pages = {11-14},
       adsurl = {https://ui.adsabs.harvard.edu/abs/2019SMAN...27...11L}
}

@ARTICLE{2017Hunter,
       author = {{Hunter}, T.~R. and {Brogan}, C.~L. and {MacLeod}, G. and {Cyganowski}, C.~J. and {Chandler}, C.~J. and {Chibueze}, J.~O. and {Friesen}, R. and {Indebetouw}, R. and {Thesner}, C. and {Young}, K.~H.},
        title = "{An Extraordinary Outburst in the Massive Protostellar System NGC6334I-MM1: Quadrupling of the Millimeter Continuum}",
      journal = {\apjl},
         year = 2017,
        month = mar,
       volume = {837},
       number = {2},
          eid = {L29},
        pages = {L29},
          doi = {10.3847/2041-8213/aa5d0e},
archivePrefix = {arXiv},
       eprint = {1701.08637},
 primaryClass = {astro-ph.SR},
       adsurl = {https://ui.adsabs.harvard.edu/abs/2017ApJ...837L..29H}
}

@ARTICLE{2018Hunter,
       author = {{Hunter}, T.~R. and {Brogan}, C.~L. and {MacLeod}, G.~C. and {Cyganowski}, C.~J. and {Chibueze}, J.~O. and {Friesen}, R. and {Hirota}, T. and {Smits}, D.~P. and {Chandler}, C.~J. and {Indebetouw}, R.},
        title = "{The Extraordinary Outburst in the Massive Protostellar System NGC 6334I-MM1: Emergence of Strong 6.7 GHz Methanol Masers}",
      journal = {\apj},
         year = 2018,
        month = feb,
       volume = {854},
       number = {2},
          eid = {170},
        pages = {170},
          doi = {10.3847/1538-4357/aaa962},
archivePrefix = {arXiv},
       eprint = {1801.02141},
 primaryClass = {astro-ph.SR},
       adsurl = {https://ui.adsabs.harvard.edu/abs/2018ApJ...854..170H}
}

@ARTICLE{2021Hunter,
       author = {{Hunter}, T.~R. and {Brogan}, C.~L. and {De Buizer}, J.~M. and {Towner}, A.~P.~M. and {Dowell}, C.~D. and {MacLeod}, G.~C. and {Stecklum}, B. and {Cyganowski}, C.~J. and {El-Abd}, S.~J. and {McGuire}, B.~A.},
        title = "{The Extraordinary Outburst in the Massive Protostellar System NGC 6334 I-MM1: Strong Increase in Mid-Infrared Continuum Emission}",
      journal = {\apjl},
         year = 2021,
        month = may,
       volume = {912},
       number = {1},
          eid = {L17},
        pages = {L17},
          doi = {10.3847/2041-8213/abf6d9},
archivePrefix = {arXiv},
       eprint = {2104.05187},
 primaryClass = {astro-ph.GA},
       adsurl = {https://ui.adsabs.harvard.edu/abs/2021ApJ...912L..17H}
}

@ARTICLE{2021Stecklum,
       author = {{Stecklum}, B. and {Wolf}, V. and {Linz}, H. and {Caratti o Garatti}, A. and {Schmidl}, S. and {Klose}, S. and {Eisl{\"o}ffel}, J. and {Fischer}, Ch. and {Brogan}, C. and {Burns}, R.~A. and {Bayandina}, O. and {Cyganowski}, C. and {Gurwell}, M. and {Hunter}, T. and {Hirano}, N. and {Kim}, K. -T. and {MacLeod}, G. and {Menten}, K.~M. and {Olech}, M. and {Orosz}, G. and {Sobolev}, A. and {Sridharan}, T.~K. and {Surcis}, G. and {Sugiyama}, K. and {van der Walt}, J. and {Volvach}, A. and {Yonekura}, Y.},
        title = "{Infrared observations of the flaring maser source G358.93-0.03. SOFIA confirms an accretion burst from a massive young stellar object}",
      journal = {\aap},
         year = 2021,
        month = feb,
       volume = {646},
          eid = {A161},
        pages = {A161},
          doi = {10.1051/0004-6361/202039645},
archivePrefix = {arXiv},
       eprint = {2101.01812},
 primaryClass = {astro-ph.SR},
       adsurl = {https://ui.adsabs.harvard.edu/abs/2021A&A...646A.161S}
}

@ARTICLE{2015Fujisawa,
       author = {{Fujisawa}, Kenta and {Yonekura}, Yoshinori and {Sugiyama}, Koichiro and {Horiuchi}, Hikari and {Hayashi}, Takehiro and {Hachisuka}, Kazuya and {Matsumoto}, Naoko and {Niinuma}, Kotaro},
        title = "{A flare of methanol maser in S255}",
      journal = {The Astronomer's Telegram},
         year = 2015,
        month = nov,
       volume = {8286},
        pages = {1},
       adsurl = {https://ui.adsabs.harvard.edu/abs/2015ATel.8286....1F}
}

@ARTICLE{2018Szymczak,
       author = {{Szymczak}, M. and {Olech}, M. and {Sarniak}, R. and {Wolak}, P. and {Bartkiewicz}, A.},
        title = "{Monitoring observations of 6.7 GHz methanol masers}",
      journal = {\mnras},
         year = 2018,
        month = feb,
       volume = {474},
       number = {1},
        pages = {219-253},
          doi = {10.1093/mnras/stx2693},
archivePrefix = {arXiv},
       eprint = {1710.04595},
 primaryClass = {astro-ph.GA},
       adsurl = {https://ui.adsabs.harvard.edu/abs/2018MNRAS.474..219S}
}

@ARTICLE{2018MacLeod,
       author = {{MacLeod}, G.~C. and {Smits}, D.~P. and {Goedhart}, S. and {Hunter}, T.~R. and {Brogan}, C.~L. and {Chibueze}, J.~O. and {van den Heever}, S.~P. and {Thesner}, C.~J. and {Banda}, P.~J. and {Paulsen}, J.~D.},
        title = "{A masing event in NGC 6334I: contemporaneous flaring of hydroxyl, methanol, and water masers}",
      journal = {\mnras},
         year = 2018,
        month = jul,
       volume = {478},
       number = {1},
        pages = {1077-1092},
          doi = {10.1093/mnras/sty996},
archivePrefix = {arXiv},
       eprint = {1804.05308},
 primaryClass = {astro-ph.SR},
       adsurl = {https://ui.adsabs.harvard.edu/abs/2018MNRAS.478.1077M}
}

@ARTICLE{2019Brogan,
       author = {{Brogan}, C.~L. and {Hunter}, T.~R. and {Towner}, A.~P.~M. and {McGuire}, B.~A. and {MacLeod}, G.~C. and {Gurwell}, M.~A. and {Cyganowski}, C.~J. and {Brand}, J. and {Burns}, R.~A. and {Caratti o Garatti}, A. and {Chen}, X. and {Chibueze}, J.~O. and {Hirano}, N. and {Hirota}, T. and {Kim}, K. -T. and {Kramer}, B.~H. and {Linz}, H. and {Menten}, K.~M. and {Remijan}, A. and {Sanna}, A. and {Sobolev}, A.~M. and {Sridharan}, T.~K. and {Stecklum}, B. and {Sugiyama}, K. and {Surcis}, G. and {Van der Walt}, J. and {Volvach}, A.~E. and {Volvach}, L.~N.},
        title = "{Sub-arcsecond (Sub)millimeter Imaging of the Massive Protocluster G358.93-0.03: Discovery of 14 New Methanol Maser Lines Associated with a Hot Core}",
      journal = {\apjl},
         year = 2019,
        month = aug,
       volume = {881},
       number = {2},
          eid = {L39},
        pages = {L39},
          doi = {10.3847/2041-8213/ab2f8a},
archivePrefix = {arXiv},
       eprint = {1907.02470},
 primaryClass = {astro-ph.SR},
       adsurl = {https://ui.adsabs.harvard.edu/abs/2019ApJ...881L..39B}
}

@ARTICLE{2018Brogan,
       author = {{Brogan}, C.~L. and {Hunter}, T.~R. and {Cyganowski}, C.~J. and {Chibueze}, J.~O. and {Friesen}, R.~K. and {Hirota}, T. and {MacLeod}, G.~C. and {McGuire}, B.~A. and {Sobolev}, A.~M.},
        title = "{The Extraordinary Outburst in the Massive Protostellar System NGC 6334I-MM1: Flaring of the Water Masers in a North-South Bipolar Outflow Driven by MM1B}",
      journal = {\apj},
         year = 2018,
        month = oct,
       volume = {866},
       number = {2},
          eid = {87},
        pages = {87},
          doi = {10.3847/1538-4357/aae151},
archivePrefix = {arXiv},
       eprint = {1809.04178},
 primaryClass = {astro-ph.SR},
       adsurl = {https://ui.adsabs.harvard.edu/abs/2018ApJ...866...87B}
}

@ARTICLE{2018Cesaroni,
       author = {{Cesaroni}, R. and {Moscadelli}, L. and {Neri}, R. and {Sanna}, A. and {Caratti o Garatti}, A. and {Eisloffel}, J. and {Stecklum}, B. and {Ray}, T. and {Walmsley}, C.~M.},
        title = "{Radio outburst from a massive (proto)star. When accretion turns into ejection}",
      journal = {\aap},
         year = 2018,
        month = may,
       volume = {612},
          eid = {A103},
        pages = {A103},
          doi = {10.1051/0004-6361/201732238},
archivePrefix = {arXiv},
       eprint = {1802.04228},
 primaryClass = {astro-ph.SR},
       adsurl = {https://ui.adsabs.harvard.edu/abs/2018A&A...612A.103C}
}

@INPROCEEDINGS{2023Fischer,
       author = {{Fischer}, W.~J. and {Hillenbrand}, L.~A. and {Herczeg}, G.~J. and {Johnstone}, D. and {Kospal}, A. and {Dunham}, M.~M.},
        title = "{Accretion Variability as a Guide to Stellar Mass Assembly}",
    booktitle = {Protostars and Planets VII},
         year = 2023,
       editor = {{Inutsuka}, S. and {Aikawa}, Y. and {Muto}, T. and {Tomida}, K. and {Tamura}, M.},
       series = {Astronomical Society of the Pacific Conference Series},
       volume = {534},
        month = jul,
        pages = {355},
          doi = {10.48550/arXiv.2203.11257},
archivePrefix = {arXiv},
       eprint = {2203.11257},
 primaryClass = {astro-ph.SR},
       adsurl = {https://ui.adsabs.harvard.edu/abs/2023ASPC..534..355F}
}

@INPROCEEDINGS{2014Audard,
       author = {{Audard}, M. and {{\'A}brah{\'a}m}, P. and {Dunham}, M.~M. and {Green}, J.~D. and {Grosso}, N. and {Hamaguchi}, K. and {Kastner}, J.~H. and {K{\'o}sp{\'a}l}, {\'A}. and {Lodato}, G. and {Romanova}, M.~M. and {Skinner}, S.~L. and {Vorobyov}, E.~I. and {Zhu}, Z.},
        title = "{Episodic Accretion in Young Stars}",
    booktitle = {Protostars and Planets VI},
         year = 2014,
       editor = {{Beuther}, Henrik and {Klessen}, Ralf S. and {Dullemond}, Cornelis P. and {Henning}, Thomas},
        month = jan,
        pages = {387-410},
          doi = {10.2458/azu_uapress_9780816531240-ch017},
archivePrefix = {arXiv},
       eprint = {1401.3368},
 primaryClass = {astro-ph.SR},
       adsurl = {https://ui.adsabs.harvard.edu/abs/2014prpl.conf..387A}
}

@ARTICLE{1990Kenyon,
       author = {{Kenyon}, Scott J. and {Hartmann}, Lee W. and {Strom}, Karen M. and {Strom}, Stephen E.},
        title = "{An IRAS Survey of the Taurus-Auriga Molecular Cloud}",
      journal = {\aj},
         year = 1990,
        month = mar,
       volume = {99},
        pages = {869},
          doi = {10.1086/115380},
       adsurl = {https://ui.adsabs.harvard.edu/abs/1990AJ.....99..869K}
}

@ARTICLE{2021Chen,
       author = {{Chen}, Zhiwei and {Sun}, Wei and {Chini}, Rolf and {Haas}, Martin and {Jiang}, Zhibo and {Chen}, Xuepeng},
        title = "{M17 MIR: A Massive Protostar with Multiple Accretion Outbursts}",
      journal = {\apj},
         year = 2021,
        month = nov,
       volume = {922},
       number = {1},
          eid = {90},
        pages = {90},
          doi = {10.3847/1538-4357/ac2151},
archivePrefix = {arXiv},
       eprint = {2108.12554},
 primaryClass = {astro-ph.SR},
       adsurl = {https://ui.adsabs.harvard.edu/abs/2021ApJ...922...90C}
}

@ARTICLE{2024Zhou,
       author = {{Zhou}, Wei and {Chen}, Zhiwei and {Jiang}, Zhibo and {Feng}, Haoran and {Jiang}, Yu},
        title = "{M17 MIR: A Massive Star Is Forming via Episodic Mass Accretion}",
      journal = {\apjl},
         year = 2024,
        month = jul,
       volume = {969},
       number = {1},
          eid = {L6},
        pages = {L6},
          doi = {10.3847/2041-8213/ad55c7},
archivePrefix = {arXiv},
       eprint = {2406.04980},
 primaryClass = {astro-ph.SR},
       adsurl = {https://ui.adsabs.harvard.edu/abs/2024ApJ...969L...6Z}
}

@ARTICLE{2025Chen,
       author = {{Chen}, Zhiwei and {Johnstone}, Doug and {Contreras Pe{\~n}a}, Carlos and {Lee}, Jeong-Eun and {Liu}, Sheng-Yuan and {Herczeg}, Gregory and {Mairs}, Steve and {Park}, Geumsook and {Kim}, Kee-Tae and {Kim}, Mi-Ryang and {Qiu}, Keping and {Wang}, Yao-Te and {Zhang}, Xu and {Reiter}, Megan and {JCMT Transient Team}},
        title = "{Submillimeter and Mid-infrared Variability of Young Stellar Objects in the M17 H II Region}",
      journal = {\aj},
         year = 2025,
        month = aug,
       volume = {170},
       number = {2},
          eid = {125},
        pages = {125},
          doi = {10.3847/1538-3881/ade988},
archivePrefix = {arXiv},
       eprint = {2506.08389},
 primaryClass = {astro-ph.GA},
       adsurl = {https://ui.adsabs.harvard.edu/abs/2025AJ....170..125C}
}

@ARTICLE{2011Offner,
       author = {{Offner}, Stella S.~R. and {McKee}, Christopher F.},
        title = "{The Protostellar Luminosity Function}",
      journal = {\apj},
         year = 2011,
        month = jul,
       volume = {736},
       number = {1},
          eid = {53},
        pages = {53},
          doi = {10.1088/0004-637X/736/1/53},
archivePrefix = {arXiv},
       eprint = {1105.0671},
 primaryClass = {astro-ph.SR},
       adsurl = {https://ui.adsabs.harvard.edu/abs/2011ApJ...736...53O}
}

@ARTICLE{2017Meyer,
       author = {{Meyer}, D.~M. -A. and {Vorobyov}, E.~I. and {Kuiper}, R. and {Kley}, W.},
        title = "{On the existence of accretion-driven bursts in massive star formation}",
      journal = {\mnras},
         year = 2017,
        month = jan,
       volume = {464},
       number = {1},
        pages = {L90-L94},
          doi = {10.1093/mnrasl/slw187},
archivePrefix = {arXiv},
       eprint = {1609.03402},
 primaryClass = {astro-ph.SR},
       adsurl = {https://ui.adsabs.harvard.edu/abs/2017MNRAS.464L..90M}
}

@ARTICLE{2019Meyer,
       author = {{Meyer}, D.~M. -A. and {Vorobyov}, E.~I. and {Elbakyan}, V.~G. and {Stecklum}, B. and {Eisl{\"o}ffel}, J. and {Sobolev}, A.~M.},
        title = "{Burst occurrence in young massive stellar objects}",
      journal = {\mnras},
         year = 2019,
        month = feb,
       volume = {482},
       number = {4},
        pages = {5459-5476},
          doi = {10.1093/mnras/sty2980},
archivePrefix = {arXiv},
       eprint = {1811.00574},
 primaryClass = {astro-ph.SR},
       adsurl = {https://ui.adsabs.harvard.edu/abs/2019MNRAS.482.5459M}
}

@ARTICLE{2021Elbakyan,
       author = {{Elbakyan}, Vardan G. and {Nayakshin}, Sergei and {Vorobyov}, Eduard I. and {Caratti o Garatti}, Alessio and {Eisl{\"o}ffel}, Jochen},
        title = "{Accretion bursts in high-mass protostars: A new test bed for models of episodic accretion}",
      journal = {\aap},
         year = 2021,
        month = jul,
       volume = {651},
          eid = {L3},
        pages = {L3},
          doi = {10.1051/0004-6361/202140871},
archivePrefix = {arXiv},
       eprint = {2106.08734},
 primaryClass = {astro-ph.SR},
       adsurl = {https://ui.adsabs.harvard.edu/abs/2021A&A...651L...3E}
}

@ARTICLE{2023Burns,
       author = {{Burns}, R.~A. and {Uno}, Y. and {Sakai}, N. and {Blanchard}, J. and {Rosli}, Z. and {Orosz}, G. and {Yonekura}, Y. and {Tanabe}, Y. and {Sugiyama}, K. and {Hirota}, T. and {Kim}, Kee-Tae and {Aberfelds}, A. and {Volvach}, A.~E. and {Bartkiewicz}, A. and {Caratti o Garatti}, A. and {Sobolev}, A.~M. and {Stecklum}, B. and {Brogan}, C. and {Phillips}, C. and {Ladeyschikov}, D.~A. and {Johnstone}, D. and {Surcis}, G. and {MacLeod}, G.~C. and {Linz}, H. and {Chibueze}, J.~O. and {Brand}, J. and {Eisl{\"o}ffel}, J. and {Hyland}, L. and {Uscanga}, L. and {Olech}, M. and {Durjasz}, M. and {Bayandina}, O. and {Breen}, S. and {Ellingsen}, S.~P. and {van den Heever}, S.~P. and {Hunter}, T.~R. and {Chen}, X.},
        title = "{A Keplerian disk with a four-arm spiral birthing an episodically accreting high-mass protostar}",
      journal = {Nature Astronomy},
         year = 2023,
        month = may,
       volume = {7},
        pages = {557-568},
          doi = {10.1038/s41550-023-01899-w},
archivePrefix = {arXiv},
       eprint = {2304.14740},
 primaryClass = {astro-ph.SR},
       adsurl = {https://ui.adsabs.harvard.edu/abs/2023NatAs...7..557B}
}

@ARTICLE{2020Burns,
       author = {{Burns}, R.~A. and {Sugiyama}, K. and {Hirota}, T. and {Kim}, Kee-Tae and {Sobolev}, A.~M. and {Stecklum}, B. and {MacLeod}, G.~C. and {Yonekura}, Y. and {Olech}, M. and {Orosz}, G. and {Ellingsen}, S.~P. and {Hyland}, L. and {Caratti o Garatti}, A. and {Brogan}, C. and {Hunter}, T.~R. and {Phillips}, C. and {van den Heever}, S.~P. and {Eisl{\"o}ffel}, J. and {Linz}, H. and {Surcis}, G. and {Chibueze}, J.~O. and {Baan}, W. and {Kramer}, B.},
        title = "{A heatwave of accretion energy traced by masers in the G358-MM1 high-mass protostar}",
      journal = {Nature Astronomy},
         year = 2020,
        month = jan,
       volume = {4},
        pages = {506-510},
          doi = {10.1038/s41550-019-0989-3},
archivePrefix = {arXiv},
       eprint = {2304.14739},
 primaryClass = {astro-ph.SR},
       adsurl = {https://ui.adsabs.harvard.edu/abs/2020NatAs...4..506B}
}

@ARTICLE{2017Herczeg,
       author = {{Herczeg}, Gregory J. and {Johnstone}, Doug and {Mairs}, Steve and {Hatchell}, Jennifer and {Lee}, Jeong-Eun and {Bower}, Geoffrey C. and {Chen}, Huei-Ru Vivien and {Aikawa}, Yuri and {Yoo}, Hyunju and {Kang}, Sung-Ju and {Kang}, Miju and {Chen}, Wen-Ping and {Williams}, Jonathan P. and {Bae}, Jaehan and {Dunham}, Michael M. and {Vorobyov}, Eduard I. and {Zhu}, Zhaohuan and {Rao}, Ramprasad and {Kirk}, Helen and {Takahashi}, Satoko and {Morata}, Oscar and {Lacaille}, Kevin and {Lane}, James and {Pon}, Andy and {Scholz}, Aleks and {Samal}, Manash R. and {Bell}, Graham S. and {Graves}, Sarah and {Lee}, E. 'lisa M. and {Parsons}, Harriet and {He}, Yuxin and {Zhou}, Jianjun and {Kim}, Mi-Ryang and {Chapman}, Scott and {Drabek-Maunder}, Emily and {Chung}, Eun Jung and {Eyres}, Stewart P.~S. and {Forbrich}, Jan and {Hillenbrand}, Lynne A. and {Inutsuka}, Shu-ichiro and {Kim}, Gwanjeong and {Kim}, Kyoung Hee and {Kuan}, Yi-Jehng and {Kwon}, Woojin and {Lai}, Shih-Ping and {Lalchand}, Bhavana and {Lee}, Chang Won and {Lee}, Chin-Fei and {Long}, Feng and {Lyo}, A. -Ran and {Qian}, Lei and {Scicluna}, Peter and {Soam}, Archana and {Stamatellos}, Dimitris and {Takakuwa}, Shigehisa and {Tang}, Ya-Wen and {Wang}, Hongchi and {Wang}, Yiren},
        title = "{How Do Stars Gain Their Mass? A JCMT/SCUBA-2 Transient Survey of Protostars in Nearby Star-forming Regions}",
      journal = {\apj},
         year = 2017,
        month = nov,
       volume = {849},
       number = {1},
          eid = {43},
        pages = {43},
          doi = {10.3847/1538-4357/aa8b62},
archivePrefix = {arXiv},
       eprint = {1709.02052},
 primaryClass = {astro-ph.SR},
       adsurl = {https://ui.adsabs.harvard.edu/abs/2017ApJ...849...43H}
}

@ARTICLE{2024Mairs,
       author = {{Mairs}, Steve and {Lee}, Seonjae and {Johnstone}, Doug and {Broughton}, Colton and {Lee}, Jeong-Eun and {Herczeg}, Gregory J. and {Bell}, Graham S. and {Chen}, Zhiwei and {Contreras-Pe{\~n}a}, Carlos and {Francis}, Logan and {Hatchell}, Jennifer and {Kim}, Mi-Ryang and {Liu}, Sheng-Yuan and {Park}, Geumsook and {Qiu}, Keping and {Wang}, Yao-Te and {Zhang}, Xu and {JCMT Transient Team}},
        title = "{The JCMT Transient Survey: Six Year Summary of 450/850 {\ensuremath{\mu}}m Protostellar Variability and Calibration Pipeline Version 2.0}",
      journal = {\apj},
         year = 2024,
        month = may,
       volume = {966},
       number = {2},
          eid = {215},
        pages = {215},
          doi = {10.3847/1538-4357/ad35b6},
archivePrefix = {arXiv},
       eprint = {2401.03549},
 primaryClass = {astro-ph.IM},
       adsurl = {https://ui.adsabs.harvard.edu/abs/2024ApJ...966..215M}
}

@ARTICLE{2004Goedhart,
       author = {{Goedhart}, S. and {Gaylard}, M.~J. and {van der Walt}, D.~J.},
        title = "{Long-term monitoring of 6.7-GHz methanol masers}",
      journal = {\mnras},
         year = 2004,
        month = dec,
       volume = {355},
       number = {2},
        pages = {553-584},
          doi = {10.1111/j.1365-2966.2004.08340.x},
       adsurl = {https://ui.adsabs.harvard.edu/abs/2004MNRAS.355..553G}
}

@ARTICLE{2019Chen,
       author = {{Chen}, Xi and {Li}, Jing-Jing and {Zhang}, Bo and {Ellingsen}, Simon P. and {Xu}, Ye and {Ren}, Zhi-Yuan and {Shen}, Zhi-Qiang and {Sobolev}, Andrej M.},
        title = "{High-mass Star Formation in the nearby Region G352.630-1.067. I. Parallax}",
      journal = {\apj},
         year = 2019,
        month = feb,
       volume = {871},
       number = {2},
          eid = {198},
        pages = {198},
          doi = {10.3847/1538-4357/aaf862},
       adsurl = {https://ui.adsabs.harvard.edu/abs/2019ApJ...871..198C}
}

@ARTICLE{2013Motogi,
       author = {{Motogi}, K. and {Sorai}, K. and {Niinuma}, K. and {Sugiyama}, K. and {Honma}, M. and {Fujisawa}, K.},
        title = "{Intermittent maser flare around the high-mass young stellar object G353.273 + 0.641 - II. Detection of a radio and molecular jet}",
      journal = {\mnras},
         year = 2013,
        month = jan,
       volume = {428},
       number = {1},
        pages = {349-353},
          doi = {10.1093/mnras/sts035},
archivePrefix = {arXiv},
       eprint = {1209.4313},
 primaryClass = {astro-ph.SR},
       adsurl = {https://ui.adsabs.harvard.edu/abs/2013MNRAS.428..349M}
}

@ARTICLE{2018Johnstone,
       author = {{Johnstone}, Doug and {Herczeg}, Gregory J. and {Mairs}, Steve and {Hatchell}, Jennifer and {Bower}, Geoffrey C. and {Kirk}, Helen and {Lane}, James and {Bell}, Graham S. and {Graves}, Sarah and {Aikawa}, Yuri and {Chen}, Huei-Ru Vivien and {Chen}, Wen-Ping and {Kang}, Miju and {Kang}, Sung-Ju and {Lee}, Jeong-Eun and {Morata}, Oscar and {Pon}, Andy and {Scicluna}, Peter and {Scholz}, Aleks and {Takahashi}, Satoko and {Yoo}, Hyunju and {JCMT Transient Team}},
        title = "{The JCMT Transient Survey: Stochastic and Secular Variability of Protostars and Disks In the Submillimeter Region Observed over 18 Months}",
      journal = {\apj},
         year = 2018,
        month = feb,
       volume = {854},
       number = {1},
          eid = {31},
        pages = {31},
          doi = {10.3847/1538-4357/aaa764},
archivePrefix = {arXiv},
       eprint = {1801.03537},
 primaryClass = {astro-ph.SR},
       adsurl = {https://ui.adsabs.harvard.edu/abs/2018ApJ...854...31J}
}

@ARTICLE{2019Park,
       author = {{Park}, Geumsook and {Kim}, Kee-Tae and {Johnstone}, Doug and {Kang}, Sung-ju and {Liu}, Tie and {Mairs}, Steve and {Choi}, Minho and {Lee}, Jeong-Eun and {Sanhueza}, Patricio and {Juvela}, Mika and {Kang}, Miju and {Eden}, David and {Soam}, Archana and {Montillaud}, Julien and {Fuller}, Gary A. and {Koch}, Patrick M. and {Lee}, Chang Won and {Stamatellos}, Dimitris and {Rawlings}, Jonathan and {Kim}, Gwanjeong and {Zhang}, Chuan-Peng and {Kwon}, Woojin and {Yoo}, Hyunju},
        title = "{Submillimeter Continuum Variability in Planck Galactic Cold Clumps}",
      journal = {\apjs},
         year = 2019,
        month = jun,
       volume = {242},
       number = {2},
          eid = {27},
        pages = {27},
          doi = {10.3847/1538-4365/ab1eae},
archivePrefix = {arXiv},
       eprint = {1905.12147},
 primaryClass = {astro-ph.GA},
       adsurl = {https://ui.adsabs.harvard.edu/abs/2019ApJS..242...27P}
}

@ARTICLE{2020Liu,
       author = {{Liu}, Sheng-Yuan and {Su}, Yu-Nung and {Zinchenko}, Igor and {Wang}, Kuo-Song and {Meyer}, Dominique M.-A. and {Wang}, Yuan and {Hsieh}, I.-Ta},
        title = "{ALMA View of the Infalling Envelope around a Massive Protostar in S255IR SMA1}",
      journal = {\apj},
         year = 2020,
        month = dec,
       volume = {904},
       number = {2},
          eid = {181},
        pages = {181},
          doi = {10.3847/1538-4357/abc0ec},
archivePrefix = {arXiv},
       eprint = {2010.09199},
 primaryClass = {astro-ph.GA},
       adsurl = {https://ui.adsabs.harvard.edu/abs/2020ApJ...904..181L}
}

@ARTICLE{2020Uchiyama,
       author = {{Uchiyama}, Mizuho and {Yamashita}, Takuya and {Sugiyama}, Koichiro and {Nakaoka}, Tatsuya and {Kawabata}, Miho and {Itoh}, Ryosuke and {Yamanaka}, Masayuki and {Akitaya}, Hiroshi and {Kawabata}, Koji and {Yonekura}, Yoshinori and {Saito}, Yu and {Motogi}, Kazuhito and {Fujisawa}, Kenta},
        title = "{Near-infrared monitoring of the accretion outburst in the massive young stellar object S255-NIRS3}",
      journal = {\pasj},
         year = 2020,
        month = feb,
       volume = {72},
       number = {1},
          eid = {4},
        pages = {4},
          doi = {10.1093/pasj/psz122},
archivePrefix = {arXiv},
       eprint = {1910.07691},
 primaryClass = {astro-ph.SR},
       adsurl = {https://ui.adsabs.harvard.edu/abs/2020PASJ...72....4U}
}

@ARTICLE{2021Chibueze,
       author = {{Chibueze}, James O. and {MacLeod}, Gordon C. and {Vorster}, Jakobus M. and {Hirota}, Tomoya and {Brogan}, Crystal L. and {Hunter}, Todd R. and {van Rooyen}, Ruby},
        title = "{The Extraordinary Outburst in the Massive Protostellar System NGC 6334 I-MM1: Spatiokinematics of Water Masers during a Contemporaneous Flare Event}",
      journal = {\apj},
         year = 2021,
        month = feb,
       volume = {908},
       number = {2},
          eid = {175},
        pages = {175},
          doi = {10.3847/1538-4357/abd474},
archivePrefix = {arXiv},
       eprint = {2101.11913},
 primaryClass = {astro-ph.SR},
       adsurl = {https://ui.adsabs.harvard.edu/abs/2021ApJ...908..175C}
}

@ARTICLE{2022Hirota,
       author = {{Hirota}, Tomoya and {Wolak}, Pawel and {Hunter}, Todd R. and {Brogan}, Crystal L. and {Bartkiewicz}, Anna and {Durjasz}, Michal and {Kobak}, Agnieszka and {Olech}, Mateusz and {Szymczak}, Marian and {Burns}, Ross A. and {Aberfelds}, Artis and {Baek}, Giseon and {Brand}, Jan and {Breen}, Shari and {Byun}, Do-Young and {Caratti o Garatti}, Alessio and {Chen}, Xi and {Chibueze}, James O. and {Cyganowski}, Claudia and {Eisl{\"o}ffel}, Jochen and {Ellingsen}, Simon and {Hirano}, Naomi and {Hu}, Bo and {Kang}, Ji-hyun and {Kim}, Jeong-Sook and {Kim}, Jungha and {Kim}, Kee-Tae and {Kim}, Mi Kyoung and {Kramer}, Busaba and {Lee}, Jeong-Eun and {Linz}, Hendrik and {Liu}, Tie and {MacLeod}, Gordon and {McCarthy}, Tiege P. and {Menten}, Karl and {Motogi}, Kazuhito and {Oh}, Chung-Sik and {Orosz}, Gabor and {Sobolev}, Andrey M. and {Stecklum}, Bringfried and {Sugiyama}, Koichiro and {Sunada}, Kazuyoshi and {Uscanga}, Lucero and {van den Heever}, Fanie and {Volvach}, Alexandr E. and {Volvach}, Larisa N. and {Wu}, Yuan Wei and {Yonekura}, Yoshinori},
        title = "{Millimeter methanol emission in the high-mass young stellar object G24.33+0.14}",
      journal = {\pasj},
         year = 2022,
        month = oct,
       volume = {74},
       number = {5},
        pages = {1234-1262},
          doi = {10.1093/pasj/psac067},
       adsurl = {https://ui.adsabs.harvard.edu/abs/2022PASJ...74.1234H}
}

@ARTICLE{2024Wolf,
       author = {{Wolf}, V. and {Stecklum}, B. and {Caratti o Garatti}, A. and {Boley}, P.~A. and {Fischer}, Ch. and {Harries}, T. and {Eisl{\"o}ffel}, J. and {Linz}, H. and {Ahmadi}, A. and {Kobus}, J. and {Haubois}, X. and {Matter}, A. and {Cruzalebes}, P.},
        title = "{The accretion burst of the massive young stellar object G323.46{\ensuremath{-}}0.08}",
      journal = {\aap},
         year = 2024,
        month = aug,
       volume = {688},
          eid = {A8},
        pages = {A8},
          doi = {10.1051/0004-6361/202449891},
archivePrefix = {arXiv},
       eprint = {2405.10427},
 primaryClass = {astro-ph.SR},
       adsurl = {https://ui.adsabs.harvard.edu/abs/2024A&A...688A...8W}
}

@ARTICLE{2007Cyganowski,
       author = {{Cyganowski}, C.~J. and {Brogan}, C.~L. and {Hunter}, T.~R.},
        title = "{Evidence for a Massive Protocluster in S255N}",
      journal = {\aj},
         year = 2007,
        month = jul,
       volume = {134},
       number = {1},
        pages = {346-358},
          doi = {10.1086/518740},
archivePrefix = {arXiv},
       eprint = {0704.0988},
 primaryClass = {astro-ph},
       adsurl = {https://ui.adsabs.harvard.edu/abs/2007AJ....134..346C}
}

@ARTICLE{2013Palau,
       author = {{Palau}, Aina and {Fuente}, Asunci{\'o}n and {Girart}, Josep M. and {Estalella}, Robert and {Ho}, Paul T.~P. and {S{\'a}nchez-Monge}, {\'A}lvaro and {Fontani}, Francesco and {Busquet}, Gemma and {Commer{\c{c}}on}, Benoit and {Hennebelle}, Patrick and {Boissier}, J{\'e}r{\'e}mie and {Zhang}, Qizhou and {Cesaroni}, Riccardo and {Zapata}, Luis A.},
        title = "{Early Stages of Cluster Formation: Fragmentation of Massive Dense Cores down to <\raisebox{-0.5ex}\textasciitilde 1000 AU}",
      journal = {\apj},
         year = 2013,
        month = jan,
       volume = {762},
       number = {2},
          eid = {120},
        pages = {120},
          doi = {10.1088/0004-637X/762/2/120},
archivePrefix = {arXiv},
       eprint = {1211.2666},
 primaryClass = {astro-ph.GA},
       adsurl = {https://ui.adsabs.harvard.edu/abs/2013ApJ...762..120P}
}

@ARTICLE{2014Monge,
       author = {{S{\'a}nchez-Monge}, {\'A}. and {Beltr{\'a}n}, M.~T. and {Cesaroni}, R. and {Etoka}, S. and {Galli}, D. and {Kumar}, M.~S.~N. and {Moscadelli}, L. and {Stanke}, T. and {van der Tak}, F.~F.~S. and {Vig}, S. and {Walmsley}, C.~M. and {Wang}, K.-S. and {Zinnecker}, H. and {Elia}, D. and {Molinari}, S. and {Schisano}, E.},
        title = "{A necklace of dense cores in the high-mass star forming region G35.20-0.74 N: ALMA observations}",
      journal = {\aap},
         year = 2014,
        month = sep,
       volume = {569},
          eid = {A11},
        pages = {A11},
          doi = {10.1051/0004-6361/201424032},
archivePrefix = {arXiv},
       eprint = {1406.4081},
 primaryClass = {astro-ph.GA},
       adsurl = {https://ui.adsabs.harvard.edu/abs/2014A&A...569A..11S}
}

@ARTICLE{2009Evans,
       author = {{Evans}, II, Neal J. and {Dunham}, Michael M. and {J{\o}rgensen}, Jes K. and {Enoch}, Melissa L. and {Mer{\'\i}n}, Bruno and {van Dishoeck}, Ewine F. and {Alcal{\'a}}, Juan M. and {Myers}, Philip C. and {Stapelfeldt}, Karl R. and {Huard}, Tracy L. and {Allen}, Lori E. and {Harvey}, Paul M. and {van Kempen}, Tim and {Blake}, Geoffrey A. and {Koerner}, David W. and {Mundy}, Lee G. and {Padgett}, Deborah L. and {Sargent}, Anneila I.},
        title = "{The Spitzer c2d Legacy Results: Star-Formation Rates and Efficiencies; Evolution and Lifetimes}",
      journal = {\apjs},
         year = 2009,
        month = apr,
       volume = {181},
       number = {2},
        pages = {321-350},
          doi = {10.1088/0067-0049/181/2/321},
archivePrefix = {arXiv},
       eprint = {0811.1059},
 primaryClass = {astro-ph},
       adsurl = {https://ui.adsabs.harvard.edu/abs/2009ApJS..181..321E}
}

@ARTICLE{2018Kristensen,
       author = {{Kristensen}, L.~E. and {Dunham}, M.~M.},
        title = "{Protostellar half-life: new methodology and estimates}",
      journal = {\aap},
         year = 2018,
        month = oct,
       volume = {618},
          eid = {A158},
        pages = {A158},
          doi = {10.1051/0004-6361/201731584},
archivePrefix = {arXiv},
       eprint = {1807.11262},
 primaryClass = {astro-ph.SR},
       adsurl = {https://ui.adsabs.harvard.edu/abs/2018A&A...618A.158K}
}

@ARTICLE{2010Dunham,
       author = {{Dunham}, Michael M. and {Evans}, II, Neal J. and {Terebey}, Susan and {Dullemond}, Cornelis P. and {Young}, Chadwick H.},
        title = "{Evolutionary Signatures in the Formation of Low-Mass Protostars. II. Toward Reconciling Models and Observations}",
      journal = {\apj},
         year = 2010,
        month = feb,
       volume = {710},
       number = {1},
        pages = {470-502},
          doi = {10.1088/0004-637X/710/1/470},
archivePrefix = {arXiv},
       eprint = {0912.5229},
 primaryClass = {astro-ph.GA},
       adsurl = {https://ui.adsabs.harvard.edu/abs/2010ApJ...710..470D}
}

@ARTICLE{2015Safron,
       author = {{Safron}, Emily J. and {Fischer}, William J. and {Megeath}, S. Thomas and {Furlan}, Elise and {Stutz}, Amelia M. and {Stanke}, Thomas and {Billot}, Nicolas and {Rebull}, Luisa M. and {Tobin}, John J. and {Ali}, Babar and {Allen}, Lori E. and {Booker}, Joseph and {Watson}, Dan M. and {Wilson}, T.~L.},
        title = "{Hops 383: an Outbursting Class 0 Protostar in Orion}",
      journal = {\apjl},
         year = 2015,
        month = feb,
       volume = {800},
       number = {1},
          eid = {L5},
        pages = {L5},
          doi = {10.1088/2041-8205/800/1/L5},
archivePrefix = {arXiv},
       eprint = {1501.00492},
 primaryClass = {astro-ph.SR},
       adsurl = {https://ui.adsabs.harvard.edu/abs/2015ApJ...800L...5S}
}

@ARTICLE{2025Sheehan,
       author = {{Sheehan}, Patrick D. and {Johnstone}, Doug and {Contreras Pe{\~n}a}, Carlos and {Lee}, Seonjae and {Herczeg}, Gregory and {Lee}, Jeong-Eun and {Mairs}, Steve and {Tobin}, John J. and {Yun}, Hyeong-Sik and {The JCMT Transient Team}},
        title = "{Submillimeter Variability in the Envelope and Warped Protostellar Disk of the Class 0 Protostar HOPS 358}",
      journal = {\apj},
         year = 2025,
        month = apr,
       volume = {982},
       number = {2},
          eid = {176},
        pages = {176},
          doi = {10.3847/1538-4357/adaf9b},
archivePrefix = {arXiv},
       eprint = {2502.15887},
 primaryClass = {astro-ph.SR},
       adsurl = {https://ui.adsabs.harvard.edu/abs/2025ApJ...982..176S}
}

@ARTICLE{2015Vorobyov,
       author = {{Vorobyov}, Eduard I. and {Basu}, Shantanu},
        title = "{Variable Protostellar Accretion with Episodic Bursts}",
      journal = {\apj},
         year = 2015,
        month = jun,
       volume = {805},
       number = {2},
          eid = {115},
        pages = {115},
          doi = {10.1088/0004-637X/805/2/115},
archivePrefix = {arXiv},
       eprint = {1503.07888},
 primaryClass = {astro-ph.SR},
       adsurl = {https://ui.adsabs.harvard.edu/abs/2015ApJ...805..115V}
}

@ARTICLE{2016Hosokawa,
       author = {{Hosokawa}, Takashi and {Hirano}, Shingo and {Kuiper}, Rolf and {Yorke}, Harold W. and {Omukai}, Kazuyuki and {Yoshida}, Naoki},
        title = "{Formation of Massive Primordial Stars: Intermittent UV Feedback with Episodic Mass Accretion}",
      journal = {\apj},
         year = 2016,
        month = jun,
       volume = {824},
       number = {2},
          eid = {119},
        pages = {119},
          doi = {10.3847/0004-637X/824/2/119},
archivePrefix = {arXiv},
       eprint = {1510.01407},
 primaryClass = {astro-ph.GA},
       adsurl = {https://ui.adsabs.harvard.edu/abs/2016ApJ...824..119H}
}

@ARTICLE{2019Francis,
       author = {{Francis}, Logan and {Johnstone}, Doug and {Dunham}, Michael M. and {Hunter}, Todd R. and {Mairs}, Steve},
        title = "{Identifying Variability in Deeply Embedded Protostars with ALMA and CARMA}",
      journal = {\apj},
         year = 2019,
        month = feb,
       volume = {871},
       number = {2},
          eid = {149},
        pages = {149},
          doi = {10.3847/1538-4357/aaf972},
archivePrefix = {arXiv},
       eprint = {1902.00588},
 primaryClass = {astro-ph.SR},
       adsurl = {https://ui.adsabs.harvard.edu/abs/2019ApJ...871..149F}
}

@ARTICLE{2018Liu,
       author = {{Liu}, Hauyu Baobab and {Dunham}, Michael M. and {Pascucci}, Ilaria and {Bourke}, Tyler L. and {Hirano}, Naomi and {Longmore}, Steven and {Andrews}, Sean and {Carrasco-Gonz{\'a}lez}, Carlos and {Forbrich}, Jan and {Galv{\'a}n-Madrid}, Roberto and {Girart}, Josep M. and {Green}, Joel D. and {Ju{\'a}rez}, Carmen and {K{\'o}sp{\'a}l}, {\'A}gnes and {Manara}, Carlo F. and {Palau}, Aina and {Takami}, Michihiro and {Testi}, Leonardo and {Vorobyov}, Eduard I.},
        title = "{A 1.3 mm SMA survey of 29 variable young stellar objects}",
      journal = {\aap},
         year = 2018,
        month = apr,
       volume = {612},
          eid = {A54},
        pages = {A54},
          doi = {10.1051/0004-6361/201731951},
archivePrefix = {arXiv},
       eprint = {1710.08686},
 primaryClass = {astro-ph.SR},
       adsurl = {https://ui.adsabs.harvard.edu/abs/2018A&A...612A..54L}
}

@ARTICLE{2023Vargas,
       author = {{Vargas-Gonz{\'a}lez}, J. and {Forbrich}, J. and {Rivilla}, V.~M. and {Menten}, K.~M. and {G{\"u}del}, M. and {Hacar}, A.},
        title = "{A systematic survey of millimetre-wavelength flaring variability of young stellar objects in the Orion Nebula Cluster}",
      journal = {\mnras},
         year = 2023,
        month = jun,
       volume = {522},
       number = {1},
        pages = {56-69},
          doi = {10.1093/mnras/stad926},
archivePrefix = {arXiv},
       eprint = {2303.15516},
 primaryClass = {astro-ph.SR},
       adsurl = {https://ui.adsabs.harvard.edu/abs/2023MNRAS.522...56V}
}

@ARTICLE{2020Wendeborn,
       author = {{Wendeborn}, John and {Espaillat}, Catherine C. and {Mac{\'\i}as}, Enrique and {Feh{\'e}r}, Orsolya and {K{\'o}sp{\'a}l}, {\'A}. and {Hartmann}, Lee and {Zhu}, Zhaohuan and {Dunham}, Michael M. and {Kounkel}, Marina},
        title = "{A Study of Millimeter Variability in FUor Objects}",
      journal = {\apj},
         year = 2020,
        month = jul,
       volume = {897},
       number = {1},
          eid = {54},
        pages = {54},
          doi = {10.3847/1538-4357/ab9304},
archivePrefix = {arXiv},
       eprint = {2005.10371},
 primaryClass = {astro-ph.SR},
       adsurl = {https://ui.adsabs.harvard.edu/abs/2020ApJ...897...54W}
}

@ARTICLE{2025Laznevoi,
       author = {{Laznevoi}, S.~I. and {Akimkin}, V.~V. and {Pavlyuchenkov}, Ya. N. and {Il'in}, V.~B. and {K{\'o}sp{\'a}l}, {\'A}. and {{\'A}brah{\'a}m}, P.},
        title = "{Time-dependent response of protoplanetary disk temperature to an FU Ori-type luminosity outburst}",
      journal = {\aap},
         year = 2025,
        month = aug,
       volume = {700},
          eid = {L24},
        pages = {L24},
          doi = {10.1051/0004-6361/202554962},
archivePrefix = {arXiv},
       eprint = {2508.04686},
 primaryClass = {astro-ph.EP},
       adsurl = {https://ui.adsabs.harvard.edu/abs/2025A&A...700L..24L}
}

@ARTICLE{2003Minier,
       author = {{Minier}, V. and {Ellingsen}, S.~P. and {Norris}, R.~P. and {Booth}, R.~S.},
        title = "{The protostellar mass limit for 6.7 GHz methanol masers.  I. A low-mass YSO survey}",
      journal = {\aap},
         year = 2003,
        month = jun,
       volume = {403},
        pages = {1095-1100},
          doi = {10.1051/0004-6361:20030465},
       adsurl = {https://ui.adsabs.harvard.edu/abs/2003A&A...403.1095M}
}

@ARTICLE{2008Xu,
       author = {{Xu}, Y. and {Li}, J.~J. and {Hachisuka}, K. and {Pandian}, J.~D. and {Menten}, K.~M. and {Henkel}, C.},
        title = "{A high-sensitivity 6.7 GHz methanol maser survey toward H2O sources}",
      journal = {\aap},
         year = 2008,
        month = jul,
       volume = {485},
       number = {3},
        pages = {729-734},
          doi = {10.1051/0004-6361:200809472},
archivePrefix = {arXiv},
       eprint = {0803.2232},
 primaryClass = {astro-ph},
       adsurl = {https://ui.adsabs.harvard.edu/abs/2008A&A...485..729X}
}

@ARTICLE{2025OYCAT,
       author = {{Contreras Pe{\~n}a}, Carlos and {Lee}, Jeong-Eun and {Herczeg}, Gregory and {Johnstone}, Doug and {{\'A}brah{\'a}m}, P{\'e}ter and {Antoniucci}, Simone and {Audard}, Marc and {Ashraf}, Mizna and {Baek}, Giseon and {Garatti}, Alessio Caratti o. and {Carvalho}, Adolfo and {Cieza}, Lucas and {Cruz-Sa{\'e}nz de Miera}, Fernando and {Eisl{\"o}ffel}, Jochen and {Froebrich}, Dirk and {Giannini}, Teresa and {Green}, Joel and {Ghosh}, Arpan and {Guo}, Zhen and {Hillenbrand}, Lynne and {Hodapp}, Klaus and {Jheonn}, Hyunwook and {Jose}, Jessy and {Kim}, Young-Jun and {K{\'o}sp{\'a}l}, {\'A}gnes and {Lee}, Ho-Gyu and {Lucas}, Philip W. and {Magakian}, Tigran and {Nagy}, Zs{\'o}fia and {Naylor}, Tim and {Ninan}, Joe P. and {Peneva}, S. and {Reipurth}, Bo and {Scholz}, Alexander and {Semkov}, E. and {Sicilia-Aguilar}, Aurora and {Singh}, Koshvendra and {Siwak}, Michal and {Stecklum}, Bringfried and {Szab{\'o}}, Zs{\'o}fia Marianna and {Wolf}, Verena and {Yoon}, Sung-Yong},
        title = "{The Outbursting YSOs Catalogue (OYCAT)}",
      journal = {Journal of Korean Astronomical Society},
         year = 2025,
        month = sep,
       volume = {58},
        pages = {209-230},
          doi = {10.5303/JKAS.2025.58.2.209},
archivePrefix = {arXiv},
       eprint = {2509.24876},
 primaryClass = {astro-ph.SR},
       adsurl = {https://ui.adsabs.harvard.edu/abs/2025JKAS...58..209C}
}

@ARTICLE{2025Contreras,
       author = {{Contreras Pe{\~n}a}, Carlos and {Lee}, Jeong-Eun and {Lee}, Ho-Gyu and {Herczeg}, Gregory and {Johnstone}, Doug and {Liu}, Hanpu and {Lucas}, Philip W. and {Guo}, Zhen and {Kuhn}, Michael A. and {Smith}, Leigh C. and {Ashraf}, Mizna and {Jose}, Jessy and {Yoon}, Sung-Yong and {Yoon}, Sung-Chul},
        title = "{``Oh FUors, Where Art Thou?'': A Search for Long-lasting Young Stellar Object Outbursts Hiding in Infrared Surveys}",
      journal = {\apj},
         year = 2025,
        month = jul,
       volume = {987},
       number = {1},
          eid = {23},
        pages = {23},
          doi = {10.3847/1538-4357/add25f},
archivePrefix = {arXiv},
       eprint = {2504.21237},
 primaryClass = {astro-ph.SR},
       adsurl = {https://ui.adsabs.harvard.edu/abs/2025ApJ...987...23C}
}

@ARTICLE{2024Park,
       author = {{Park}, Geumsook and {Johnstone}, Doug and {Pe{\~n}a}, Carlos Contreras and {Lee}, Jeong-Eun and {Liu}, Sheng-Yuan and {Herczeg}, Gregory and {Mairs}, Steve and {Chen}, Zhiwei and {Hatchell}, Jennifer and {Kim}, Kee-Tae and {Kim}, Mi-Ryang and {Qiu}, Keping and {Wang}, Yao-Te and {Zhang}, Xu and {The JCMT Transient Team}},
        title = "{Submillimeter and Mid-Infrared Variability of Young Stellar Objects in the M17SWex Intermediate-mass Star-forming Region}",
      journal = {\aj},
         year = 2024,
        month = sep,
       volume = {168},
       number = {3},
          eid = {122},
        pages = {122},
          doi = {10.3847/1538-3881/ad5e6e},
archivePrefix = {arXiv},
       eprint = {2407.03445},
 primaryClass = {astro-ph.GA},
       adsurl = {https://ui.adsabs.harvard.edu/abs/2024AJ....168..122P}
}

@ARTICLE{2022Francis,
       author = {{Francis}, Logan and {Johnstone}, Doug and {Lee}, Jeong-Eun and {Herczeg}, Gregory J. and {Long}, Feng and {Mairs}, Steve and {Contreras Pe{\~n}a}, Carlos and {Moriarty-Schieven}, Gerald and {JCMT Transient Team}},
        title = "{Accretion Burst Echoes as Probes of Protostellar Environments and Episodic Mass Assembly}",
      journal = {\apj},
         year = 2022,
        month = sep,
       volume = {937},
       number = {1},
          eid = {29},
        pages = {29},
          doi = {10.3847/1538-4357/ac8a9e},
archivePrefix = {arXiv},
       eprint = {2208.13568},
 primaryClass = {astro-ph.SR},
       adsurl = {https://ui.adsabs.harvard.edu/abs/2022ApJ...937...29F}
}

@unpublished{Wang_prep,
  author = {Wang, X. and others},
  title = {Work in preparation},
  note = {in prep.},
  year = {in prep.}
}

@unpublished{Zhang_prep,
  author = {Zhang, X. and others},
  title = {Work in preparation},
  note = {in prep.},
  year = {in prep.}
}

@ARTICLE{2024Xu,
       author = {{Xu}, Fengwei and {Wang}, Ke and {Liu}, Tie and {Zhu}, Lei and {Garay}, Guido and {Liu}, Xunchuan and {Goldsmith}, Paul and {Zhang}, Qizhou and {Sanhueza}, Patricio and {Qin}, Shengli and {He}, Jinhua and {Juvela}, Mika and {Tej}, Anandmayee and {Liu}, Hongli and {Li}, Shanghuo and {Morii}, Kaho and {Zhang}, Siju and {Zhou}, Jianwen and {Stutz}, Amelia and {Evans}, Neal J. and {Kim}, Kee-Tae and {Liu}, Shengyuan and {Mardones}, Diego and {Li}, Guangxing and {Bronfman}, Leonardo and {Tatematsu}, Ken'ichi and {Lee}, Chang Won and {Lu}, Xing and {Mai}, Xiaofeng and {Jiao}, Sihan and {Chibueze}, James O. and {Su}, Keyun and {T{\'o}th}, Viktor L.},
        title = "{The ALMA-QUARKS Survey. II. The ACA 1.3 mm Continuum Source Catalog and the Assembly of Dense Gas in Massive Star-Forming Clumps}",
      journal = {Research in Astronomy and Astrophysics},
         year = 2024,
        month = jun,
       volume = {24},
       number = {6},
          eid = {065011},
        pages = {065011},
          doi = {10.1088/1674-4527/ad3dc3},
archivePrefix = {arXiv},
       eprint = {2404.02275},
 primaryClass = {astro-ph.GA},
       adsurl = {https://ui.adsabs.harvard.edu/abs/2024RAA....24f5011X}
}

@ARTICLE{2020Francis,
       author = {{Francis}, Logan and {Johnstone}, Doug and {Herczeg}, Gregory and {Hunter}, Todd R. and {Harsono}, Daniel},
        title = "{On the Accuracy of the ALMA Flux Calibration in the Time Domain and across Spectral Windows}",
      journal = {\aj},
         year = 2020,
        month = dec,
       volume = {160},
       number = {6},
          eid = {270},
        pages = {270},
          doi = {10.3847/1538-3881/abbe1a},
archivePrefix = {arXiv},
       eprint = {2010.02186},
 primaryClass = {astro-ph.IM},
       adsurl = {https://ui.adsabs.harvard.edu/abs/2020AJ....160..270F}
}

@ARTICLE{2019Yang,
       author = {{Yang}, A.~Y. and {Thompson}, M.~A. and {Tian}, W.~W. and {Bihr}, S. and {Beuther}, H. and {Hindson}, L.},
        title = "{A search for hypercompact H II regions in the Galactic Plane}",
      journal = {\mnras},
         year = 2019,
        month = jan,
       volume = {482},
       number = {2},
        pages = {2681-2696},
          doi = {10.1093/mnras/sty2811},
archivePrefix = {arXiv},
       eprint = {1809.00404},
 primaryClass = {astro-ph.GA},
       adsurl = {https://ui.adsabs.harvard.edu/abs/2019MNRAS.482.2681Y}
}

@INPROCEEDINGS{2005Kurtz,
       author = {{Kurtz}, Stan},
        title = "{Hypercompact HII regions}",
    booktitle = {Massive Star Birth: A Crossroads of Astrophysics},
         year = 2005,
       editor = {{Cesaroni}, R. and {Felli}, M. and {Churchwell}, E. and {Walmsley}, M.},
       series = {IAU Symposium},
       volume = {227},
        month = jan,
        pages = {111-119},
          doi = {10.1017/S1743921305004424},
       adsurl = {https://ui.adsabs.harvard.edu/abs/2005IAUS..227..111K}
}

@ARTICLE{2017Mairs,
       author = {{Mairs}, Steve and {Lane}, James and {Johnstone}, Doug and {Kirk}, Helen and {Lacaille}, Kevin and {Bower}, Geoffrey C. and {Bell}, Graham S. and {Graves}, Sarah and {Chapman}, Scott and {JCMT Transient Team}},
        title = "{The JCMT Transient Survey: Data Reduction and Calibration Methods}",
      journal = {\apj},
         year = 2017,
        month = jul,
       volume = {843},
       number = {1},
          eid = {55},
        pages = {55},
          doi = {10.3847/1538-4357/aa7844},
archivePrefix = {arXiv},
       eprint = {1706.01897},
 primaryClass = {astro-ph.IM},
       adsurl = {https://ui.adsabs.harvard.edu/abs/2017ApJ...843...55M}
}

@ARTICLE{2005Vorobyov,
       author = {{Vorobyov}, E.~I. and {Basu}, Shantanu},
        title = "{The Origin of Episodic Accretion Bursts in the Early Stages of Star Formation}",
      journal = {\apjl},
         year = 2005,
        month = nov,
       volume = {633},
       number = {2},
        pages = {L137-L140},
          doi = {10.1086/498303},
archivePrefix = {arXiv},
       eprint = {astro-ph/0510014},
 primaryClass = {astro-ph},
       adsurl = {https://ui.adsabs.harvard.edu/abs/2005ApJ...633L.137V}
}

@ARTICLE{2011Machida,
       author = {{Machida}, Masahiro N. and {Inutsuka}, Shu-ichiro and {Matsumoto}, Tomoaki},
        title = "{Recurrent Planet Formation and Intermittent Protostellar Outflows Induced by Episodic Mass Accretion}",
      journal = {\apj},
         year = 2011,
        month = mar,
       volume = {729},
       number = {1},
          eid = {42},
        pages = {42},
          doi = {10.1088/0004-637X/729/1/42},
archivePrefix = {arXiv},
       eprint = {1101.1997},
 primaryClass = {astro-ph.SR},
       adsurl = {https://ui.adsabs.harvard.edu/abs/2011ApJ...729...42M}
}

@ARTICLE{2012Nayakshin,
       author = {{Nayakshin}, Sergei and {Lodato}, Giuseppe},
        title = "{Fu Ori outbursts and the planet-disc mass exchange}",
      journal = {\mnras},
         year = 2012,
        month = oct,
       volume = {426},
       number = {1},
        pages = {70-90},
          doi = {10.1111/j.1365-2966.2012.21612.x},
archivePrefix = {arXiv},
       eprint = {1110.6316},
 primaryClass = {astro-ph.EP},
       adsurl = {https://ui.adsabs.harvard.edu/abs/2012MNRAS.426...70N}
}

@ARTICLE{2021Meyer,
       author = {{Meyer}, D.~M.-A. and {Vorobyov}, E.~I. and {Elbakyan}, V.~G. and {Eisl{\"o}ffel}, J. and {Sobolev}, A.~M. and {St{\"o}hr}, M.},
        title = "{Parameter study for the burst mode of accretion in massive star formation}",
      journal = {\mnras},
         year = 2021,
        month = jan,
       volume = {500},
       number = {4},
        pages = {4448-4468},
          doi = {10.1093/mnras/staa3528},
archivePrefix = {arXiv},
       eprint = {2011.05017},
 primaryClass = {astro-ph.SR},
       adsurl = {https://ui.adsabs.harvard.edu/abs/2021MNRAS.500.4448M}
}

@ARTICLE{2018Meyer,
        author = {{Meyer}, D.~M.-A. and {Kuiper}, R. and {Kley}, W. and 
{Johnston}, K.~G. and {Vorobyov}, E.},
         title = "{Forming spectroscopic massive protobinaries by disc 
fragmentation}",
       journal = {\mnras},
          year = 2018,
         month = jan,
        volume = {473},
        number = {3},
         pages = {3615-3637},
           doi = {10.1093/mnras/stx2551},
archivePrefix = {arXiv},
        eprint = {1710.01162},
  primaryClass = {astro-ph.SR},
        adsurl = {https://ui.adsabs.harvard.edu/abs/2018MNRAS.473.3615M}
}

@ARTICLE{2022Meyer,
        author = {{Meyer}, D.~M.-A. and {Vorobyov}, E.~I. and 
{Elbakyan}, V.~G. and {Kraus}, S. and {Liu}, S.-Y. and {Nayakshin}, S. 
and {Sobolev}, A.~M.},
         title = "{The burst mode of accretion in massive star formation 
with stellar inertia}",
       journal = {\mnras},
          year = 2022,
         month = dec,
        volume = {517},
        number = {4},
         pages = {4795-4812},
           doi = {10.1093/mnras/stac2956},
archivePrefix = {arXiv},
        eprint = {2210.09662},
  primaryClass = {astro-ph.SR},
        adsurl = {https://ui.adsabs.harvard.edu/abs/2022MNRAS.517.4795M}
}

@ARTICLE{2022AstropyCollaboration,
       author = {{Astropy Collaboration} and {Price-Whelan}, Adrian M. and {Lim}, Pey Lian and {Earl}, Nicholas and {Starkman}, Nathaniel and {Bradley}, Larry and {Shupe}, David L. and {Patil}, Aarya A. and {Corrales}, Lia and {Brasseur}, C.~E. and {N{\"o}the}, Maximilian and {Donath}, Axel and {Tollerud}, Erik and {Morris}, Brett M. and {Ginsburg}, Adam and {Vaher}, Eero and {Weaver}, Benjamin A. and {Tocknell}, James and {Jamieson}, William and {van Kerkwijk}, Marten H. and {Robitaille}, Thomas P. and {Merry}, Bruce and {Bachetti}, Matteo and {G{\"u}nther}, H. Moritz and {Aldcroft}, Thomas L. and {Alvarado-Montes}, Jaime A. and {Archibald}, Anne M. and {B{\'o}di}, Attila and {Bapat}, Shreyas and {Barentsen}, Geert and {Baz{\'a}n}, Juanjo and {Biswas}, Manish and {Boquien}, M{\'e}d{\'e}ric and {Burke}, D.~J. and {Cara}, Daria and {Cara}, Mihai and {Conroy}, Kyle E. and {Conseil}, Simon and {Craig}, Matthew W. and {Cross}, Robert M. and {Cruz}, Kelle L. and {D'Eugenio}, Francesco and {Dencheva}, Nadia and {Devillepoix}, Hadrien A.~R. and {Dietrich}, J{\"o}rg P. and {Eigenbrot}, Arthur Davis and {Erben}, Thomas and {Ferreira}, Leonardo and {Foreman-Mackey}, Daniel and {Fox}, Ryan and {Freij}, Nabil and {Garg}, Suyog and {Geda}, Robel and {Glattly}, Lauren and {Gondhalekar}, Yash and {Gordon}, Karl D. and {Grant}, David and {Greenfield}, Perry and {Groener}, Austen M. and {Guest}, Steve and {Gurovich}, Sebastian and {Handberg}, Rasmus and {Hart}, Akeem and {Hatfield-Dodds}, Zac and {Homeier}, Derek and {Hosseinzadeh}, Griffin and {Jenness}, Tim and {Jones}, Craig K. and {Joseph}, Prajwel and {Kalmbach}, J. Bryce and {Karamehmetoglu}, Emir and {Ka{\l}uszy{\'n}ski}, Miko{\l}aj and {Kelley}, Michael S.~P. and {Kern}, Nicholas and {Kerzendorf}, Wolfgang E. and {Koch}, Eric W. and {Kulumani}, Shankar and {Lee}, Antony and {Ly}, Chun and {Ma}, Zhiyuan and {MacBride}, Conor and {Maljaars}, Jakob M. and {Muna}, Demitri and {Murphy}, N.~A. and {Norman}, Henrik and {O'Steen}, Richard and {Oman}, Kyle A. and {Pacifici}, Camilla and {Pascual}, Sergio and {Pascual-Granado}, J. and {Patil}, Rohit R. and {Perren}, Gabriel I. and {Pickering}, Timothy E. and {Rastogi}, Tanuj and {Roulston}, Benjamin R. and {Ryan}, Daniel F. and {Rykoff}, Eli S. and {Sabater}, Jose and {Sakurikar}, Parikshit and {Salgado}, Jes{\'u}s and {Sanghi}, Aniket and {Saunders}, Nicholas and {Savchenko}, Volodymyr and {Schwardt}, Ludwig and {Seifert-Eckert}, Michael and {Shih}, Albert Y. and {Jain}, Anany Shrey and {Shukla}, Gyanendra and {Sick}, Jonathan and {Simpson}, Chris and {Singanamalla}, Sudheesh and {Singer}, Leo P. and {Singhal}, Jaladh and {Sinha}, Manodeep and {Sip{\H{o}}cz}, Brigitta M. and {Spitler}, Lee R. and {Stansby}, David and {Streicher}, Ole and {{\v{S}}umak}, Jani and {Swinbank}, John D. and {Taranu}, Dan S. and {Tewary}, Nikita and {Tremblay}, Grant R. and {de Val-Borro}, Miguel and {Van Kooten}, Samuel J. and {Vasovi{\'c}}, Zlatan and {Verma}, Shresth and {de Miranda Cardoso}, Jos{\'e} Vin{\'\i}cius and {Williams}, Peter K.~G. and {Wilson}, Tom J. and {Winkel}, Benjamin and {Wood-Vasey}, W.~M. and {Xue}, Rui and {Yoachim}, Peter and {Zhang}, Chen and {Zonca}, Andrea and {Astropy Project Contributors}},
        title = "{The Astropy Project: Sustaining and Growing a Community-oriented Open-source Project and the Latest Major Release (v5.0) of the Core Package}",
      journal = {\apj},
         year = 2022,
        month = aug,
       volume = {935},
       number = {2},
          eid = {167},
        pages = {167},
          doi = {10.3847/1538-4357/ac7c74},
archivePrefix = {arXiv},
       eprint = {2206.14220},
 primaryClass = {astro-ph.IM},
       adsurl = {https://ui.adsabs.harvard.edu/abs/2022ApJ...935..167A}
}

@ARTICLE{2018AstropyCollaboration,
       author = {{Astropy Collaboration} and {Price-Whelan}, A.~M. and {Sip{\H{o}}cz}, B.~M. and {G{\"u}nther}, H.~M. and {Lim}, P.~L. and {Crawford}, S.~M. and {Conseil}, S. and {Shupe}, D.~L. and {Craig}, M.~W. and {Dencheva}, N. and {Ginsburg}, A. and {VanderPlas}, J.~T. and {Bradley}, L.~D. and {P{\'e}rez-Su{\'a}rez}, D. and {de Val-Borro}, M. and {Aldcroft}, T.~L. and {Cruz}, K.~L. and {Robitaille}, T.~P. and {Tollerud}, E.~J. and {Ardelean}, C. and {Babej}, T. and {Bach}, Y.~P. and {Bachetti}, M. and {Bakanov}, A.~V. and {Bamford}, S.~P. and {Barentsen}, G. and {Barmby}, P. and {Baumbach}, A. and {Berry}, K.~L. and {Biscani}, F. and {Boquien}, M. and {Bostroem}, K.~A. and {Bouma}, L.~G. and {Brammer}, G.~B. and {Bray}, E.~M. and {Breytenbach}, H. and {Buddelmeijer}, H. and {Burke}, D.~J. and {Calderone}, G. and {Cano Rodr{\'\i}guez}, J.~L. and {Cara}, M. and {Cardoso}, J.~V.~M. and {Cheedella}, S. and {Copin}, Y. and {Corrales}, L. and {Crichton}, D. and {D'Avella}, D. and {Deil}, C. and {Depagne}, {\'E}. and {Dietrich}, J.~P. and {Donath}, A. and {Droettboom}, M. and {Earl}, N. and {Erben}, T. and {Fabbro}, S. and {Ferreira}, L.~A. and {Finethy}, T. and {Fox}, R.~T. and {Garrison}, L.~H. and {Gibbons}, S.~L.~J. and {Goldstein}, D.~A. and {Gommers}, R. and {Greco}, J.~P. and {Greenfield}, P. and {Groener}, A.~M. and {Grollier}, F. and {Hagen}, A. and {Hirst}, P. and {Homeier}, D. and {Horton}, A.~J. and {Hosseinzadeh}, G. and {Hu}, L. and {Hunkeler}, J.~S. and {Ivezi{\'c}}, {\v{Z}}. and {Jain}, A. and {Jenness}, T. and {Kanarek}, G. and {Kendrew}, S. and {Kern}, N.~S. and {Kerzendorf}, W.~E. and {Khvalko}, A. and {King}, J. and {Kirkby}, D. and {Kulkarni}, A.~M. and {Kumar}, A. and {Lee}, A. and {Lenz}, D. and {Littlefair}, S.~P. and {Ma}, Z. and {Macleod}, D.~M. and {Mastropietro}, M. and {McCully}, C. and {Montagnac}, S. and {Morris}, B.~M. and {Mueller}, M. and {Mumford}, S.~J. and {Muna}, D. and {Murphy}, N.~A. and {Nelson}, S. and {Nguyen}, G.~H. and {Ninan}, J.~P. and {N{\"o}the}, M. and {Ogaz}, S. and {Oh}, S. and {Parejko}, J.~K. and {Parley}, N. and {Pascual}, S. and {Patil}, R. and {Patil}, A.~A. and {Plunkett}, A.~L. and {Prochaska}, J.~X. and {Rastogi}, T. and {Reddy Janga}, V. and {Sabater}, J. and {Sakurikar}, P. and {Seifert}, M. and {Sherbert}, L.~E. and {Sherwood-Taylor}, H. and {Shih}, A.~Y. and {Sick}, J. and {Silbiger}, M.~T. and {Singanamalla}, S. and {Singer}, L.~P. and {Sladen}, P.~H. and {Sooley}, K.~A. and {Sornarajah}, S. and {Streicher}, O. and {Teuben}, P. and {Thomas}, S.~W. and {Tremblay}, G.~R. and {Turner}, J.~E.~H. and {Terr{\'o}n}, V. and {van Kerkwijk}, M.~H. and {de la Vega}, A. and {Watkins}, L.~L. and {Weaver}, B.~A. and {Whitmore}, J.~B. and {Woillez}, J. and {Zabalza}, V. and {Astropy Contributors}},
        title = "{The Astropy Project: Building an Open-science Project and Status of the v2.0 Core Package}",
      journal = {\aj},
         year = 2018,
        month = sep,
       volume = {156},
       number = {3},
          eid = {123},
        pages = {123},
          doi = {10.3847/1538-3881/aabc4f},
archivePrefix = {arXiv},
       eprint = {1801.02634},
 primaryClass = {astro-ph.IM},
       adsurl = {https://ui.adsabs.harvard.edu/abs/2018AJ....156..123A}
}

@unpublished{Jiao_prep,
  author = {Jiao, YF. and others},
  title = {Work in preparation},
  note = {in prep.},
  year = {in prep.}
}

@unpublished{Liu_prep,
  author = {Liu, SY. and others},
  title = {Work in preparation},
  note = {in prep.},
  year = {in prep.}
}

@ARTICLE{2001Gordon,
       author = {{Gordon}, M.~A. and {Holder}, B.~P. and {Jisonna}, Jr., L.~J. and {Jorgenson}, R.~A. and {Strelnitski}, V.~S.},
        title = "{3 Year Monitoring of Millimeter-Wave Radio Recombination Lines from MWC 349}",
      journal = {\apj},
         year = 2001,
        month = sep,
       volume = {559},
       number = {1},
        pages = {402-418},
          doi = {10.1086/322328},
       adsurl = {https://ui.adsabs.harvard.edu/abs/2001ApJ...559..402G}
}

\appendix
\section{Observing Parameters}
\label{appendix:observing_parameters}
\setcounter{table}{0}
\renewcommand{\thetable}{A\arabic{table}}
\startlongtable 
\begin{deluxetable*}{rlrrrrlllll} 
\renewcommand{\arraystretch}{0.76} 
\setlength{\tabcolsep}{5pt}
\tablecaption{Observing Parameters}
\label{tab:almaobs}
\tablehead{
\colhead{ID} &
\colhead{Source Name\tablenotemark{a}} &
\colhead{R.A. (J2000)} &
\colhead{Decl. (J2000)} &
\multicolumn{2}{c}{Calibrators} &
\colhead{Min./Max. BL} &
\colhead{Obs. Date} &
\colhead{Beam Size\tablenotemark{b}} &
\colhead{$D$\tablenotemark{c}} \\
\cline{5-6}
\colhead{} &
\colhead{} &
\colhead{(h:m:s)} &
\colhead{(d:m:s)} &
\colhead{Phase} &
\colhead{Bandpass/Flux} &
\colhead{(m/m)} &
\colhead{(yyyy-mm-dd)} &
\colhead{($\arcsec\times\arcsec$)} &
\colhead{(kpc)}
}
\startdata
\multirow{3}{*}{1} & G294.52-1.62 & 11:35:34.06 & -63:14:49.30 & J1123-6417 & J1107-4449 & 15.1/783.5  & 2022-06-02 & \multirow{3}{*}{\(0.62 \times 0.45\)} & \multirow{3}{*}{1.40} \\
\phn&  &  &  &  &  & 15.3/1210.6 & 2022-08-21 &  &  \\
\phn& I11332-6258 & 11:35:32.23 & -63:14:46.80 & J1047-6217 & J1107-4449 & 15.1/1397.8 & 2024-05-31 &  &  \\
\hline
\multirow{2}{*}{2} & I12320-6122 & 12:34:53.38 & -61:39:46.90 & J1337-6509 & J1617-5848 & 15.3/2516.8 & 2023-05-02 & \multirow{2}{*}{\(0.34 \times 0.27\)} & \multirow{2}{*}{4.17} \\
\phn&  &  &  &  &  & 15.1/1397.8 & 2024-06-02 &  &  \\
\hline
\multirow{2}{*}{3} & I12326-6245 & 12:35:34.81 & -63:02:32.10 & J1337-6509 & J1617-5848 & 15.3/2516.8 & 2023-05-02 & \multirow{2}{*}{\(0.35 \times 0.27\)} & \multirow{2}{*}{4.21} \\
\phn&  &  &  &  &  & 15.1/1397.8 & 2024-06-02 &  &  \\
\hline
\multirow{2}{*}{4} & I12383-6128 & 12:41:17.32 & -61:44:38.60 & J1337-6509 & J1617-5848 & 15.3/2516.8 & 2023-05-02 & \multirow{2}{*}{\(0.34 \times 0.27\)} & \multirow{2}{*}{4.12} \\
\phn&  &  &  &  &  & 15.1/1397.8 & 2024-06-02 &  &  \\
\hline
\multirow{2}{*}{5} & I12572-6316\_1 & 13:00:24.03 & -63:32:31.90 & J1337-6509 & J1617-5848 & 15.3/2516.8 & 2023-05-02 & \multirow{2}{*}{\(0.35 \times 0.27\)} & \multirow{2}{*}{11.63} \\
\phn&  &  &  &  &  & 15.1/1397.8 & 2024-06-02 &  &  \\
\hline
\multirow{2}{*}{6} & I12572-6316\_2 & 13:00:28.73 & -63:32:37.30 & J1337-6509 & J1617-5848 & 15.3/2516.8 & 2023-05-02 & \multirow{2}{*}{\(0.34 \times 0.27\)} & \multirow{2}{*}{11.63} \\
\phn&  &  &  &  &  & 15.1/1397.8 & 2024-06-02 &  &  \\
\hline
\multirow{3}{*}{7} & G305.21+0.21 & 13:11:13.78 & -62:34:41.90 & J1337-6509 & J1427-4206 & 15.1/783.5  & 2022-06-02 & \multirow{3}{*}{\(0.62 \times 0.43\)} & \multirow{3}{*}{3.11} \\
\phn& I13079-6218\_1 & 13:11:13.73 & -62:34:40.20 & J1337-6509 & J1617-5848 & 15.3/2516.8 & 2023-05-02 &  &  \\
\phn&  &  &  &  &  & 15.1/1397.8 & 2024-06-02 &  &  \\
\hline
\multirow{2}{*}{8} & I13079-6218\_2 & 13:11:09.50 & -62:34:39.70 & J1337-6509 & J1617-5848 & 15.3/2516.8 & 2023-05-02 & \multirow{2}{*}{\(0.34 \times 0.27\)} & \multirow{2}{*}{3.11} \\
\phn&  &  &  &  &  & 15.1/1397.8 & 2024-06-02 &  &  \\
\hline
\multirow{2}{*}{9} & I13080-6229 & 13:11:14.28 & -62:44:58.30 & J1337-6509 & J1617-5848 & 15.3/2516.8 & 2023-05-02 & \multirow{2}{*}{\(0.34 \times 0.27\)} & \multirow{2}{*}{2.68} \\
\phn&  &  &  &  &  & 15.1/1397.8 & 2024-06-02 &  &  \\
\hline
\multirow{2}{*}{10} & I13111-6228 & 13:14:26.49 & -62:44:28.30 & J1337-6509 & J1617-5848 & 15.3/2516.8 & 2023-05-02 & \multirow{2}{*}{\(0.34 \times 0.27\)} & \multirow{2}{*}{2.97} \\
\phn&  &  &  &  &  & 15.1/1397.8 & 2024-06-02 &  &  \\
\hline
\multirow{2}{*}{11} & I13134-6242 & 13:16:42.99 & -62:58:29.30 & J1337-6509 & J1617-5848 & 15.3/2516.8 & 2023-05-02 & \multirow{2}{*}{\(0.34 \times 0.27\)} & \multirow{2}{*}{4.93} \\
\phn&  &  &  &  &  & 15.1/1397.8 & 2024-06-02 &  &  \\
\hline
\multirow{2}{*}{12} & I13140-6226 & 13:17:15.90 & -62:42:27.00 & J1337-6509 & J1617-5848 & 15.3/2516.8 & 2023-05-02 & \multirow{2}{*}{\(0.34 \times 0.27\)} & \multirow{2}{*}{4.88} \\
\phn&  &  &  &  &  & 15.1/1397.8 & 2024-06-02 &  &  \\
\hline
\multirow{2}{*}{13} & I13291-6229\_1 & 13:32:31.77 & -62:45:11.80 & J1337-6509 & J1617-5848 & 15.3/2516.8 & 2023-05-02 & \multirow{2}{*}{\(0.34 \times 0.28\)} & \multirow{2}{*}{2.66} \\
\phn&  &  &  &  &  & 15.1/1397.8 & 2024-06-02 &  &  \\
\hline
\multirow{2}{*}{14} & I13291-6229\_2 & 13:32:34.58 & -62:45:27.00 & J1337-6509 & J1617-5848 & 15.3/2516.8 & 2023-05-02 & \multirow{2}{*}{\(0.34 \times 0.27\)} & \multirow{2}{*}{2.66} \\
\phn&  &  &  &  &  & 15.1/1397.8 & 2024-06-02 &  &  \\
\hline
\multirow{2}{*}{15} & I13291-6249 & 13:32:31.23 & -63:05:21.80 & J1337-6509 & J1617-5848 & 15.3/2516.8 & 2023-05-02 & \multirow{2}{*}{\(0.34 \times 0.27\)} & \multirow{2}{*}{7.72} \\
\phn&  &  &  &  &  & 15.1/1397.8 & 2024-06-02 &  &  \\
\hline
\multirow{2}{*}{16} & I13295-6152 & 13:32:53.49 & -62:07:49.30 & J1337-6509 & J1617-5848 & 15.3/2516.8 & 2023-05-02 & \multirow{2}{*}{\(0.34 \times 0.27\)} & \multirow{2}{*}{3.27} \\
\phn&  &  &  &  &  & 15.1/1397.8 & 2024-06-02 &  &  \\
\hline
\multirow{2}{*}{17} & G309.92+0.48 & 13:50:41.91 & -61:35:10.20 & J1337-6509 & J1427-4206 & 15.1/783.5  & 2022-06-02 & \multirow{2}{*}{\(0.65 \times 0.44\)} & \multirow{2}{*}{5.17} \\
\phn& I13471-6120 & 13:50:42.10 & -61:35:14.90 & J1408-5712 & J1617-5848 & 15.1/1397.8 & 2024-06-04 &  &  \\
\hline
\multirow{4}{*}{18} & G328.81+0.63 & 15:55:48.71 & -52:43:06.40 & J1603-4904 & J1617-5848 & 15.1/783.5  & 2022-05-30 & \multirow{4}{*}{\(0.53 \times 0.45\)} & \multirow{4}{*}{2.56} \\
\phn&  &  &  &  &  & 15.1/1301.6 & 2022-08-09 &  &  \\
\phn& I15520-5234 & 15:55:48.39 & -52:43:09.80 & J1603-4904 & J1617-5848 & 15.1/1397.8 & 2024-06-02 &  &  \\
\phn&  &  &  &  &  & 15.1/1397.8 & 2024-06-03 &  &  \\
\hline
\multirow{4}{*}{19} & G331.13-0.24 & 16:10:59.68 & -51:50:15.50 & J1603-4904 & J1617-5848 & 15.1/783.5  & 2022-05-30 & \multirow{4}{*}{\(0.53 \times 0.51\)} & \multirow{4}{*}{5.29} \\
\phn&  &  &  &  &  & 15.1/1301.6 & 2022-08-09 &  &  \\
\phn& I16071-5142 & 16:10:59.01 & -51:50:21.60 & J1603-4904 & J1617-5848 & 15.1/1397.8 & 2024-06-02 &  &  \\
\phn&  &  &  &  &  & 15.1/1397.8 & 2024-06-03 &  &  \\
\hline
\multirow{4}{*}{20} & G331.28-0.19 & 16:11:26.48 & -51:41:56.90 & J1603-4904 & J1617-5848 & 15.1/783.5  & 2022-05-30 & \multirow{4}{*}{\(0.52 \times 0.50\)} & \multirow{4}{*}{5.31} \\
\phn&  &  &  &  &  & 15.1/1301.6 & 2022-08-09 &  &  \\
\phn& I16076-5134 & 16:11:27.2 & -51:41:56.90 & J1603-4904 & J1617-5848 & 15.1/1397.8 & 2024-06-02 &  &  \\
\phn&  &  &  &  &  & 15.1/1397.8 & 2024-06-03 &  &  \\
\hline
\multirow{4}{*}{21} & G336.99-0.03 & 16:35:33.43 & -47:31:11.60 & J1631-4345 & J1617-5848 & 15.1/783.5  & 2022-05-30 & \multirow{4}{*}{\(0.68 \times 0.46\)} & \multirow{4}{*}{7.95} \\
\phn& I16318-4724 & 16:35:33.20 & -47:31:11.30 & J1650-5044 & J1617-5848 & 15.1/1397.8 & 2024-06-02 &  &  \\
\phn&  &  &  &  &  & 15.1/1397.8 & 2024-06-03 &  &  \\
\phn&  &  &  &  &  & 15.1/1397.8 & 2024-06-03 &  &  \\
\hline
\multirow{4}{*}{22} & G339.88-1.26 & 16:52:04.83 & -46:08:34.40 & J1631-4345 & J1617-5848 & 15.1/783.5  & 2022-05-30 & \multirow{4}{*}{\(0.66 \times 0.45\)} & \multirow{4}{*}{2.17} \\
\phn& I16484-4603 & 16:52:03.99 & -46:08:24.60 & J1650-5044 & J1617-5848 & 15.1/1397.8 & 2024-06-02 &  &  \\
\phn&  &  &  &  &  & 15.1/1397.8 & 2024-06-03 &  &  \\
\phn&  &  &  &  &  & 15.1/1397.8 & 2024-06-03 &  &  \\
\enddata
\tablenotetext{a}{Each source is listed with two names from different surveys. The first name follows the Galactic coordinate-based naming convention used in the MaMMOtH survey, while the second name follows the IRAS-based designation used in the QUARKS survey. Both are retained for cross-identification.}
\tablenotetext{b}{Synthesized beam sizes after final smoothing. See Sect.~\ref{sub:continuum_imaging} for details.}
\tablenotetext{c}{Distances are adopted from \citet{Liu2024QUARKS-I}.}
\end{deluxetable*}

\section{Image Alignment and Comparison of Difference Maps}
\label{appendix:alignment}
\setcounter{figure}{0}
\renewcommand{\thefigure}{B\arabic{figure}}
\setcounter{table}{0}
\renewcommand{\thetable}{B\arabic{table}}

\begin{figure}[ht]
    \centering
    \includegraphics[angle=0, width=0.75\textwidth]{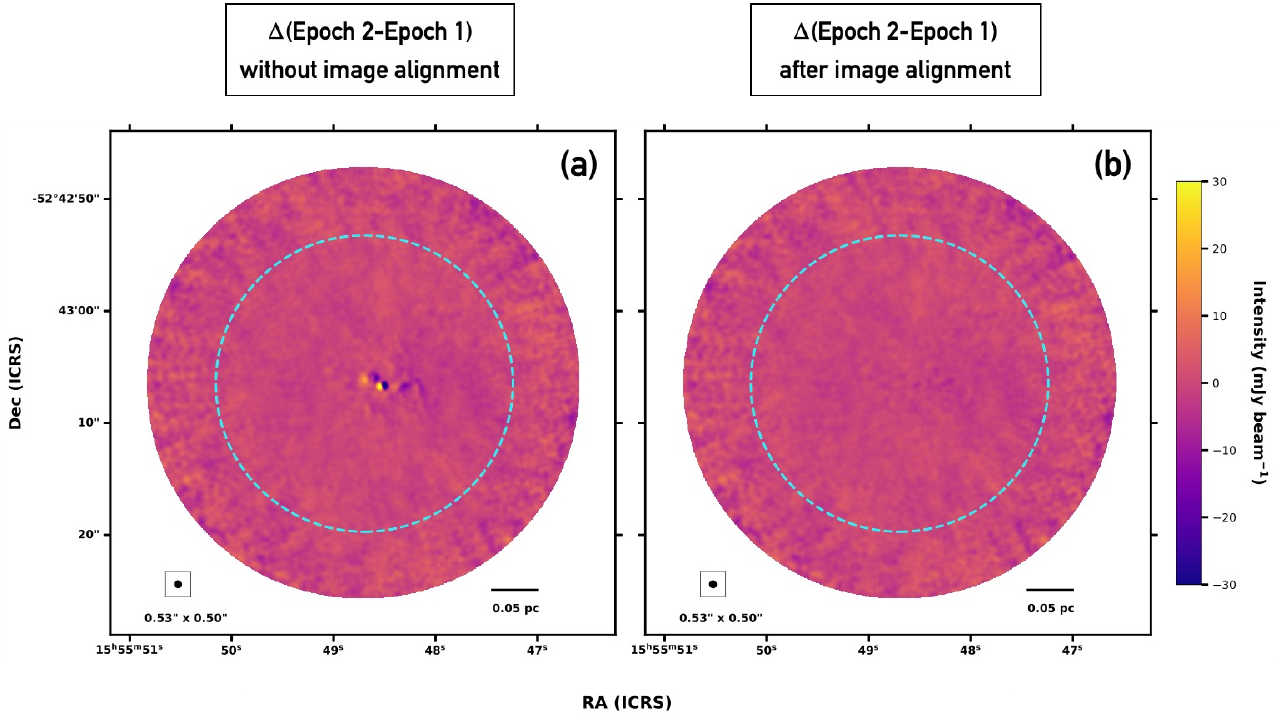}
    \caption{
    Comparison of continuum difference maps constructed before and after image alignment. Shown is the source I15520–5234, observed with ALMA as part of the MaMMOtH survey, using the difference maps from the first two observing epochs. The difference maps are constructed using primary beam corrected images. The left panel shows the difference map constructed before alignment, defined as the Epoch 2 image (observed on 2022 August 09) minus the Epoch 1 image (observed on 2022 May 30). The right panel shows the corresponding difference map after image alignment. The cyan dashed circle in each panel marks the 0.5 primary beam FWHM region of the MaMMOtH survey ($\sim$13.24$\arcsec$), which is also used as the reference area for image alignment. Details of the alignment procedure are described in Sect.~\ref{sub:core_extraction}.
    }
    \label{fig:alignment}

\end{figure}

\begin{deluxetable}{ccccc}
\tablewidth{0pt}
\setlength{\tabcolsep}{18pt}
\tablecaption{Measured Angular Offsets Between Epochs
\label{tab:offsets}}
\tablehead{
\colhead{Source Name} &
\colhead{Reference Epoch} &
\colhead{Compared Epoch} &
\colhead{$\Delta x$ (pix)} &
\colhead{$\Delta y$ (pix)}
}
\startdata
I11332-6258 & 2022-06-02 & 2022-08-21 & +1.00 & +1.00 \\
            & 2022-06-02 & 2024-05-31 & 0.00 & 0.00 \\
I12320-6122 & 2023-05-02 & 2024-06-02 & +0.28 & $-0.11$ \\
I12326-6245 & 2023-05-02 & 2024-06-02 & +0.24 & +0.22 \\
I12383-6128 & 2023-05-02 & 2024-06-02 & +0.65 & $-0.28$ \\
I12572-6316\_1 & 2023-05-02 & 2024-06-02 & +0.26 & $-0.29$ \\
I12572-6316\_2 & 2023-05-02 & 2024-06-02 & +0.42 & $-0.43$ \\
I13079-6218\_1 & 2022-06-02 & 2023-05-02 & $-1.00$ & $-1.00$ \\
               & 2022-06-02 & 2024-06-02 & 0.00 & $-1.00$ \\
I13079-6218\_2 & 2023-05-02 & 2024-06-02 & $-0.02$ & +0.01 \\
I13080-6229 & 2023-05-02 & 2024-06-02 & +0.15 & +0.13 \\
I13111-6228 & 2023-05-02 & 2024-06-02 & +0.28 & $-0.33$ \\
I13134-6242 & 2023-05-02 & 2024-06-02 & +0.00 & +0.12 \\
I13140-6226 & 2023-05-02 & 2024-06-02 & $-0.05$ & $-0.22$ \\
I13291-6229\_1 & 2023-05-02 & 2024-06-02 & $-0.04$ & +0.28 \\
I13291-6229\_2 & 2023-05-02 & 2024-06-02 & +0.23 & $-0.07$ \\
I13291-6249 & 2023-05-02 & 2024-06-02 & +0.07 & +0.02 \\
I13295-6152 & 2023-05-02 & 2024-06-02 & +0.54 & $-0.17$ \\
I13471-6120 & 2022-06-02 & 2024-06-04 & +1.00 & 0.00 \\
I15520-5234 & 2022-05-30 & 2022-08-09 & +3.00 & 0.00 \\
            & 2022-05-30 & 2024-06-02 & +2.00 & +1.00 \\
            & 2022-05-30 & 2024-06-03 & +3.00 & 0.00 \\
I16071-5142 & 2022-05-30 & 2022-08-09 & 0.00 & +1.00 \\
            & 2022-05-30 & 2024-06-02 & $-1.00$ & +1.00 \\
            & 2022-05-30 & 2024-06-03 & $-1.00$ & +2.00 \\
I16076-5134 & 2022-05-30 & 2022-08-09 & +2.00 & 0.00 \\
            & 2022-05-30 & 2024-06-02 & +1.00 & +1.00 \\
            & 2022-05-30 & 2024-06-03 & +1.00 & +1.00 \\
I16318-4724 & 2022-05-30 & 2024-06-02 & 0.00 & $-1.00$ \\
            & 2022-05-30 & 2024-06-03 & $-1.00$ & 0.00 \\
            & 2022-05-30 & 2024-06-03 & $-1.00$ & $-1.00$ \\
I16484-4603 & 2022-05-30 & 2024-06-02 & $-1.00$ & 0.00 \\
            & 2022-05-30 & 2024-06-03 & $-2.00$ & 0.00 \\
            & 2022-05-30 & 2024-06-03 & $-2.00$ & 0.00 \\
\enddata
\tablecomments{
The first epoch is adopted as the reference epoch, and all spatial offsets are measured relative to this reference. The pixel scale of the images is 0.05$\arcsec$ per pixel.}
\end{deluxetable}
\clearpage

\newpage
\section{Peak Intensities}
\label{appendix:peakintensity_parameters}
\setcounter{table}{0}
\renewcommand{\thetable}{C\arabic{table}}
\setlength{\tabcolsep}{5pt}
\startlongtable
\begin{deluxetable*}{cccccccc}
\tablecaption{Measured Peak Intensities and Coordinates of Condensations}
\label{table:epoch_data}
\tablehead{
\colhead{Source Name} &
\colhead{Condensation ID} &
\colhead{R.A. (J2000)} &
\colhead{Decl. (J2000)} &
\colhead{$I_{\mathrm{max},1}$} &
\colhead{$I_{\mathrm{max},2}$} &
\colhead{$I_{\mathrm{max},3}$} &
\colhead{$I_{\mathrm{max},4}$} \\
& &
\colhead{(h:m:s)} &
\colhead{(d:m:s)} &
\colhead{(mJy\,beam$^{-1}$)} &
\colhead{(mJy\,beam$^{-1}$)} &
\colhead{(mJy\,beam$^{-1}$)} &
\colhead{(mJy\,beam$^{-1}$)}
}
\colnumbers
\startdata
 I11332-6258 & 7  & 11:35:32.30 & -63:14:43.23 & 24.36 & 26.20 & 25.96 & \nodata \\
 I12320-6122 & 5  & 12:34:52.59 & -61:39:57.17 & 16.41 & 19.56 & \nodata & \nodata \\
 I13111-6228 & 10 & 13:14:26.37 & -62:44:30.25 & 14.36 & 24.77 & \nodata & \nodata \\
 I16076-5134 & 6  & 16:11:26.54 & -51:41:57.48 & 64.67 & 62.65 & 68.18 & 65.66 \\
 I16484-4603 & 1  & 16:52:04.67 & -46:08:34.29 & 103.42 & 97.06 & 94.84 & 93.92 \\
\hline
 I11332-6258 & 1 & 11:35:32.74 & -63:14:50.49 & 19.94 & 19.33 & 19.03 & \nodata \\
 I11332-6258 & 2 & 11:35:32.18 & -63:14:48.03 & 2.01 & 2.02 & 1.44 & \nodata \\
 I11332-6258 & 3 & 11:35:32.61 & -63:14:47.23 & 2.55 & 2.79 & 2.63 & \nodata \\
 I11332-6258 & 4 & 11:35:33.01 & -63:14:46.92 & 1.16 & 1.17 & 1.07 & \nodata \\
 I11332-6258 & 5 & 11:35:32.24 & -63:14:45.99 & 4.16 & 3.47 & 3.51 & \nodata \\
\enddata
\tablecomments{
The first five rows correspond to the variable sources identified in this work, and the remaining condensations are listed in their original order. `$\cdot\cdot\cdot$' Indicates that no observation was available for that epoch. Peak intensities are directly measured values and have not been relatively calibrated. Only a portion of this table is shown here to illustrate its form and content. The full table is available online.
}
\end{deluxetable*}

\section{Peak Intensity Ratio Maps}
\label{appendix:ratio_method}
\setcounter{figure}{0}
\renewcommand{\thefigure}{D\arabic{figure}}
\setcounter{table}{0}
\renewcommand{\thetable}{D\arabic{table}}

In this section, we present the set of peak intensity ratio maps for the sample. The first epoch is adopted as the reference. 
The uncertainty on the flux ratio is estimated by propagating the fiducial noise model introduced in Eq.~\ref{eq:SDfid}, 
\begin{equation}
\sigma_{\mathrm{ratio}}^2 = \left( \frac{\mathrm{SD}_{\mathrm{fid}}(f_{\mathrm{ref}})}{f_{\mathrm{ref}}} \right)^2 + 
\left( \frac{\mathrm{SD}_{\mathrm{fid}}(f_{n})}{f_{n}} \right)^2,
\end{equation}
where $f_{\mathrm{ref}}$ is the flux density at the reference epoch and $f_{n}$ the flux density at the $n$-th epoch. Due to the large number of sources, we only show some examples in this appendix. The complete set of figures is available online.

\vspace{2mm}
\begin{figure*}[ht!]
\gridline{
  \fig{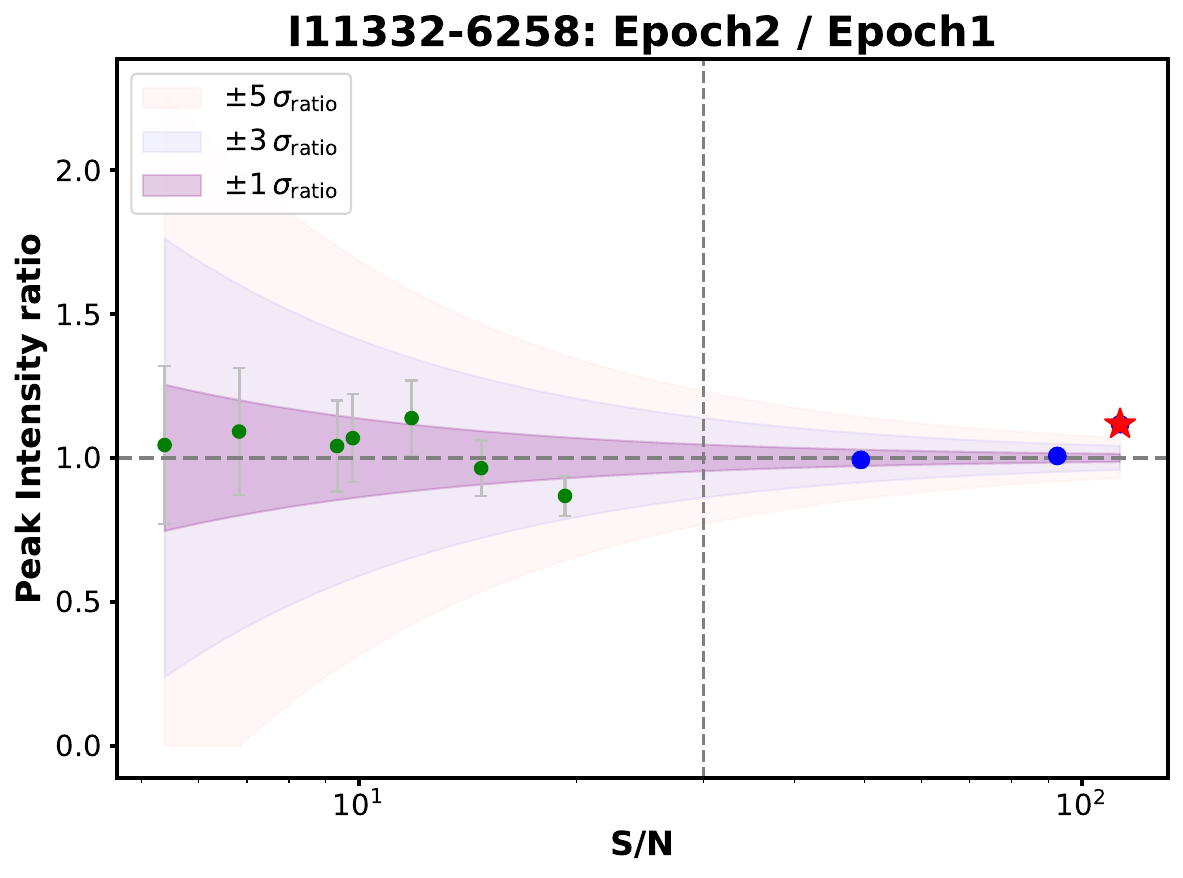}{0.32\textwidth}{}
  \fig{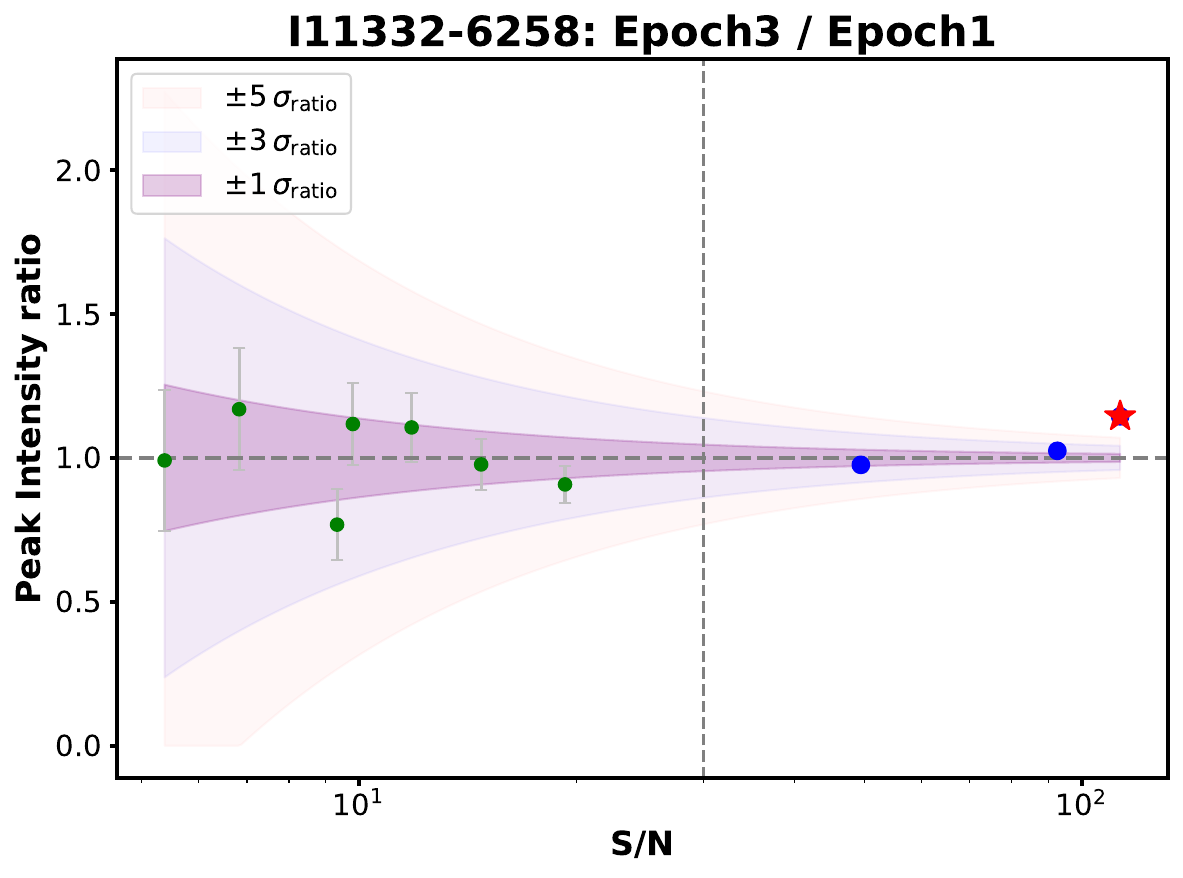}{0.32\textwidth}{}
  \fig{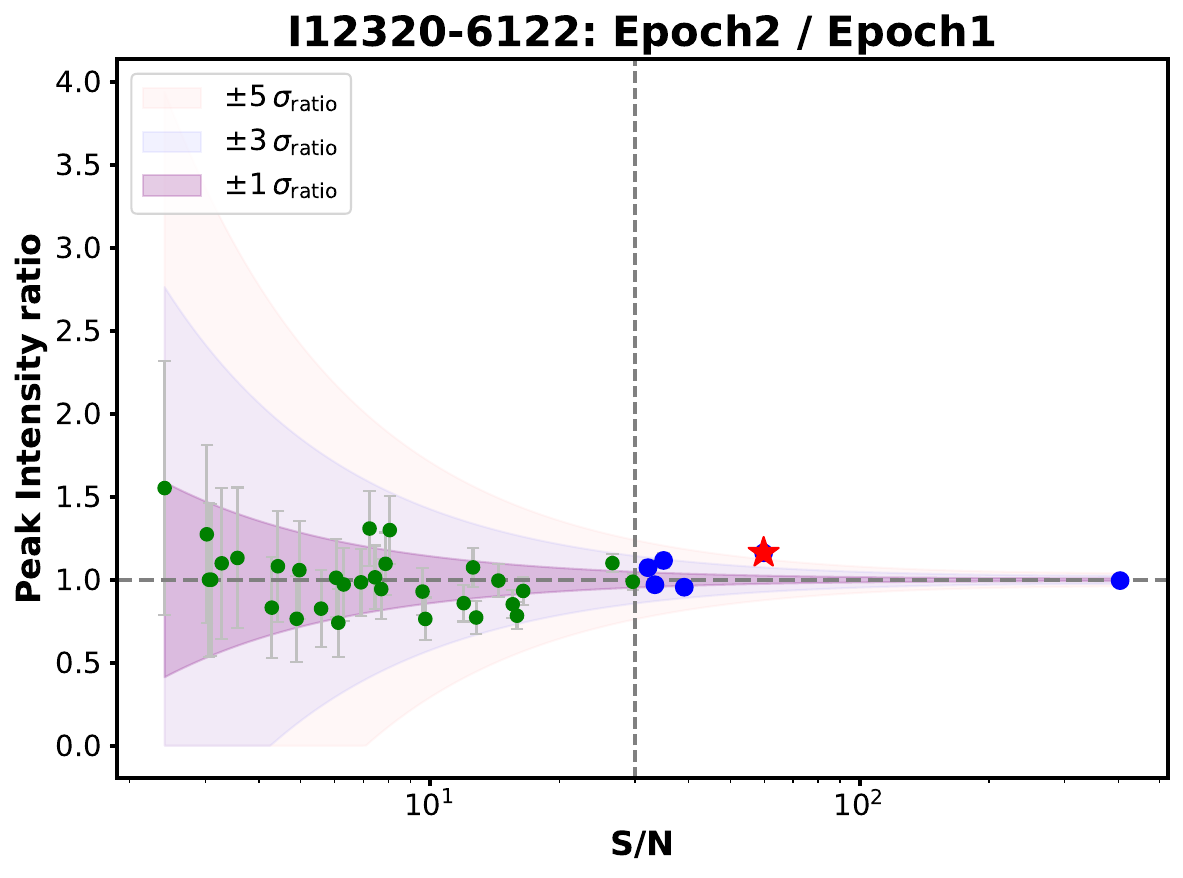}{0.32\textwidth}{}
}
\vspace{-7mm}
\caption
{Peak intensity ratio maps for the full sample (epoch $n$ relative to epoch~1). 
Shaded regions represent $\pm1\sigma_{\mathrm{ratio}}$, $\pm3\sigma_{\mathrm{ratio}}$, and $\pm5\sigma_{\mathrm{ratio}}$ intervals from the fiducial noise model. 
}
\end{figure*}

\clearpage
\section{Difference Maps for Variable Candidates}
\label{appendix:difference_map}
\setcounter{figure}{0}
\renewcommand{\thefigure}{E\arabic{figure}}

\begin{figure*}[ht!]
    \centering
    \includegraphics[angle=0, width=0.85\textwidth]{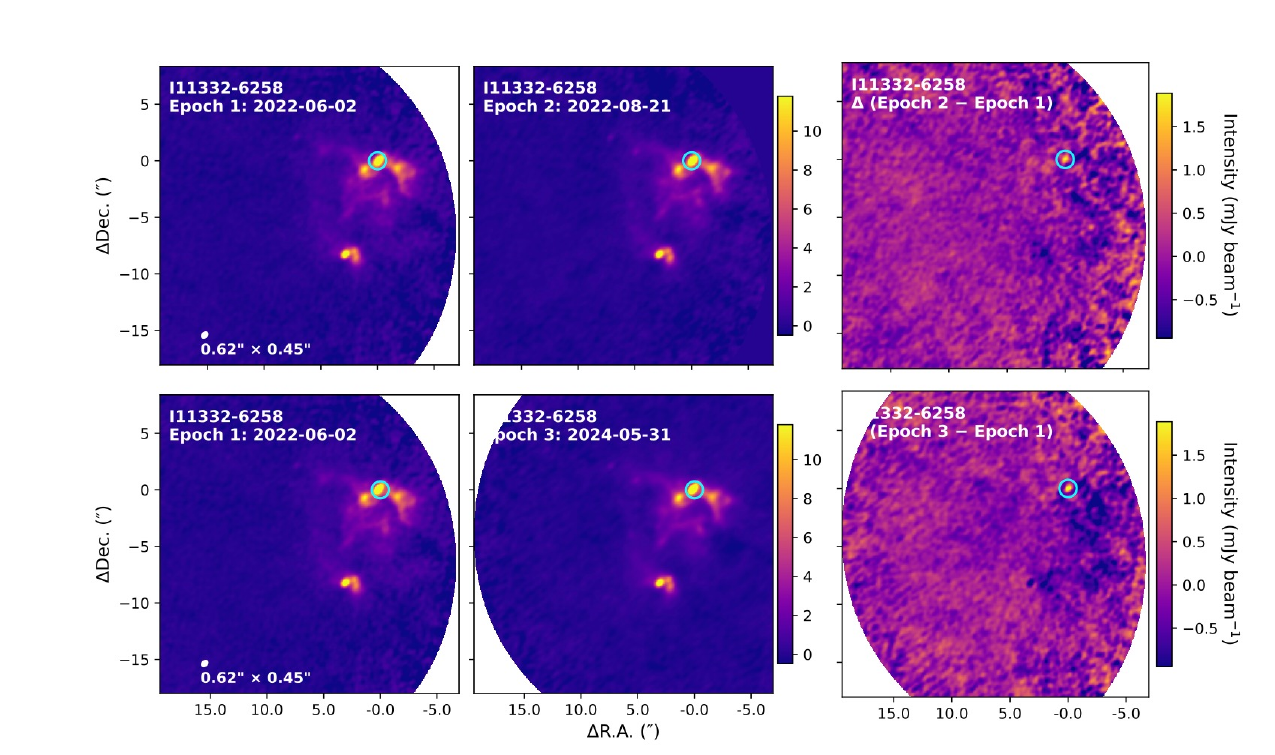} 
    \caption{
    1.3\,mm continuum images of \textbf{I11332-6258} observed with ALMA at three epochs and their corresponding difference maps. 
    \textbf{The top row} shows the Epoch~1 image (observed on 2022 June 02), the Epoch~2 image (observed on 2022 August 21), and the resulting difference map $\Delta$ (Epoch~2 - Epoch~1). 
    The rms noise level in the difference map is $\sigma_{\rm rms} = 0.32~{\rm mJy\,beam^{-1}}$, and the residual peak intensity reaches $1.92~{\rm mJy\,beam^{-1}}$, yielding ${\rm S/N} \approx 6$.
    \textbf{The bottom row} shows the Epoch~1 image, the Epoch~3 image (observed on 2024 May 31), and the resulting difference map $\Delta$ (Epoch~3 - Epoch~1). 
    The rms noise level in this difference map is $\sigma_{\rm rms} = 0.34~{\rm mJy\,beam^{-1}}$, and the residual peak intensity reaches $2.58~{\rm mJy\,beam^{-1}}$, yielding ${\rm S/N} \approx 7$.
    The cyan circle marks a radius of $0.75\arcsec$ centered on the residual peak in each difference map. 
    All panels share the same synthesized beam of $0.62\arcsec \times 0.45\arcsec$, shown as the white filled ellipse in the lower-left corner.
    }
\end{figure*}

\begin{figure*}[ht!]
    \centering
    \includegraphics[angle=0, width=0.85\textwidth]{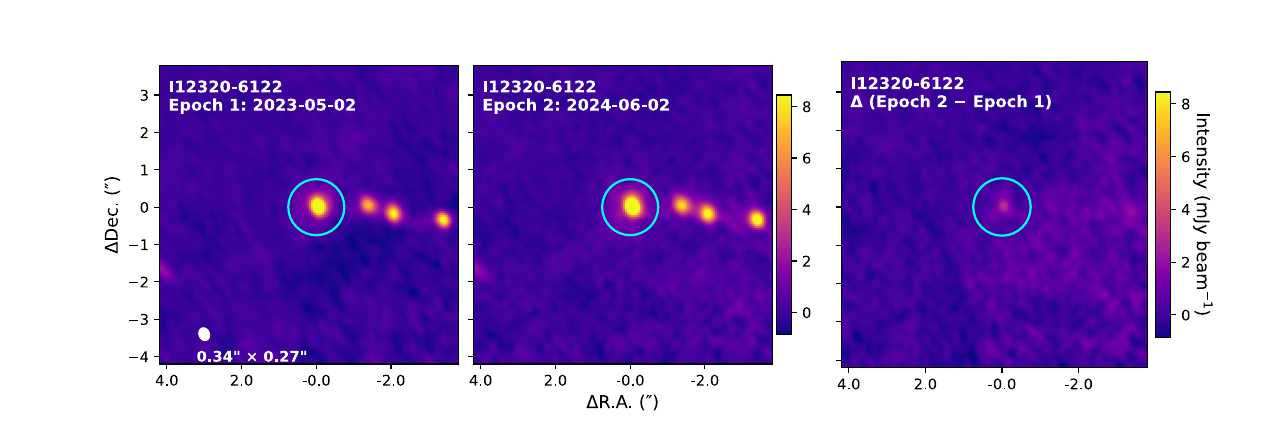} 
    \caption{
    1.3\,mm continuum images of \textbf{I12320-6122} observed with ALMA at two epochs and their corresponding difference map. 
    The left and middle panels show the Epoch~1 image (observed on 2023 May 02) and the Epoch~2 image (observed on 2024 June 02). 
    The right panel presents the difference map $\Delta$ (Epoch~2 - Epoch~1). 
    The rms noise level in the difference map is $\sigma_{\rm rms} = 0.26~{\rm mJy\,beam^{-1}}$, and the residual peak intensity reaches $3.26~{\rm mJy\,beam^{-1}}$, yielding ${\rm S/N} \approx 12$.
    The cyan circle marks a radius of $0.75\arcsec$ centered on the residual peak in the difference map. 
    All panels share the same synthesized beam of $0.34\arcsec \times 0.27\arcsec$, shown as the white filled ellipse in the lower-left corner.
    }
\end{figure*}

\begin{figure*}[ht!]
    \centering
    \includegraphics[angle=0, width=0.85\textwidth]{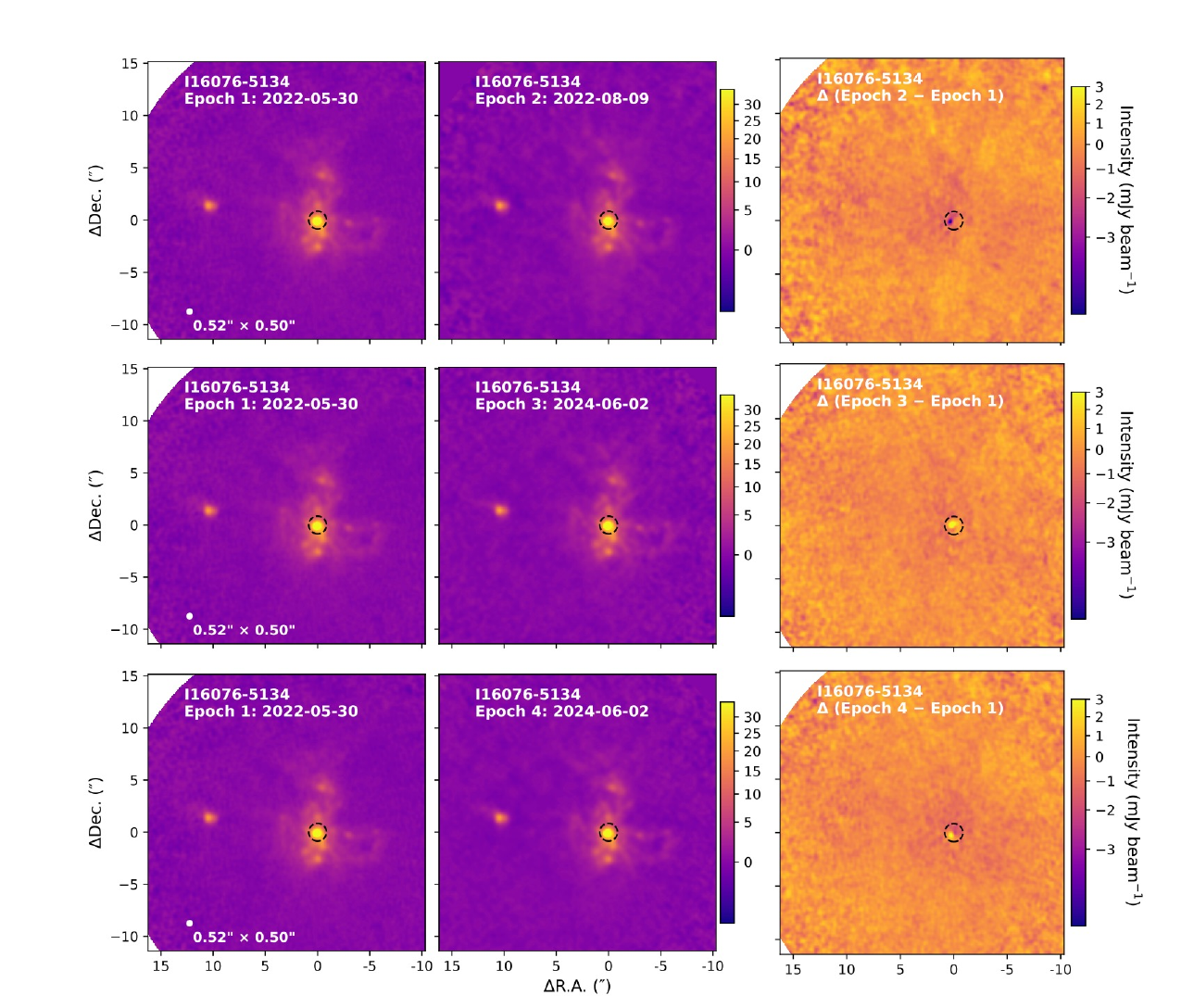} 
    \caption{
    1.3\,mm continuum images of \textbf{I16076-5134} observed with ALMA at four epochs and their corresponding difference maps.
    \textbf{The top row} shows the Epoch~1 image (observed on 2022 May 30), the Epoch~2 image (observed on 2022 August 09), and the resulting difference map $\Delta$ (Epoch~2 - Epoch~1). The rms noise level in the difference map is $\sigma_{\rm rms} = 0.52~{\rm mJy\,beam^{-1}}$, and the residual peak intensity reaches $-4.04~{\rm mJy\,beam^{-1}}$, yielding ${\rm S/N} \approx 8$.
    \textbf{The second row} shows the Epoch~1 image, the Epoch~3 image (observed on 2024 June 02), and the resulting difference map $\Delta$ (Epoch~3 - Epoch~1). 
    The rms noise level in this difference map is $\sigma_{\rm rms} = 0.49~{\rm mJy\,beam^{-1}}$, and the residual peak intensity reaches $5.36~{\rm mJy\,beam^{-1}}$, yielding ${\rm S/N} \approx 11$.
    \textbf{The third row} shows the Epoch~1 image, the Epoch~4 image (observed on 2024 June 02), and the resulting difference map $\Delta$ (Epoch~4 - Epoch~1). 
    The rms noise level in this difference map is $\sigma_{\rm rms} = 0.46~{\rm mJy\,beam^{-1}}$, and the residual peak intensity reaches $3.29~{\rm mJy\,beam^{-1}}$, yielding ${\rm S/N} \approx 7$.
    The black circle marks a radius of $0.75\arcsec$ centered on the residual peak in each difference map. 
    All panels share the same synthesized beam of $0.52\arcsec \times 0.50\arcsec$, shown as the white filled ellipse in the lower-left corner.
    }
\end{figure*}

\begin{figure*}[ht!]
    \centering
    \includegraphics[angle=0, width=0.85\textwidth]{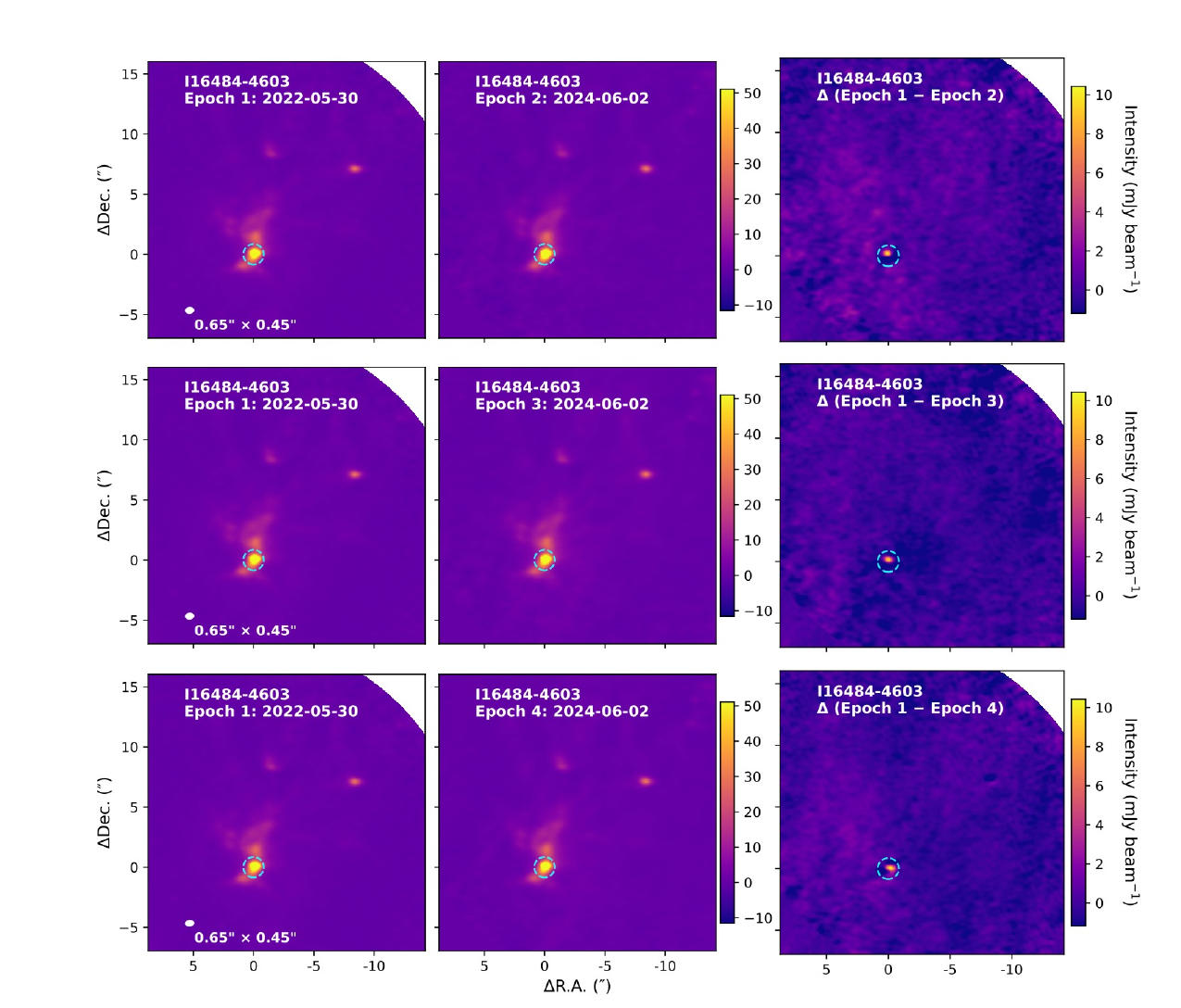} 
    \caption{
    1.3\,mm continuum images of \textbf{I16484-4603} observed with ALMA at four epochs and their corresponding difference maps. 
    \textbf{The top row} shows the Epoch~1 image (observed on 2022 May 30), the Epoch~2 image (observed on 2024 June 02), 
    and the resulting difference map $\Delta$ (Epoch~1 - Epoch~2). 
    The rms noise level in the difference map is $\sigma_{\rm rms} = 0.68~{\rm mJy\,beam^{-1}}$, and the dominant residual peak is positive, 
    reaching $9.75~{\rm mJy\,beam^{-1}}$, yielding ${\rm S/N} \approx 14$.
    \textbf{The second row} shows the Epoch~1 image, the Epoch~3 image (observed on 2024 June 02), 
    and the resulting difference map $\Delta$ (Epoch~1 - Epoch~3). 
    The rms noise level in this difference map is $\sigma_{\rm rms} = ~{\rm 0.61 mJy\,beam^{-1}}$, and the dominant residual peak is again positive, 
    reaching $9.35~{\rm mJy\,beam^{-1}}$, yielding ${\rm S/N} \approx 15$.
    \textbf{The third row} shows the Epoch~1 image, the Epoch~4 image (observed on 2024 June 02), 
    and the resulting difference map $\Delta$ (Epoch~1 - Epoch~4). 
    The rms noise level in this difference map is $\sigma_{\rm rms} = 0.58~{\rm mJy\,beam^{-1}}$, and the dominant residual peak is positive, 
    reaching $10.50~{\rm mJy\,beam^{-1}}$, yielding ${\rm S/N} \approx 18$.
    The cyan circle marks a radius of $0.75\arcsec$ centered on the residual peak in each difference map. 
    All panels share the same synthesized beam of $0.65\arcsec \times 0.45\arcsec$, shown as the white filled ellipse in the lower-left corner.
    }
\end{figure*}

\end{document}